\documentclass[english,12pt]{book}

\usepackage[latin1]{inputenc}
\usepackage[T1]{fontenc}
\usepackage[german,danish,english]{babel}

\usepackage{amsmath}
\usepackage{vmargin}
\usepackage{dsfont}
\usepackage{longtable}

\usepackage{ae,aecompl}
\usepackage{amssymb}
\usepackage{textcomp}
\usepackage{type1cm}
\usepackage{bm}
\usepackage[multiple]{footmisc}
\usepackage{multicol}
\usepackage{algorithmic}
\usepackage{algorithm}
\usepackage{ifthen}
\usepackage{layout}
\usepackage{acronym}

\usepackage{cite}
\usepackage{fancyhdr}
\usepackage{amssymb}
\usepackage{epic}
\usepackage{graphics}
\usepackage{epsfig}
\usepackage{rotating}

\usepackage{multirow}
\usepackage{hhline}
\usepackage{subfigure}
\usepackage{calc}
\usepackage{curves}
\usepackage[bf,footnotesize]{caption}
\usepackage{ae,aecompl}
\usepackage{textcomp}
\usepackage{latexsym}
\usepackage{amsfonts}

\usepackage{amsmath,amsthm}
\allowdisplaybreaks[4]
\usepackage[multiple]{footmisc}
\usepackage{graphicx}
\usepackage{index} 

\usepackage{array}
\usepackage{blkarray}
\usepackage{acronym}
\usepackage{minitoc}
\usepackage{amsthm} 

\theoremstyle{definition}

\theoremstyle{plain}

\theoremstyle{remark}

\newcommand*{\takenat}[2]{\left.#1\vphantom{{}_{#2}}\right|{}_{#2}}

\makeatletter

\newcommand{\pd}{\partial}
\newcommand{\with}{\qquad\text{with}\qquad}
\newcommand{\andx}{\qquad\text{and}\qquad}
\newcommand{\unit}{\mathds{1}} 
\newcommand{\defi}{\overset{\text{def}}{=}}
\newcommand{\trace}{\text{Tr}}
\newcommand{\tr}{\text{tr}}

\newcommand{\lrx}[1]{\left({#1}\right)}
\newcommand{\lry}[1]{\left[{#1}\right]}

\renewcommand{\chaptermark}[1]%
{\markboth{\chaptername\ \thechapter\ #1}{}}
\renewcommand{\sectionmark}[1]%
{\markright{\thesection\ #1}}

\begin{document}
\pagestyle{empty}

\renewcommand\thefootnote{$\star$}

\noindent{\LARGE\bf Field theory on a Non--commutative Plane: a Non--perturbative Study
\footnote{Slightly modified PhD thesis, accepted at Humboldt--Universit\"at zu Berlin in March 2003, defended
in June 2003.}}\\

\renewcommand\thefootnote{\arabic{footnote}}

\vspace{.5cm}
\noindent{\bf Frank Hofheinz}\\
Institut f\"ur Physik, Humboldt--Universit\"at zu Berlin\\
Newtonstr.\ 15, D--12489 Berlin, Germany\\
E--mail: hofheinz@physik.hu-berlin.de\\

\vspace{.5cm}
\noindent{\bf Abstract:}
The 2d gauge theory on the lattice is equivalent to the twisted
Eguchi--Kawai model, which we simulated at $N$ ranging from 25 to 515.
We observe a clear large $N$ scaling for the 1-- and 2--point function
of Wilson loops, as well as the 2--point function of Polyakov lines.
The 2--point functions agree with a universal wave function
renormalization.  The large $N$ double scaling limit corresponds to
the continuum limit of non--commutative gauge theory, so the observed
large $N$ scaling demonstrates the non--perturbative renormalizability
of this non--commutative field theory.  The area law for the Wilson
loops holds at small physical area as in commutative 2d planar gauge
theory, but at large areas we find an oscillating behavior instead. In
that regime the phase of the Wilson loop grows linearly with the area.
This agrees with the Aharonov--Bohm effect in the presence of a
constant magnetic field, identified with the inverse
non--commutativity parameter.

Next we investigate the 3d $\lambda \phi^4$ model with two
non--commutative coordinates and explore its phase diagram.  Our
results agree with a conjecture by Gubser and Sondhi in $d=4$, who
predicted that the ordered regime splits into a uniform phase and a
phase dominated by stripe patterns.  We further present results for
the correlators and the dispersion relation. In non--commutative field
theory the Lorentz invariance is explicitly broken, which leads to a
deformation of the dispersion relation. In one loop perturbation
theory this deformation involves an additional infrared divergent
term.  Our data agree with this perturbative result.

We also confirm the recent observation by Ambj\o rn and Catterall that
stripes occur even in $d=2$, although they imply the spontaneous
breaking of the translation symmetry.\\


\cleardoublepage
\pagestyle{plain}

\selectlanguage{english}
\pagenumbering{arabic}
\pagestyle{fancy}
\fancyhead{}
\fancyhead[LE,RO]{\bfseries\thepage}
\fancyhead[LO]{\bfseries Contents}
\fancyhead[RE]{\bfseries Contents}
\tableofcontents
\cleardoublepage


\chapter*{1 Introduction}
\label{introduction}

\setcounter{chapter}{1}
\addcontentsline{toc}{chapter}{1 \hspace{.25cm}Introduction}
\pagestyle{fancy}
\fancyhead{}
\fancyhead[LE,RO]{\bfseries\thepage}
\fancyhead[LO]{\bfseries 1 Introduction}
\fancyhead[RE]{\bfseries 1 Introduction}

The ideas of non--commutative space--time and field theories defined
on such a structure started already in the year 1947. At this time the
concept of renormalization was not yet well established and therefore
ultraviolet divergences in quantum field theory still caused serious
problems. To solve these problems or at least weaken them Snyder
introduced the quantized space--time \cite{Snyder:1947qz} (see also
\cite{Yang:1947}).

The plan was to define quantum field theories on a space--time which
is smeared out at very small length scales. This means that in
addition to Heisenberg's uncertainty relation between coordinates
and momenta there is a uncertainty between different coordinates.

As in the quantization of the classical phase space, space--time can be
quantized by replacing the usual coordinates $x_\mu$ by Hermitian
operators $\hat{x}_\mu$, obeying the commutator relation
\begin{equation}
  \label{eq:intro-commutator}
  \lry{\hat{x}_\mu,\hat{x}_\nu}=i\Theta_{\mu\nu}\,.
\end{equation}
The non--commutativity tensor $\Theta_{\mu\nu}$ is a real--valued
antisymmetric $d\times d$ matrix and $d$ is the space--time dimension.
Since the coordinate operators do not commute they cannot be
diagonalized simultaneously and therefore induce the uncertainty
relation
\begin{equation}
  \label{eq:intro-uncertainty}
  \Delta x_\mu \Delta x_\nu \geq\frac{1}{2} |\Theta_{\mu\nu}|\,.
\end{equation}
This uncertainty implies a quantum structure of space--time and due to
the lack of points in space--time it then represents an effective
ultraviolet cut--off.

Much later, in 1996, it was shown by Filk \cite{Filk:1996dm} that in
field theory on a non--commutative plane the divergences of
commutative field theory still occur. In addition to those divergences
the authors of Refs.\ \cite{Minwalla:1999px,Varilly:1998gq} found in 1999
that
there is a mixing of ultraviolet and infrared divergences.\\

The concept of quantized space--time has not been followed further in
the early days of quantum field theory, since the renormalization
technique became more and more successful. It came up again first in
the 80's, when Connes formulated the mathematically rigorous
\cite{Connes:1987} framework of non--commu\-ta\-tive geometry. In physical
theories a non--commutative space--time first appeared in string
theory, namely in the quantization of open strings
\cite{Abouelsaood:1987gd}. In an constant magnetic background field
the boundary conditions change and the zero momentum modes of the
string do not commute anymore. Instead they obey a commutation
relation of the type (\ref{eq:intro-commutator}), where
$|\Theta_{\mu\nu}|$ is proportional to the inverse background field.

The zero momentum modes of an open string can be interpreted as the
end points of the string, which are confined to a submanifold, i.e.\ 
a D--brane. The commutator (\ref{eq:intro-commutator}) implies a
non--commutative geometry on the branes. Hence a quantized space--time
appears naturally in string
theory \cite{Seiberg:1999vs}.\\

Another field of interest, where the non--commutativity of space--time
could play an important role, is quantum gravity. It is an old believe
that for a quantized theory of gravitation space--time has to change
its nature on very small length scales. The synthesis of the
principles of quantum mechanics and of classical general relativity
leads to a space--time uncertainty
\cite{Doplicher:1994zv,Doplicher:1995tu},  which implies that any
theory of quantum gravity will not be local in the conventional sense
\cite{Douglas:2001ba}.
Such effects could be modeled by a non--commutative space--time.\\

Also in condensed matter physics the concept of non--commutative
space--time is applied. The theory of electrons in a magnetic field
projected to the lowest Landau level can be naturally described by a
non--commutative field theory
\cite{Girvin:1987fp,Bellissard1994,Girvin:1999,Susskind:2001fb,Polychronakos:2001uw},
where $|\Theta_{\mu\nu}|$ is again proportional to the inverse
magnetic field. These ideas are relevant for the quantum Hall effect.
For the integer quantum Hall effect, a non--commutative treatment
serves as an alternative description to standard condensed matter
techniques. This is already remarkable, since it is the first
application of non--commutativity geometry that provides
phenomenological results. However, with these methods only the known
results are reproduced; it does not provide new insight in the nature
of the integer quantum Hall effect.

This may be different in the case of the {\em fractional} quantum Hall
effect. That effect is not well understood from the theoretical point
of view. Here a non--commutative field theory is considered by many
researchers as the most promising candidate for a description.\\

One may also try to study the non--commutative analog of pure
Yang--Mills theories or of QED and QCD. Such theories can be
considered as an extended standard model and a study of them could
allow for an experimental verification or falsification. Hence results
from these extended theories may provide sensible tests of a quantized
space--time.\\

The above described applications of non--commutative field theory
suffer from the ultraviolet/infrared (UV/IR) mixing. This effect
causes still severe problems in a perturbative treatment. Our
goal is to study non--commutative field theories on a
completely non--perturbative level.

This work represents the first non--perturbative study of
non--commutative field theories.  As usual when entering a new topic
we studied toy models, which share important properties of the full
theory. In this work we studied field theory in lower dimensions and
we focused on basic properties of these theories. The results
presented here are published in Refs.\
\cite{Bietenholz:2002ch,Bietenholz:2002vj,Bietenholz:2002ev}.

In a two dimensional gauge theory we address the problem of
renormalizability. 
It is an interesting question whether this
model can be renormalized non--perturbatively.
In the three dimensional $\lambda\phi^4$ theory we studied the effects
of UV/IR mixing. Our main interest was here the phase diagram of this
theory and the question if there is a phase with spontaneously broken
translation invariance, as it had been conjectured from analytic
results.  In addition we studied the dispersion relation in
this theory, for which perturbation theory predicts a deformation due 
to the non--commutativity.\\


This work is organized as follows: in Chapter \ref{cha:NCFT} we give
an introduction to non--commutative field theory. We concentrate here
on the main differences compared to commutative field theory. In a
first Section we set up the non--commutative geometry on which we
define a scalar field theory and a pure gauge theory (Sections
\ref{scalar} and \ref{gauge}, respectively). These are the theories
studied in this work. In Section \ref{standard} we briefly
comment on the extension to the non--commutative standard model. In
addition to the motivation already given in the introduction we want
to motivate the study of non--commutative space--time from this point
of view.

Chapter \ref{cha:lattice} is dedicated to the lattice regularization.
As we already mentioned in this introduction the non--commutativity of
space--time does not cure the ultraviolet divergences, and therefore
one still has to regularize the theory. To this end we will introduce a
momentum cut--off via discretization of space--time.

In Chapter \ref{cha:TEK} and \ref{phi} we present the two models we
investigated; a two dimensional non--commutative pure gauge theory and
a three dimensional scalar field theory. The explicit construction of
the lattice actions as well as the Monte Carlo results of our studies
are presented in these Chapters. In Chapter \ref{A-2d} we show results
on a two dimensional scalar theory, and in Chapter \ref{conclusion} we
summarize our results and give an outlook.  For the sake of continuity
the details of the simulations are banned
to Appendix \ref{numerics}.\\

Note that throughout this work we always work in Euclidean
space--time.  We should mention here that in contrast to the
commutative case, it is an open question if the Euclidean version of
non--commutative field theory can be interpreted in the Minkowski
world, since there is no equivalent of the Osterwalder--Schrader
theorem \cite{Osterwalder:1975tc}. However, in non--commutative field
theory with a commuting time coordinate it is generally believed that
this interpretation exists. For a discussion of Wick rotation and the
related question of unitarity, see e.g.\ Refs.\ 
\cite{Bahns:2002vm,Rim:2002if,Liao:2002pj}.


\cleardoublepage

\lhead[\fancyplain{}{\bfseries\thepage}]%
{\fancyplain{}{\bfseries\rightmark}}
\rhead[\fancyplain{}{\bfseries\leftmark}]%
{\fancyplain{}{\bfseries\thepage}} 
\chead{}
\lfoot{}
\cfoot{}
\rfoot{}
\renewcommand{\sectionmark}[1]{\markright{\thesection\ #1}}
\renewcommand{\chaptermark}[1]%
{\markboth{\chaptername\ \thechapter\ #1}{}}


\chapter{Non--commutative field theory}
\label{cha:NCFT}

In this Chapter we give an introduction to the concept of
non--commutative space--time and field theories defined on it.  We
will work out the main differences to their commutative counterparts
and discuss the additional problems that arise in such theories.  
For a general discussion of non--commutative space--time see for
example Refs.\ \cite{Connes:1998cr,Douglas:2001ba}. We follow here the
discussion in Refs.\ \cite{Ambjorn:2000cs,Szabo:2001kg}.

\section{Non--commutative flat space--time}
\label{NC}

In this Section we discuss those features of non--commutative
geometry, which are needed to define field theories on such a
geometry. We will find two alternative formulations, in terms of Weyl
operators and in terms of functions with a deformed multiplication.

\subsection{Weyl operators}

Let us start with the commutative algebra of complex valued functions
on $d$ dimensional Euclidean space--time $\mathbb{R}^d$.  An element
of this algebra corresponds to a configuration of a scalar field,
with pointwise addition and multiplication.  We consider here functions
of sufficiently rapid decrease at infinity, so that any function $f(x)$
may be described by its Fourier transform
\begin{equation}
  \label{eq:NC-fourier}
  \tilde{f}(k)=\int d^d x\,e^{-ix_\mu k_\mu}f(x)\,.
\end{equation}

A non--commutative space--time can be defined by replacing the local
coordinates $x\in\mathbb{R}^d$ by Hermitian operators $\hat{x}_\mu$
satisfying
\begin{equation}
  \label{eq:NC-1}
    [\hat{x}_\mu,\hat{x}_\nu] = i\Theta_{\mu\nu}\,.
\end{equation}
The non--commutativity tensor $\Theta_{\mu\nu}$ is antisymmetric with
the dimension length squared and it can in general depend on
space--time.  Here we restrict ourselves to a constant
non--commutativity tensor parametrized by the non--commutativity
parameter $\theta$
\begin{equation}
  \label{eq:NC-theta-para}
  \Theta_{\mu\nu}=\theta\,
  \begin{pmatrix}
    0&1\\-1&0
  \end{pmatrix}
  \otimes\unit_{d/2}\,.
\end{equation}
Here we assume the space--time dimension $d$ to be even. The $\hat{x}_\mu$
generate a non--commuta\-tive and associative algebra of operators.
Elements of this algebra, the Weyl operators $W[f]$, can be
constructed by a formal Fourier transform involving the operators
$\hat{x}_\mu$ and the ordinary Fourier transform of $f(x)$
\begin{equation}
  \label{eq:NC-operator}
  W[f]=\int\frac{d^dk}{\lrx{2\pi}^d}\,\tilde{f}(k)\,e^{ik_\mu\hat{x}_\mu}\,.
\end{equation}
Combining equations (\ref{eq:NC-fourier}) and (\ref{eq:NC-operator}) we
find an explicit map $\Delta(x)$ between operators and fields
\begin{equation}
  \label{eq:NC-delta}
  W[f]=\int d^dx\,f(x)\,\Delta(x)\qquad\text{with}\qquad
  \Delta(x)=\int\frac{d^dk}{\lrx{2\pi}^d}\,e^{ik_\mu\hat{x}_\mu}\,e^{-ik_\mu x_\mu}\,,
\end{equation}
where $\Delta(x)$ is a Hermitian operator that can be understood as a
mixed basis for operators of fields.

We may define a linear and anti--Hermitian derivative $\hat{\pd}_\mu$
on the algebra of Weyl operators by the commutator relations
\begin{equation}
  \label{eq:NC-derivative}
  [\hat{\pd}_\mu,\hat{x}_\nu]=\delta_{\mu\nu},\qquad[\hat{\pd}_\mu,\hat{\pd}_\nu]=ic_{\mu\nu}\,,
\end{equation}
where $c_{\mu\nu}$ is a real valued c--number.  With this definition
of the derivative one can show that
\begin{equation}
  \label{eq:NC-delta-1}
  [\hat{\pd}_\mu,\Delta(x)] = -\pd_\mu\Delta(x)\,.
\end{equation}
Together with equation (\ref{eq:NC-delta}) and integration by
parts one obtains that the derivative of Weyl operators is equal to
the Weyl operator of the usual derivative of the functions
\begin{equation}
  \label{eq:NC-derivative-1}
  [\hat{\pd}_\mu,W[f]]=\int d^dx\; \pd_\mu f(x) \Delta(x) = W[\pd_\mu f]\,.
\end{equation}
Any global translation $x+v$ with $v\in\mathbb{R}^d$ can
be obtained with the unitary operators $\exp(v_\mu\hat{\pd}_\mu)$
\begin{equation}
  \label{eq:NC-translation}
  \Delta(x+v)=e^{v_\mu\hat{\pd}_\mu}\Delta(x)e^{-v_\nu\hat{\pd}_\nu}\,.
\end{equation}
This follows directly from the commutator relation
(\ref{eq:NC-delta-1}). This property implies that any trace
$\text{Tr}\Delta(x)$, with $\text{Tr}$ defined on the algebra of Weyl
operators, is independent of $x\in\mathbb{R}^d$. Together with
equation (\ref{eq:NC-delta-1}) it follows that the trace $\text{Tr}$
is unambiguously given by an integration over space--time
\begin{equation}
  \label{eq:NC-trace}
  \text{Tr}\,W[f]=\int d^dx \,f(x),
\end{equation}
with the normalization $\text{Tr}\Delta(x)=1$.

In Ref.\ \cite{Szabo:2001kg} it is shown that if $\Theta$ is
invertible (which implies that the dimension $d$ of space--time has to
be even) the product of two operators $\Delta$ at distinct points
can be computed as follows
\begin{equation}
  \label{eq:NC-delta3}
  \Delta(x)\Delta(y)=\frac{1}{\pi^d\det{\Theta}}\int d^dz\,\Delta(z)
  e^{-2i(\Theta^{-1})_{\mu\nu}(x-z)_\mu(y-z)_\nu}\,.
\end{equation}
Together with the normalization of the trace the operators
$\Delta(x)$ form an orthonormal set for $x\in\mathbb{R}^d$,
\begin{equation}
  \label{eq:NC-delta-2}
  \text{Tr}\lrx{\Delta(x)\Delta(y)}=\delta^d(x-y)\,.
\end{equation}
With this property of $\Delta(x)$ we can
define the inverse map to (\ref{eq:NC-delta})
\begin{equation}
  \label{eq:NC-inverse-map}
  f(x)=\text{Tr}\lrx{W[f]\Delta(x)}.
\end{equation}
This one--to--one correspondence can be thought of as an analog of the
operator--state correspondence in quantum mechanics.

\subsection{The star--product}

Let us now consider the product of two Weyl operators $W[f]W[g]$
corresponding to the two functions $f(x_1)$ and $g(x_2)$. We want to
transform this product into the coordinate space representation with the
help of the inverse map (\ref{eq:NC-inverse-map}),
\begin{equation}
  \label{eq:star-inverse}
  h(x)=\trace\lrx{W[f]W[g]\Delta(x)}\,.
\end{equation}
To achieve this we rewrite the product in terms of the map
(\ref{eq:NC-delta})
\begin{equation}
  \label{eq:star-inverse2}
  \begin{split}
    W[f]W[g]&=\int d^dx_1\int d^dx_2\,f(x_1)g(x_2)\Delta(x_1)\Delta(x_2)\\
    &=\frac{1}{\pi^d\det{\Theta}}\int d^dx_1\int d^dx_2\int d^dx_3\\
    &\phantom{\frac{1}{\pi^d\det{\Theta}}}
    \times f(x_1)g(x_2)\Delta(x_3)e^{-2i(\Theta^{-1})_{\mu\nu}(x_1-x_3)_\mu(x_2-x_3)_\nu}\,,
  \end{split}
\end{equation}
where in the second line equation (\ref{eq:NC-delta3}) was used.
Multiplying both sides with $\Delta(x)$ from the right and taking the
trace leads to
\begin{equation}
  \label{eq:star-product}
  \begin{split}
  h(x)&=\trace\lrx{W[f]W[g]\Delta(x)}\\
  &=\frac{1}{\pi^d\det{\Theta}}\int d^dx_1\int d^dx_2
  f(x_1)g(x_2)e^{-2i(\Theta^{-1})_{\mu\nu}(x_1-x)_\mu(x_2-x)_\nu}\\
  &=f(x)\exp\lrx{\frac{i}{2}\overset{\leftarrow}{\pd_\mu}\Theta_{\mu\nu}
    \overset{\rightarrow}{\pd_\nu}}g(x) \defi f(x)\star g(x)\;,
  \end{split}
\end{equation}
where we used the completeness relation (\ref{eq:NC-delta-2}). 
Using this product we obtain
\begin{equation}
  \label{eq:star-inverse3}
  W[f]W[g]=W[f\star g]\,,
\end{equation}
i.e.\ the product of Weyl operators is equal to the Weyl operator of
the star--products of functions in coordinate space.  With the
star--product we can rewrite the commutation relation between
space--time operators (\ref{eq:NC-1}) in terms space--time coordinates
\begin{equation}
  \label{eq:NC-comm-star}
  \lry{x_\mu,x_\nu}_\star=x_\mu\star x_\nu-x_\nu\star x_\mu=i\Theta_{\mu\nu}\,.
\end{equation}

The star--product is associative but non--commutative. For
$\Theta_{\mu\nu}=0$ it reduces to the ordinary product of functions.
It can be thought of as a deformation of the algebra of functions on
$\mathbb{R}^d$ to a non--commutative algebra, with the same elements
and addition law, but with a different multiplication law given by
(\ref{eq:star-product}). Note that the commutator of a function $f(x)$
with the coordinates $x_\mu$ can be used to generate derivatives
\begin{equation}
  \label{eq:NC-note}
  x_\mu\star f(x) - f(x)\star x_\mu = i\Theta_{\mu\nu}\pd_\nu f(x)\,.
\end{equation}

Due to the cyclicity of the trace defined in equation
(\ref{eq:NC-trace}) the integral
\begin{equation}
  \label{eq:NC-cycle}
  \trace\lrx{W[f_1]W[f_2]\dots W[f_n]}=\int d^dx f_1(x)\star f_2(x)\star\dots\star f_n(x)
\end{equation}
is invariant under cyclic permutations of the functions $f_i(x)$. In
particular the integral of the star--product of two
functions is identical to the integral of the ordinary product
of two functions\\
\begin{equation}
  \label{eq:NC-cycle2}
  \int d^dx f_1(x)\star f_2(x)=\int d^dx f_1(x)f_2(x)\,.
\end{equation}
We have now two ways to encode non--commutative space--time:

\begin{itemize}
\item we can use ordinary products in the non--commutative
  $C^*$--algebra of Weyl operators,
\item or we may deform the ordinary product of the
  commutative $C^*$--algebra of functions in $\mathbb{R}^d$ to the
  non--commutative star--product.
\end{itemize}

\subsection{The non--commutative torus}
\label{sec:NC-torus}

In this Subsection we briefly discuss the case when space--time is a
$d$--dimensional torus $\mathbb{T}^d$ instead of $\mathbb{R}^d$.
For a more detailed discussion, see \cite{Ambjorn:2000cs,Szabo:2001kg}.

Let us consider functions $f(x)$ on a periodic torus
\begin{equation}
  \label{eq:NCT-period}
  f(x+\Sigma_{\mu\nu}\,\hat{\mu})=f(x)\with \nu=1,\dots,d\,.
\end{equation}
$\hat{\mu}$ is the unit vector in the $\mu$ direction and
$\Sigma_{\mu\nu}$ is the $d\times d$ period matrix of the torus.
Due to this periodicity the momenta $k_\mu$ in
(\ref{eq:NC-fourier}) are discretized according to
\begin{equation}
  \label{eq:NCT-momenta}
  k_\mu=2\pi\lrx{\Sigma^{-1}}_{\nu\mu}m_\nu\with m_\nu\in\mathbb{Z}\,.
\end{equation}
Using the discrete version of Fourier transform (\ref{eq:NC-fourier})
we can define a mapping from fields to operators in the same way as we
did in flat $\mathbb{R}^d$. The result is
\begin{equation}
  \label{eq:NCT-mapping}
  \Delta(x)=\frac{1}{|\det{\Sigma}|}\sum_{\vec{m}\in\mathbb{Z}^d}\lrx{\prod_{\nu=1}^d\lrx{\hat{Z}_\nu}^{m_\nu}}
  e^{-\pi i\sum_{\nu<\rho}\tilde{\Theta}_{\nu\rho}m_\nu m_\rho}e^{-2\pi i (\Sigma^{-1})_{\nu\mu}m_\nu x_\mu}\,,
\end{equation}
where we introduced the dimensionless non-commutativity tensor
$\tilde{\Theta}$
\begin{equation}
  \label{eq:NCT-theta}
  \tilde{\Theta}_{\rho\sigma}=2\pi\lrx{\Sigma^{-1}}_{\rho\mu}\Theta_{\mu\nu}\lrx{\Sigma^{-1}}_{\sigma\nu}
\end{equation}
and the operators $\hat{Z}_\nu$
\begin{equation}
  \label{eq:NCT-Z}
  \hat{Z}_\nu=e^{2\pi i(\Sigma^{-1})_{\nu\mu}\hat{x}_\mu}
  \with \hat{Z}_\mu\hat{Z}_\nu=e^{-2\pi i \tilde{\Theta}_{\mu\nu}}\hat{Z}_\nu\hat{Z}_\mu\,.
\end{equation}
With this mapping we find the one--to--one correspondence
(\ref{eq:NC-delta}) and (\ref{eq:NC-inverse-map}) also on the
non--commutative torus. These definitions will reappear in Section
\ref{sec:lat-dsp}, where we discuss the discrete torus.


\section{Non--commutative scalar field theory}
\label{scalar}

Having defined the algebra of functions of non--commutative
space--time we are able to define a scalar field theory on
this geometry. At this point we make a change of notation
and introduce the short--hand notation for Weyl operators
$W[f]\rightarrow \hat{f}$.

\subsection{Non--commutative scalar action}
\label{sec:sc-ncsa}

We start our discussion with the action of an Euclidean commutative
$\lambda\phi^4$ theory,
\begin{equation}
  \label{eq:sc-1-c_action}
  S[\phi]=\int d^dx \lrx{\frac{1}{2}\,\pd_\mu\phi(x)\pd_\mu\phi(x)
    +\frac{m^2}{2}\phi^2(x)+\frac{\lambda}{4}\phi^4(x)}\,,
\end{equation}
where $\phi$ is a real valued scalar field and $d$ is the dimension of space--time.\\

To transform an ordinary scalar field theory to a non--commutative
field theory we can use one of the procedures described in the last
Section. Either we may use the Weyl quantization via Hermitian operators
$\hat{\phi}$, or we use the deformation of the product into the
star--product (\ref{eq:star-product}).

The quantum field theory written in terms of Weyl operators $\hat{\phi}$,
corresponding to a real scalar field $\phi(x)$ on $\mathbb{R}^d$,
reads
\begin{equation}
  \label{eq:sc-action1}
  \begin{split}
    Z&=\int d\hat{\phi}\exp\lrx{-S[\hat{\phi}]}\\
    S[\hat{\phi}]&=\trace\lrx{\frac{1}{2}[\hat{\pd_\mu},\hat{\phi}]^2+\frac{m^2}{2}\hat{\phi}^2
      +\frac{\lambda}{4}\hat{\phi}^4}\,,
  \end{split}
\end{equation}
where the kinetic term is a direct consequence of equation
(\ref{eq:NC-derivative-1}) (it involves a sum over $\mu$). The measure
$d\hat{\phi}$ is here the ordinary path integral
measure for scalar fields $\mathcal{D}\phi$.

This theory may be formulated in coordinate space by applying the map
(\ref{eq:NC-inverse-map}) to the action (\ref{eq:sc-action1}) and using
equation (\ref{eq:star-inverse3}),
\begin{equation}
  \label{eq:sc-action2}
  \begin{split}
    S[\phi]&=\int d^dx\lry{\frac{1}{2}\lrx{\pd_\mu\phi(x)}^2+\frac{m^2}{2}\phi(x)^2
      +\frac{\lambda}{4}\phi(x)\star\phi(x)\star\phi(x)\star\phi(x)}\;.
  \end{split}
\end{equation}
the kinetic term and the mass term do {\em not} contain the
star--product, because of the property (\ref{eq:NC-cycle2}) of the
star--product.  As a consequence the commutative and the
non--commutative theory coincide for free fields.  The difference
arises due to the self--interaction term
\begin{equation}
  \label{eq:sc-interaction}
  \begin{split}
      \trace(\hat{\phi}^4)&=\int d^dx\, {\phi}(x)\star{\phi}(x)\star{\phi}(x)\star{\phi}(x)\\
      &=\int\lrx{\prod_{a=1}^4\frac{d^dk^a}{(2\pi)^d}}(2\pi)^d
      \delta^d\lrx{\sum_{a=1}^4k^a}\lrx{\prod_{a=1}^4\tilde{\phi}(k^a)}
      V(k_1,k_2,k_3,k_4)\,,
  \end{split}
\end{equation}
with the interaction vertex $V$ in momentum space
\begin{equation}
  \label{eq:sc-vertex}
  V(k_1,k_2,k_3,k_4) = e^{-\frac{i}{2}\Theta_{\mu\nu}\sum_{a<b}k_\mu^ak_\nu^b}\,.
\end{equation}
This vertex contains a momentum dependent phase factor and it is
therefore non--local.

\subsection{UV/IR mixing}
\label{sec:sc-uvir}

We discuss this important difference of non--commutative field
theories compared to commutative theories at the example of one loop
mass renormalization of the 4d $\lambda\phi^4$ theory, given by
equation (\ref{eq:sc-action2}). To this end we consider the one particle
irreducible two--point function
\begin{equation}
  \label{eq:UV-irr}
  \Gamma(p)=\langle\tilde{\phi}(p)\tilde{\phi}(-p)\rangle=\sum_{n=0}^\infty \lambda^{n}\Gamma^{(n)}(p)\,.
\end{equation}
At lowest order the two--point function is given by $\Gamma^{(0)}(p)=p^2+m^2$. The
one loop contribution splits topologically into two parts, one planar
and one non--planar diagram 
\begin{eqnarray}
  \label{eq:NC-pert1}
    \Gamma^{(1)}_\text{p}&=&\frac{1}{3}\int\frac{d^4k}{(2\pi)^4}\frac{1}{k^2+m^2}\,,\\
  \label{eq:NC-pert2}
    \Gamma^{(1)}_\text{np}(p)&=&\frac{1}{6}
    \int\frac{d^4k}{(2\pi)^4}\frac{\exp{(ik_\mu p_\nu\Theta_{\mu\nu})}}{k^2+m^2}\,.
\end{eqnarray}
In Refs.\ \cite{Filk:1996dm,Minwalla:1999px} it is shown that the
contribution of planar diagrams to non--commutative perturbation
theory is proportional to the commutative case (to all orders).
Therefore the planar divergences may be absorbed into the bare
parameters, if and only if the corresponding commutative theory is
renormalizable. This already disproves the expectation that
non--commutative quantum field theory would not require
renormalization.
\begin{figure}[htbp]
  \centering
  \unitlength=1.00mm \linethickness{0.4pt}
  \begin{picture}(120.00,20.00)
    \thinlines
    \put(9.00,0.00){\line(1,0){30.00}}
    \put(17.00,0.00){\vector(1,0){0.50}}
    \put(24.00,7.20){\circle{20.00}}
    \put(24.00,14.2){\vector(1,0){.5}}
    \put(12.00,3.00){\makebox(0,0)[l] {$\scriptstyle{p}$}}
    \put(34.00,6.00){\makebox(0,0)[l] {$\scriptstyle{k}$}}
    \put(17,-5){planar}

    \put(60.00,7.20){\line(1,0){18.00}}
    \put(65.00,7.20){\vector(1,0){.50}}
    \put(85.00,7.20){\line(1,0){8.00}}
    \put(75.00,7.20){\circle{20.00}}
    \put(75.00,14.2){\vector(1,0){.5}}
    \put(63.00,10.00){\makebox(0,0)[l]{$\scriptstyle{p}$}}
    \put(82.00,14.00){\makebox(0,0)[l]{$\scriptstyle{k}$}}
    \put(65,-5){non--planar}
    
  \end{picture}\\
  \vspace{.5cm}
  \caption{The planar and non-planar one loop contribution to two--point function (\ref{eq:UV-irr}).}
  \label{fig:UV-planar}
\end{figure}
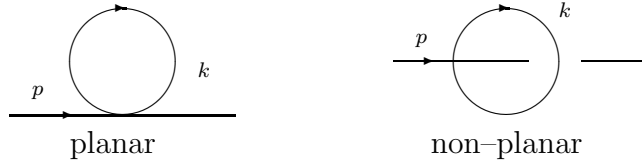

In the case of the non--planar diagrams the situation is different.
Rewriting the denominator in equation (\ref{eq:NC-pert2}) in terms of
a Schwinger parameter
\begin{equation}
  \label{eq:NC-schwinger}
  \frac{1}{k^2+m^2}=\int_0^\infty d\alpha\,e^{-\alpha(k^2+m^2)}\,,
\end{equation}
and introducing a momentum cut--off $\Lambda$ by multiplying the
resulting integrand in (\ref{eq:NC-pert2}) with a Pauli--Villars
regulator $\exp(-1/(\Lambda^2\alpha))$, leads in $d$ dimensions to
\cite{Szabo:2001kg}
\begin{equation}
  \label{eq:NC-full}
  \Gamma^{(1)}_{\text{np}}(p)=\frac{m^{\frac{d-2}{2}}}{6(2\pi)^{d/2}}
  \lrx{\frac{4}{\Lambda^2}-p_\mu\Theta^2_{\mu\nu}p_\nu}^{\frac{2-d}{4}}
  K_{\frac{d-2}{2}}\lrx{m\sqrt{\frac{4}{\Lambda^2}-p_\mu\Theta^2_{\mu\nu}p_\nu}}\,,
\end{equation}
where $K_n(x)$ is the irregular modified Bessel function of order $n$.
In $d=4$ the leading divergences of equation (\ref{eq:NC-full}) are given by
\begin{equation}
  \label{eq:NC-pert3}
  \Gamma^{(1)}_{\text{np}}(p)=\frac{1}{96\pi^2}
  \lrx{\Lambda^2_{\text{eff}}-m^2\log\lrx{\frac{\Lambda^2_{\text{eff}}}{m^2}}+O(1)}\,.
\end{equation}
Here we introduced the effective cut--off $\Lambda_{\text{eff}}$
given by
\begin{equation}
  \label{eq:NC-cutoff}
  \Lambda^2_{\text{eff}}=\frac{1}{\frac{1}{\Lambda^2}-p_\mu\Theta^2_{\mu\nu}p_\nu}
  =\frac{1}{\frac{1}{\Lambda^2}+\theta^2p^2}\,.
\end{equation}
The two--point function $\Gamma^{(1)}_{\text{np}}$ remains finite in
the limit $\Lambda\to\infty$, because it is effectively regulated by
the non--commutative space--time.  The complete one loop corrected
propagator then reads
\begin{equation}
  \label{eq:scalar-prop}
  \Gamma(p)=p^2+m^2+2\lambda\Gamma^{(1)}_\text{np}(0)+\lambda\Gamma^{(1)}_\text{np}(p)+O(\lambda^2)\,,
\end{equation}
where we used
\begin{equation}
  \label{eq:uv-ir}
  \Gamma_\text{p}^{(1)}=2\Gamma_\text{np}^{(1)}(p=0)\,.
\end{equation}
The UV limit ($\Lambda\to\infty$) does not commute with the IR limit
($p\to0$) or with the limit $\theta\to0$.  At small momenta or small
non--commutativity parameter the two--point function reads
\begin{equation}
  \label{eq:scalar-prop2}
  \Gamma(p)\simeq p^2+m^2+3\lambda\Gamma^{(1)}_\text{np}(0)+O(\lambda^2)\,.
\end{equation}
Taking now the UV limit leads to the standard mass renormalization of
the $\lambda\phi^4$ theory.  Taking these limits vice versa, the
effective cut--off is given by
\begin{equation}
  \label{eq:NC-eff-cutoff1}
  \Lambda^2_{\text{eff}}=\frac{1}{\theta^2p^2}
\end{equation}
and $\Lambda_{\text{eff}}$ diverges --- and therefore also
$\Gamma^{(1)}_\text{np}(p)$ --- either in the limit 
\footnote{Note that the limit $\theta\to 0$ in the non--commutative
  action (\ref{eq:sc-action2}) leads to the commutative action, since in
  this limit the star--product turns into the usual product. In the
  quantized theory, after the cut--off $\Lambda$ is removed, the limit
  $\theta\to0$ does not lead to the commutative theory, as equation
  (\ref{eq:scalar-leading-div}) shows.}
$\theta\to0$ or in the infrared limit when the incoming momentum $p$
goes to zero. We may absorb the planar one loop contribution of
$\Gamma(p)$ by defining the renormalized mass through
$M^2_\text{eff}=m^2+2\lambda\Gamma^{(1)}_{\text{np}}(0)$. Removing the
cut--off while keeping $M^2_\text{eff}$ fixed, then leads to a finite
$\Gamma(p)$ for finite incoming momenta $p$.  For zero momentum
$\Gamma(p)$ diverges and the divergence at one loop is given by
\begin{equation}
  \label{eq:scalar-leading-div}
  \Gamma(p)=p^2+M^2_\text{eff}+\xi\frac{\lambda}{\theta^2p^2} + \text{subleading terms}\,
\end{equation}
with $\xi=\frac{1}{96\pi^2}$.  Here we see that a non--zero
non--commutativity tensor $\Theta_{\mu\nu}$ replaces the standard
ultraviolet divergence with a singular infrared behavior. This mixing
between high
and low energy effects is called {\em UV/IR mixing}.\\

The long distance behavior of the spatial correlators is controlled by
the pole in the upper half plane nearest to the real axis, as in the
commutative case \cite{Douglas:2001ba}.  Due to the additional term in
equation (\ref{eq:scalar-leading-div}) the poles of the propagator are
now at
\begin{equation}
  \label{eq:salar-poles}
  p^2={-\frac{M^2_{\text{eff}}}{2}\pm\frac{1}{2\theta}\sqrt{M^4_{\text{eff}}\,\theta^2-4\lambda^2}}\,.
\end{equation}
In the weak coupling limit this pole is here at
$p=i\sqrt{\xi\lambda}/(M_{\text{eff}}\,\theta)$ and not at
$p=iM_{\text{eff}}$ as in the commutative case.  This can be
interpreted a a new mode with mass
$m_2=\sqrt{\xi\lambda}/(M_{\text{eff}}\,\theta)$. By definition
$m_2$ is much smaller than $M_{\text{eff}}$ and therefore dominates
the behavior of the spatial correlators. In the commutative theory
these correlation functions decay exponentially if $M_{\text{eff}}>0$.
Here we obtain at small $\lambda$ a power--like decay of the
correlators, leading to the long range correlations
\cite{Minwalla:1999px}. At large $\lambda$ the decay is again
exponential, but now with a decay constant $\propto \sqrt{\lambda}$.


\subsection{Phase structure of non--commutative $\lambda\phi^4$ theory}
\label{gubser-sondhi}

The UV/IR mixing is one of the most interesting properties of
non--commu\-ta\-tive field theory and has no counterpart in the
commutative case. A number of new effects and also problems occur due to
this term. In particular the phase diagram of the non--commutative
$\lambda\phi^4$ model is changed,
which we want to discuss here.\\

The low momentum singularity of $\Gamma(p)$, discussed in the last
Subsection, has a large impact on the phase diagram. Since a phase
transition should involve the momenta that minimize $\Gamma(p)$, it is
not likely that the low momentum modes will participate in a phase
transition. If there is a phase transition at all it will be driven by
non--zero momentum modes. Then the IR divergence leads to an
oscillation in the sign of the correlator
$\langle\phi(0)\star\phi(x)\rangle$, indicating a new type of ordered
phase.\\

Gubser and Sondhi studied the phase diagram of 4d $\lambda\phi^4$
theory \cite{Gubser:2000cd} within the framework of a one loop
self--consistent Hartree--Fock approximation \cite{Brazovkii1975}. We
do not discuss their calculation, but summarize their results on the
phase diagram.

At small non--commutativity parameter $\theta$ they obtained an Ising
type (second order) phase transition leading to an uniformly ordered
phase with $\langle\phi\rangle\neq0$. At sufficiently large $\theta$
the minimum of $\Gamma(p)$ is not at $p=0$. The phase
transition is now driven by modes $p\neq0$ and it is of first order.
\begin{figure}[htbp]
  \centering \includegraphics[width=.65\linewidth]{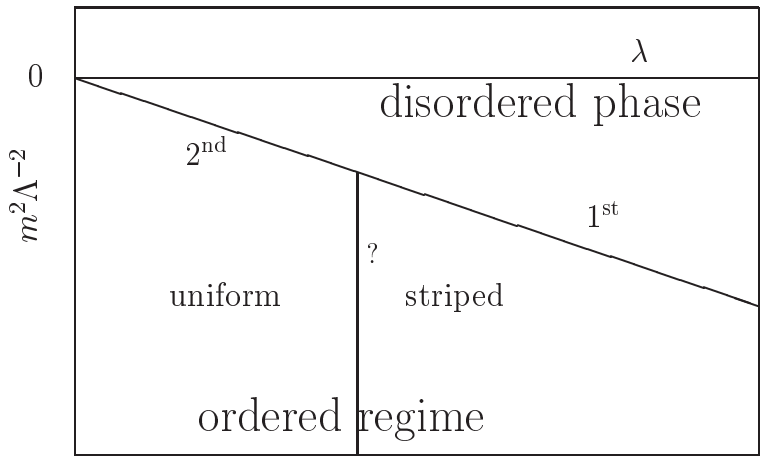}
  \caption{The phase diagram of non--commutative 4d $\lambda\phi^4$ theory conjectured by
    Gubser and Sondhi \cite{Gubser:2000cd} in the $m^2$ -- $\lambda$ plane.}
  \label{fig:GS-phase}
\end{figure}
This leads to an ordered phase where the translation invariance is
broken spontaneously. In this phase $\langle\phi\rangle$ varies in
space, which implies the ground state to involve some non--uniform
patterns like stripes. These patterns depend on the momentum mode
which drives the phase transition.  In Ref.\ \cite{Gubser:2000cd}
these results are summarized in a qualitative phase diagram in the
$m^2\Lambda^{-1}$--$\theta\Lambda^2$ plane, where $\Lambda$ is a
momentum cut--off.

In Section \ref{sec:sc-ncsa} we have seen that only the interaction
term depends on the parameter $\theta$. Therefore increasing the
coupling $\lambda$ also amplifies effects of non--commutativity.
According to Ref.\ \cite{Gubser:2000cd} the phase diagram in the
$m^2\Lambda^{-1}$--$\lambda$ plane is then given by Figure
\ref{fig:GS-phase}.

A similar phase structure was conjectured in three dimensions.  In two
dimensions it was argued that a striped phase does not occur. Gubser
and Sondhi worked with an action of the Brazovskiian form
\cite{Brazovkii1975}, which is local. Hence the Mermin--Wagner theorem
\cite{Mermin:1966fe,Hohenberg1967,Coleman:1973ci} applies, which
states that spontaneous breakdown of a continuous symmetry is not
possible in two dimensions.
We come back to this point in Chapter \ref{A-2d}.\\

In another approach renormalization group techniques were used to
study the phase diagram of the $\lambda\phi^4$ model
\cite{Chen:2001an}.  Chen and Wu obtained in $d=4-\epsilon$ a new IR stable
fixed point, i.e.\ the non--commutative counterpart of the
Wilson--Fisher fixed point.  This fixed point is stable, and therefore
a striped phase exists, when $\theta>12/\sqrt{\epsilon}$. In contrast
to the results in Ref.\ \cite{Gubser:2000cd}, this implies that in
$d=4$ there is no striped phase.  Since we studied the 3d model, we
will not address this controversy in this work.


\section{Non--commutative gauge theory}
\label{gauge}

In this Section we extend our considerations to gauge theories defined
on a non--commutative plane.

\subsection{Star--gauge invariant action} 
\label{sec:gau-nca}

To define a Yang--Mills theory on a non--commutative plane we have to
generalize the map (\ref{eq:NC-delta}) in Section \ref{NC} to the
algebra of $n\times n$ matrix valued functions. Let $A_\mu(x)$ be a
Hermitian gauge field on $\mathbb{R}^d$, which corresponds to the
unitary gauge group $U(n)$. We can introduce the Weyl operators
corresponding to $A_\mu(x)$ by taking the trace of the tensor product
of $\Delta(x)$ and the gauge field
\begin{equation}
  \label{eq:gau-operators}
  \hat{A}_\mu=\int d^dx \,\Delta(x)\otimes A_\mu(x)\,,
\end{equation}
where $\Delta(x)$ is defined in equation (\ref{eq:NC-delta}). Based
on equation (\ref{eq:NC-derivative-1}) a non--commutative version
of the Yang--Mills action can be defined
\begin{equation}
  \label{eq:gau-yang-mills}
  S[\hat{A}]=-\frac{1}{4g^2}\trace\,\tr_N\lrx{\lry{\hat{\pd}_\mu,\hat{A}_\nu}
    -\lry{\hat{\pd}_\nu,\hat{A}_\mu}-i\lry{\hat{A}_\mu,\hat{A}_\nu}}^2\,,
\end{equation}
where the term in brackets is the operator analog of the field
strength tensor.  Here $\trace$ is the operator trace
(\ref{eq:NC-trace}) and $\tr_N $ denotes the trace in color space.
This action is invariant under transformations of the form
\begin{equation}
  \label{eq:gau-trans}
  \hat{A}_\mu\to\hat{G}\,\hat{A}_\mu\,\hat{G}^\dagger-i\hat{G}\lry{\hat{\pd}_\mu,\hat{G}}\,,
\end{equation}
where $\hat{G}$ is an arbitrary unitary element of the algebra of
matrix valued operators, i.\ e.\
\begin{equation}
  \label{eq:gau-unitary}
  \hat{G}\,\hat{G}^\dagger=\hat{G}^\dagger\,\hat{G}=\hat{\unit}\otimes\unit_n\,.
\end{equation}
The symbol $\hat{\unit}$ is here the identity on the ordinary Weyl
algebra and $\unit_n$ is a $n\times n$ unit matrix.\\

To set up the action in coordinate space we can construct an inverse
map of (\ref{eq:gau-operators}). By mapping the product of matrix
valued Weyl operators to coordinate space, using this inverse map,
again the star--product (\ref{eq:star-product}) appears.
The Yang--Mills action in coordinate space then reads
\begin{equation}
  \label{eq:gau-yang-mills2}
  S[A]=-\frac{1}{4g^2}\int d^dx\,\tr_N \lrx{F_{\mu\nu}(x)\star F_{\mu\nu}(x)}\,,
\end{equation}
where we introduced the non--commutative field strength tensor
$F_{\mu\nu}$ given by
\begin{equation}
  \label{eq:gau-tensor}
  F_{\mu\nu}(x)=\pd_\mu A_\nu(x)-\pd_\nu A_\mu(x)
  -i\lry{A_\mu(x),A_\nu(x)}_\star\,.
\end{equation}
The index '$\star$' indicates that the products in this commutator are
star--products. From equation (\ref{eq:gau-tensor}) we see that
already for the simple gauge group $U(1)$ we have a Yang--Mills type
structure.  Therefore there exist three and four point gauge
interactions and non--commutative $U(1)$ theory is
asymptotically free.\\

The invariance under unitary transformations in operator space
translates here into an invariance of the action
(\ref{eq:gau-yang-mills2}) under {\em star--gauge} transformations
given by
\begin{equation}
  \label{eq:gau-star-gauge}
  A_\mu(x)\to G(x)\star A_\mu(x)\star G(x)^\dagger -iG(x)\star\pd_\mu G(x)^\dagger\,,
\end{equation}
where $G(x)$ is a {\em star--unitary} matrix field,
\begin{equation}
  \label{eq:gau-star-unitary}
  G(x)\star G(x)^\dagger = G(x)^\dagger\star G(x) = \unit_n\,.
\end{equation}
Equation (\ref{eq:gau-star-unitary}) is equivalent to the unitarity
condition (\ref{eq:gau-unitary}).\\

So far we considered non--commutative $U(n)$ theories which reduce to
the ordinary $U(n)$ theories in the limit $\theta\to0$ (on the
classical level). In Ref.\ \cite{Matsubara:2000gr} it was shown that
for other gauge groups like $SU(n)$ this cannot be realized on
non--commutative flat space.
\footnote{We refer to a constant non--commutativity tensor
  $\Theta_{\mu\nu}$.}
The $U(n)$ group is closed under the star--product; the product of two
star--unitary matrix fields is again star--unitary. In contrast to
$U(n)$ the special unitary group $SU(n)$ is not closed, since in
general
\begin{equation}
  \label{eq:gau-sun}
  \det{(G)}\star\det{(H)}\neq\det{(G\star H)}\,.
\end{equation}
The $U(1)$ and the $SU(n)$ sectors in the decomposition
\begin{equation}
  \label{eq:gau-decomposition}
  U(n)=U(1)\otimes SU(n)/\mathbb{Z}_n
\end{equation}
do not decouple in the non--commutative case, because the $U(1)$
photon interacts with the $SU(n)$ gluons \cite{Armoni:2000xr}.

\subsection{Star--gauge invariant observables}
\label{gau-SGO}

To construct star--gauge invariant observables we consider an arbitrary
oriented smooth contour $C_v$ in space--time, which connects the points
$x$ and $x+v$. The holonomy of the gauge field over this contour is
described by the non--commutative parallel transporter
\begin{equation}
  \label{eq:SGO-parallel-transporter}
  \mathcal{U}(x;C_v)=\text{P} \exp_\star\lrx{i\int_{C_v} d\xi_\mu A_\mu(x+\xi)}\,,
\end{equation}
where P indicates path ordering and $\xi$ parameterizes the contour.
The index '$\star$' at the exponential function indicates that in the
expansion of this function the star--product has to be used. The
parallel transporter (\ref{eq:SGO-parallel-transporter}) is a $n\times
n$ star--unitary matrix field and transforms under the star--gauge
transformation (\ref{eq:gau-star-gauge}) like
\begin{equation}
  \label{eq:SGO-gauge-trans}
  \mathcal{U}(x;C_v)\to G(x)\star\mathcal{U}(x;C_v)\star G^\dagger(x+v)\,.
\end{equation}
A remarkable fact in non--commutative field theory is that space-time
translations can be arranged by (star--) multiplication with plane
waves
\begin{equation}
  \label{eq:SGO-translation}
  G(x+v)=e^{ik_\mu x_\mu}\star G(x)\star e^{-ik_\rho x_\rho}
  \qquad\text{with}\qquad k_\mu=\lrx{\Theta^{-1}}_{\mu\nu}v_\nu\,,
\end{equation}
where we assume $\Theta$ to be invertible. That this equation holds
can easily be shown by expanding the exponential functions and using
equation (\ref{eq:NC-note}). With the definition of the
non--commutative parallel transporter and equation
(\ref{eq:SGO-translation}) we can associate a star--gauge invariant
observable with any arbitrary contour $C_v$ by
\begin{equation}
  \label{eq:SGO-obs}
  \mathcal{O}(C_v)=\int d^dx\,\tr_N\lrx{\mathcal{U}(x;C_v)\star e^{ik_\mu x_\mu}}\,.
\end{equation}
It is straightforward to show the invariance under the star-gauge
transformation (\ref{eq:SGO-gauge-trans}) by using equation
(\ref{eq:SGO-translation}) and the cyclicity of the trace over the
star--product.

In commutative gauge theory gauge invariant observables can only be
constructed from closed loops. In contrast to that, equation
(\ref{eq:SGO-obs}) shows that in non--commu\-ta\-tive gauge theory we
can find star--gauge invariant observables associated with {\em open}
contours.  The vector $k$ in equation (\ref{eq:SGO-obs}) can be
regarded as the total momentum of the open loop. This is again a
manifestation of the UV/IR mixing phenomenon, discussed in Section
\ref{sec:sc-uvir}.  If we increase the momentum $k_\mu$ in a given
direction, the contour will extent in
the other directions according to $\Theta_{\mu\nu}k_\nu$.\\

This completes our introduction to non--commutative field theories in
the continuum. We showed how to define a scalar field theory and a
pure gauge theory on a non--commutative plane, and we discussed the
main differences compared to the commutative case.  This sets up the
framework for our numerical studies to be presented in Chapters
\ref{cha:TEK} and \ref{phi}. In the next Section we will discuss
further properties and problems of non--commutative field theory.


\section{Phenomenological implications of a quantized space--time}
\label{standard}

In this Section we present phenomenological consequences of a
$\theta$--deformed space--time. To this end we discuss briefly some
aspects of the non--commutative standard model.

\subsubsection{Gauge fields coupled to matter fields}

To set up the non--commutative standard model we have to extend our
considerations in Section \ref{gauge} to the case where the gauge
field couples to matter fields. We start from the action of a
free Dirac field
\begin{equation}
  \label{eq:mot-free-action}
  S[\bar{\psi},\psi]=\int d^dx \,\bar{\psi}(x)\star\lrx{\gamma_\mu\pd_\mu+m}\psi(x)\,,
\end{equation}
where we extended the commutative theory to a non--commutative theory
by replacing the usual products of fields with star--products. The
Grassmann valued fermionic fields are represented by $\psi(x)$.  To
obtain an action that is invariant under the star--gauge
transformations
\begin{equation}
  \label{eq:mot-fermion}
  \psi(x)\to G(x)\star\psi(x)\andx \bar{\psi}(x)\to\bar{\psi}(x)\star G^\dagger(x)\,,
\end{equation}
with a star--unitary $n\times n$ matrix field $G(x)$, we have to
modify the kinetic term of the action (\ref{eq:mot-free-action}).  We
follow here Ref.\ \cite{Hayakawa:1999yt} and introduce, in analogy to
the commutative case, the covariant derivative
\begin{equation}
  \label{eq:mot-covariant-derivative}
  D_\mu\psi(x)=\pd_\mu\psi(x)+igA_\mu(x)\star\psi(x)\,,
\end{equation}
where $A_\mu(x)$ is the gauge field that generates the unitary
group $U(n)$.  The derivative (\ref{eq:mot-covariant-derivative}) is
covariant under the star--gauge transformation
(\ref{eq:gau-star-gauge}),
\begin{equation}
  \label{eq:mot-covariant-transform}
  D_\mu\to G(x)\star D_\mu\star G^\dagger(x)\,.
\end{equation}
Replacing the derivative in (\ref{eq:mot-free-action}) by the
covariant derivative leads to the star--gauge invariant fermion action
\begin{equation}
  \label{eq:mot-matter-action}
  S_\text{fermion}[A,\bar{\psi},\psi]=\int d^dx\lrx{\bar{\psi}\star\lry{\gamma_\mu D_\mu+m}\psi}\,.
\end{equation}
The complete action is then given by the sum of the gauge action
(\ref{eq:gau-yang-mills2}) and the fermion action
(\ref{eq:mot-matter-action})\\
\begin{equation}
  \label{eq:mot-action-complete}
  S[A,\bar{\psi},\psi] = S_\text{YM}[A]+S_\text{fermion}[A,\bar{\psi},\psi]\,.
\end{equation}

As we mentioned in the last Section, it was so far not possible to
formulate a non--commutative field theory for the special unitary
group $SU(n)$. The gauge group is restricted to $U(n)$, which is
(for $n>1$) not a gauge group of the standard model. There are
attempts to modify the space--time non--commutativity in order to take
also $SU(n)$ gauge groups into account \cite{Jurco:2000ja,Chu:2001if},
but this is an ongoing field of research. However, we may consider the
star--unitary $U(1)$ gauge field coupled to fermions as an extension
of commutative QED.

\subsubsection{About renormalizability}

The UV/IR mixing poses severe problems for the renormalization of
perturbation theory, which are not overcome yet.  For instance, scalar
fields can become unstable --- tachionic --- due to IR effects as the
non--commutativity is switched on.  The attempts to renormalize
perturbation theory include methods known from standard field theory
\cite{Chepelev:1999tt}, the application of Wilson's renormalization
group technique \cite{Sarkar:2002pb,Griguolo:2001ez,Griguolo:2001wg},
controlling the IR divergences in the framework of supersymmetry
\cite{Girotti:2000gc,Girotti:2001ku} and the Hartree--Fock method
\cite{Gubser:2000cd}.

In spite of some plausibility arguments in favor
of perturbative renormalizability, it is an open question if
non--commutative quantum field theories do really have finite UV and IR
limits. There are even claims that basic non--commutative field
theories, like non--commutative QED, are not renormalizable
\cite{Wulkenhaar:2001sq}.\\

\subsubsection{Violation of Lorentz symmetry}

Since $\Theta_{\mu\nu}$ carries Lorentz indices, two distinct types
of Lorentz symmetries have to be considered \cite{Colladay:1998fq}:
the observers Lorentz transformation and the particle Lorentz
transformation. The action (\ref{eq:mot-action-complete}) as well as
the scalar action (\ref{eq:sc-action2}) are fully covariant under
rotations or boosts of the observers reference frame, because
$\Theta_{\mu\nu}$ and the fields transform covariantly. This does not
hold anymore in the case of rotations or boosts of a particle
\cite{Carroll:2001ws}; $\Theta_{\mu\nu}$ is unaffected by these
transformations.
\footnote{The non--commutativity tensor $\Theta_{\mu\nu}$ plays the role
  of an inverse background field (see introduction). From this point
  of view the Lorentz symmetry breaking occurs very naturally. The
  symmetry would be restored in a formalism that transforms
  the particle fields as well as the background field.}

The broken Lorentz symmetry implies a deformed dispersion relation,
where the deformation depends on the theory under consideration.  In
the case of the non--commutative $\lambda\phi^4$ theory the dispersion
relation is given by the poles of the irreducible two--point function
(\ref{eq:scalar-leading-div}). We focus here on the case with two
commuting and two non--commuting directions. The on--shell condition
then reads
\begin{equation}
  \label{eq:mot-disp-phi}
  E(\vec{p})^2\simeq\vec{p}^{\,2}+P^2+M^2_\text{eff}+\xi\frac{\lambda}{\theta^2\vec{p}^{\,2}}\,,
\end{equation}
where $\vec{p}$ is the momentum that corresponds to the
non--commutative plane and $P$ corresponds to the commutative space
coordinate.  We include here only the leading IR divergence in first
order of $\lambda$; in addition there is also a logarithmic
divergence.

\subsubsection{Phenomenology of non--commutative space--time}

The effects of non--commutativity on quantities that are measurable in
experiment are studied intensively by many authors. These studies may
be interpreted as an attempt to set stringent limits on the
non--commutativity parameter $\theta$, or to suggest possibly
measurable effects of non--commutative space--time. We will present
here an
incomplete list of current investigation in this field.\\

{\noindent \it Deformed dispersion relation in gauge theories}\\
  
Also in gauge theories Lorentz symmetry is broken, which leads to a
deformed dispersion relation and changes the particle propagation.
This was studied for example in Refs.\ 
\cite{Amelino-Camelia:2001cm,Matusis:2000jf}. There a dispersion
relation for photons was obtained containing a $1/\vec{p}^{\,2}$ term as in the
scalar case.  Based on one loop perturbation theory the authors of
Ref.\ \cite{Amelino-Camelia:2001cm} obtained the photon dispersion
relation
\begin{equation}
  \label{eq:mot-disp-photon}
  E(\vec{p})^2=\vec{p}^{\,2}+\zeta g^2\frac{1}{\theta^2 \vec{p}^{\,2}}\,,
\end{equation}
where they chose the time to be commutative. The coefficient $\zeta$
depends on the number of bosonic and fermionic degrees of freedom
present in the theory and $g$ is the coupling constant.

This may give rise to experimental verification of quantized
space--time.  The non--linear term in $E(\vec{p}\,)^2$ leads to a vacuum
dispersion of light, such that the speed of light depends on the
wave length. In addition the dispersion depends on the polarization
of the photons leading to birefringence.
  
The authors of Ref.\ \cite{Amelino-Camelia:1998gz} suggest to measure
this deformation of the dispersion relation by so--called {\em time of
  arrival measurements}. In these measurements the time delay between
photons with different wave length, emitted simultaneously up to a
known precision, is measured. The delay will depend on the time the
photons are traveling. This effect is (if existent at all) very small.
Therefore the sources of the photons should be far distant to
accumulate the delay to a measurable effect.

There are attempts by experimentalists to study this effect.  Already
in $1998$ high precision measurements
\footnote{The
  accuracy in this measurements was $\delta T/T\approx 10^{-12}$,
  where $\delta T$ is the time delay and $T$ is the overall time.}
of an energy dependent speed of light were performed
\cite{Biller:1998hg}. However, up to a given sensitivity these
measurements did {\em not} show any energy dependence.

Newer experiments might give more insight. For example the HESS
project \cite{Hess}, just started to take data, has a higher
sensitivity than in Ref.\ \cite{Biller:1998hg}. Besides other
projects, this collaboration intend to make measurements related to
the vacuum dispersion \cite{Lohse2003}. Another candidate for
measuring such effects is GLAST \cite{deAngelis:2000ji}, which is
expected to start
measurements by the year 2005.\\

{\noindent \it Threshold anomalies}\\

It is still an open question why high energy photons ($E>20$TeV)
from far distant galaxies can be detected on earth \cite{Aharonian:2002cv}.
Photons in this energy range traveling over galactic distances
should interact with the cosmic microwave background, producing
electron positron pairs
\begin{equation*}
  \gamma\gamma\to e^+e^-\,.
\end{equation*}
The threshold for this reaction is approximately $E\approx 1$TeV
\cite{Greisen:1966jv,Stecker:1993ud} and it should make the
observation of high energy photons very unlikely.

A second threshold exist for cosmic high energy protons. There the
interaction of the protons with the cosmic microwave background leads
to
\begin{equation*}
  p+\gamma \to p+\pi\,.
\end{equation*}
The protons loose energy when producing pions and the threshold, i.e.\ 
the GZK thres\-hold, for this process is at approximately $E\approx
5\times 10^{19}\,$eV \cite{Zatsepin:1966jv,Greisen:1966jv} Therefore
cosmic protons with higher energy should not be observed. However,
cosmic proton
rays have been detected beyond this limit \cite{Hayashida:2000zr}.\\

In both cases non--commutative space--time could provide an
explanation of these anomalies.  The thresholds can be estimated
kinematically using the momentum and energy conservation. In a Lorentz
invariant theory the threshold momentum in the case of high energy
protons is approximately \cite{Amelino-Camelia:2000zs}
\begin{equation}
  \label{eq:mot-thres}
  p_\text{\,threshold}\approx \frac{m_p m_\pi+m_\pi^2}{4 E_\gamma}\,,
\end{equation}
where $m_p$ and $m_\pi$ are the masses of the proton and of the pion,
respectively, and $E_\gamma$ is the energy of the background photon that
interacts with the proton. In a non--commutative space--time the dispersion
relation is given by equation (\ref{eq:mot-disp-photon}). Inserting
the deformed dispersion relation into this computation
 leads to a $\theta$
dependent threshold.\\

{\noindent \it Bounds on $\theta$}\\
  
So far we discussed new effects and known problems that might be
explained by non--vanishing $\theta$. Let us now address the limits on
$\theta$ set by existing experiments and the possibility of setting
bounds on $\theta$ in near future experiments.
  
The fact that in the aforementioned time of arrival measurements no
energy dependence was measured, allows to set a {\em lower} limit on
$\theta$ (if we assume a non--zero $\theta$ from the beginning).
Note that here we find a lower limit, because the
dispersion relation (\ref{eq:mot-disp-photon}) implies that larger
values of $\theta$ correspond to a softer deformation of the standard
dispersion.
\footnote{The rather unexpected lower bound of $\theta$ is related
  to the order of UV limit and commutative limit, as we discussed in
  Section \ref{sec:sc-uvir}. This lower bound can be understood in
  the sense that {\em if} there is a non--commutative space--time,
  then $\theta$ has to be larger than a certain value. However, this
  does {\em not} exclude $\theta=0$.}
On the other hand if one wants to solve the cosmic proton threshold
anomaly within the framework of non--commutative field theory one
finds an upper bound of $\theta$.  According to Ref.\ 
\cite{Amelino-Camelia:2001cm} this leads to a rough estimation of the
a parameter $\theta$ in the range
$(10^{4}\,\text{TeV})^{-2}<\theta<(10\text{TeV})^{-2}$.\\
  
More precise bounds on $\theta$ could be obtained from accelerator
physics. Since non--commutative QCD is not well formulated yet,
these investigations are restricted to measurements described by QED.
Hence one expects the most significant results from linear colliders.
  
The idea is to compute for example cross sections for scattering
processes from the standard model and from the non--commutative standard
model \cite{Hewett:2000zp}.
\footnote{The calculations in non--commutative QED rely on the
  Feynman rules developed in Ref.\
  \cite{Armoni:2000xr,Sheikh-Jabbari:1999iw,Krajewski:1999ja,Aref'eva:2000bg}.}
This leads to different predictions for the two models, where both
predictions depend of course on the center of mass energy $\sqrt{s}$.
Measurements of these observables then may verify the non--commutative
picture or set an upper bound on $\theta$.  Measurements in existing
colliders did not show any signal of non--commutative space--time.
Therefore one is looking for search limits, which can be probed
in future colliders (TESLA \cite{Aguilar-Saavedra:2001rg}, NLC
\cite{Abe:2001nr} and CLIC \cite{Assmann:2000hg}) $\theta$.
\begin{table}[htbp]
  \centering
  \begin{tabular}{|c|c|}
    \hline\hline process & {\small search limit on $1/\sqrt{\theta}$ at $\sqrt{s}=0.5 - 5$ TeV}\\\hline
    \phantom{\LARGE A}pair production $\gamma\,\gamma\to e^-\, e^+$& $500 - 2700$ GeV \\\hline
    \phantom{\LARGE A}Compton scattering $e\,\gamma\to e\,\gamma$& $1000 - 6500$ GeV \\\hline
  \end{tabular}
  \label{tab:mot-theta-limits}
  \caption{Sensitivity to measure effects of non--commutativity in future colliders.}
\end{table}
For example, Ref.\ \cite{Godfrey:2001yy} reports search limits for the
processes listed in Table \ref{tab:mot-theta-limits} on the basis of
the design of these colliders.  Further search limits can be found for
example in Ref.\ \cite{Rizzo:2002yr}.  For recent reviews concerning
the activities in this field see
for example Ref.\ \cite{Hinchliffe:2002km,Anisimov:2001zc,Godfrey:2002yk}.\\

Note that the theories, which the above presented results are based
on, still suffer from IR divergences. They are obtained from one loop
calculations.  Higher orders in perturbation theory are not under
control yet. It is still an open question if the non--commutative
field theories are renormalizable in two or three loop calculations.
The limits on $\theta$ were obtained assuming that higher order
correction will not change the results qualitatively. However, one
cannot exclude dramatic changes coming from higher loop contributions.

\section{Summary}
\label{sec:sum}

Let us briefly summarize the effects of non--commutative space--time
on field theories defined on it. We showed that introducing
non-commutativity via the commutation relation (\ref{eq:NC-1}) leads
to a non--commutative and non--local product, i.e.\ the star--product.
Field theories may be formulated on this geometry by replacing all
products in the commutative theory by the star--product of the fields.
This results in a non--local action.  In a perturbative treatment one
discovers a mixing of high and low energy effects. The UV/IR mixing
causes serious problems in perturbative renormalization, since some of
the UV divergences in commutative theories turn into IR divergences in
the non--commutative model even in the massive case. We discussed this
new effect in one loop perturbation theory.  Already there the
divergences at low momenta cause enormous problems.  The difficulties
increase when these loops are sub--loops of a higher order
contributions.


In a $\theta$--deformed space--time the Lorentz symmetry is explicitly
broken. This leads to unexpected particle propagation like a momentum
dependent speed of light. These effects are the most likely candidates
for experimental verifications.

In non--commutative gauge theories the gauge symmetry turns into a
star--gauge symmetry.  This allows us to construct star--gauge
invariant observables that are associated with open Wilson loops. The
open loops carry a momentum proportional to the separation between the
endpoints.  This is again a UV/IR mixing effect. In non-commutative
QED the photons interact, which might again give rise for new
measurable effects.\\

As an alternative to perturbation theory we are studying
non--commu\-ta\-tive field theories in the lattice approach
\cite{Wilson:1974sk}.  This work has to be considered as the
beginning of a long term project. At the end of this project we want
of course to compute phenomenological quantities, but in the starting
phase we will study field theory in lower dimensions in order to
understand better the non--perturbative treatment.

However, already the study of these toy models gives insights into the
four dimensional theory. Since also for these models there exist
perturbative results, it is an interesting question if in a
non--perturbative study new effects arise and how far perturbation
theory can be confirmed.



\chapter{Lattice regularization}
\label{cha:lattice}

As we have seen in Section \ref{sec:sc-uvir}, introducing
non-commutativity of space--time does not cure the ultraviolet
divergences occurring in the commutative case. Therefore the field
theories still have to be regularized, by introducing a momentum
cut--off. In this Chapter we discuss the lattice regularization.  For
a more detailed discussion we refer to Refs.\ 
\cite{Ambjorn:1999ts,Ambjorn:2000nb,Ambjorn:2000cs,Szabo:2001kg,Szabo:2001qd}.

\section{Discrete non--commutative space--time}
\label{sec:lat-dsp}

On the cubic lattice the space--time points $x_\mu$ are restricted to
discrete values $x_\mu\in a\mathbb{Z}$, where $a$ is the lattice
spacing. Here  momentum space is compact and the momenta
$k_\mu$ must be identified under the shift
\begin{equation}
  \label{eq:lat-momenta}
  k_\mu \to k_\mu + \frac{2\pi}{a}\delta_{\mu\nu}\qquad\text{with}\qquad \nu=1,2,\dots,d\,.
\end{equation}
As in the continuum, non-commutativity comes into the game by replacing
the ordinary coordinates $x_\mu$ by Hermitian coordinate operators
$\hat{x}_\mu$, which satisfy the commutator relation (\ref{eq:NC-1}). As
a consequence of equation (\ref{eq:lat-momenta}), we can set up the
operator identity
\begin{equation}
  \label{eq:lat-identity}
  e^{i(k_\mu + \frac{2\pi}{a}\delta_{\mu\nu})\hat{x}_\mu}=e^{ik_\mu\hat{x}_\mu}\,.
\end{equation}
By multiplying both sides of equation (\ref{eq:lat-identity}) with 
$\exp(-ik_\rho \hat{x}_\rho)$ we find
\begin{equation}
  \label{eq:lat-identity1}
  \exp\lrx{\frac{2\pi i}{a} \hat{x}_\mu\delta_{\mu\rho}} 
  \exp\lrx{\Theta_{\mu\nu}k_\nu\frac{\pi i}{a}\delta_{\mu\rho}}= \hat{\unit}
  \qquad\text{with}\qquad \rho=1,2,\dots,d\,,
\end{equation}
where $\hat{\unit}$ is again the identity of the algebra of Weyl
operators. The usual constraint of lattice field theory that the
discretization has to be compatible with the spectrum of the position
operator leads to
\begin{equation}
  \label{eq:lat-constraint}
  e^{\frac{2\pi i}{a} \hat{x}_\mu} =\hat{\unit}  \qquad\text{for}\qquad \mu=1,2,\dots,d\,.
\end{equation}
Moreover we find an additional constraint for the momenta 
\begin{equation}
  \label{eq:lat-constraint1}
  \Theta_{\mu\nu}k_\nu\in 2a\mathbb{Z}\with \mu=1,2,\dots,d\,.
\end{equation}
Combining this result with the periodicity in momentum space
(\ref{eq:lat-momenta}) implies that $\Theta_{\mu\nu}\pi/a^2$
is an integer for all $\mu,\nu$, leading to a discrete
momentum \begin{equation}
  \label{eq:NCT-momenta1}
  k_\mu=2\pi\lrx{\Sigma^{-1}}_{\nu\mu}m_\nu\with m_\nu\in\mathbb{Z};\;\;\nu=1,\dots,d\,.
\end{equation}
The periods $\Sigma_{\mu \nu}$ are integer multiples of the lattice
spacing $a$ and the periodicity (\ref{eq:lat-momenta}) can be
expressed by the integers $m_\nu$ via $m_\nu\to m_\nu+\frac{1}{a}\Sigma_{\mu
  \nu}$.  

Due to the discrete momentum space the space--time coordinates $x_\mu$
are restricted to a periodic lattice
\begin{equation}
  \label{eq:period1}
  x_\mu\to x_\mu+\Sigma_{\mu \nu}\with \nu=1,\dots d\,.
\end{equation}
Therefore the lattice regularization forces space--time to be compact.
The discrete compactification is a non--perturbative manifestation of
the UV/IR mixing effect; the perturbative manifestation was described
in Section \ref{sec:sc-uvir}. For practical applications this means
that systematic errors resulting from the finite lattice spacing are
{\em always} entangled with finite volume errors.

The continuum limit $a\to 0$ does not commute with the commutative
limit $|\Theta|\to 0$. Taking first the continuum limit restores the
infinite space--time $\mathbb{R}^d$, while taking first the
commutative limit shrinks the space--time lattice to a single point as
one infers from equation (\ref{eq:lat-constraint1}).\\

Due to the momentum discretization (\ref{eq:NCT-momenta1}) we cannot
use the unbounded operators $\hat{x}_\mu$. Instead we have to use the
unitary coordinate operators $\hat{Z}_\nu$, which have been introduced
in Section \ref{sec:NC-torus}
\begin{equation}
  \label{eq:lat-coordinate}
  \hat{Z}_\nu=e^{2\pi i \lrx{\Sigma^{-1}}_{\nu\mu}\hat{x}_\mu}\,.
\end{equation}
These operators generate the algebra of functions on a
non--commutative torus
\footnote{Note that in equation (\ref{eq:lat-comm1}) and (\ref{eq:lat-comm-x}) {\em no}
  sum over $\mu,\nu$ is taken.}
\begin{equation}
  \label{eq:lat-comm1}
  \begin{split}
    \hat{Z}_\mu\hat{Z}_\nu&=e^{-2\pi i \tilde{\Theta}_{\mu\nu}}\hat{Z}_\nu\hat{Z}_\mu\,,
  \end{split}
\end{equation}
where the dimensionless non-commutativity tensor is defined as
\begin{equation}
  \label{eq:lat-theta}
  \tilde{\Theta}_{\rho\sigma}=2\pi\lrx{\Sigma^{-1}}_{\rho\mu}\Theta_{\mu\nu}\lrx{\Sigma^{-1}}_{\sigma\nu}\,.
\end{equation}
Moreover since we are dealing with a discrete torus, one should rather
use the shift operator $\hat{D}_\mu$
\begin{equation}
  \label{eq:lat-shift}
  \hat{D}_\mu=e^{a\,\hat{\pd}_\mu}\,,
\end{equation}
instead of the linear derivative $\hat{\pd}_\mu$ (defined in
(\ref{eq:NC-derivative})). This operator effects translations in units
of the lattice spacing $a$
\begin{equation}
  \label{eq:lat-comm-x}
  \hat{D}_\mu\hat{Z}_\nu\hat{D}_\mu^\dagger=e^{2\pi i a \lrx{\Sigma^{-1}}_{\nu\mu}}\hat{Z}_\nu\,.
\end{equation}
The algebra defined in (\ref{eq:lat-comm1}) and (\ref{eq:lat-comm-x})
replaces the algebra
based on (\ref{eq:NC-1}) and (\ref{eq:NC-derivative}).\\

Now we consider scalar functions $f(x)$ on a non--commutative discrete
torus, where we assume that these functions may be expressed by a
discrete Fourier transformation of $\tilde{f}(m)$. Lattice Weyl
operators are then defined by the formal Fourier transform
\begin{equation}
  \label{eq:lat-discrete-fourier}
  \hat{f} = \sum_{{m}} \tilde{f}({m})\, e^{2\pi i (\Sigma^{-1})_{\nu\mu}m_\nu\hat{x}_\mu}\,,
\end{equation}
analogous to the continuum (\ref{eq:NC-operator}). The relation
between the integer vector $m$ and the momentum $k$ is given by
equation (\ref{eq:NCT-momenta1}).  Replacing $\tilde{f}(m)$ in
equation (\ref{eq:lat-discrete-fourier}) by its Fourier transform and
using equation (\ref{eq:lat-comm1}) leads to an explicit map $\Delta(x)$
between operators and fields
\begin{equation}
  \label{eq:lat-mapping}
  \Delta(x)=\frac{1}{|\det{\frac{1}{a}\Sigma}|}\sum_{{m}}\lrx{\prod_{\nu=1}^d\lrx{\hat{Z}_\nu}^{m_\nu}}
  e^{-\pi i\sum_{\nu<\rho}\tilde{\Theta}_{\nu\rho}m_\nu m_\rho}e^{-2\pi i (\Sigma^{-1})_{\nu\mu}m_\nu x_\mu}\,.
\end{equation}
This map is the lattice analog of the continuum map
(\ref{eq:NCT-mapping}).  Here $x$ is a point on the space--time
lattice $(a\mathbb{Z})^d$. The map between fields and operators then reads
\begin{eqnarray}
  \label{eq:lat-map1}
  \hat{f}&=&\sum_xf(x)\Delta(x),\\ 
  \label{eq:lat-map2}
  f(x)&=&\trace\lrx{\hat{f}\Delta(x)}\,,
\end{eqnarray}
where the operator trace is uniquely given by
\begin{equation}
  \label{eq:lat-trace}
  \trace \hat{f}=\sum_x f(x)\andx \trace\Delta(x)=1\qquad \text{for all }x\in(a\mathbb{Z})^d\,.
\end{equation}

The lattice version of the star--product (\ref{eq:star-product})
can be found by applying the inverse map (\ref{eq:lat-map2}) to the
product of two Weyl operators $\hat{f}\,\hat{g}$ 
\begin{equation}
  \label{eq:lat-star-product}
  f(x)\star g(x)\defi \trace\lrx{\hat{f}\,\hat{g}\Delta(x)}=
  \frac{1}{|\det\frac{1}{a}\Sigma|}\sum_{y,z}e^{-2i\lrx{\Theta^{-1}}_{\mu\nu}y_\mu z_\nu}
  f(x+y)g(x+z)\,.
\end{equation}
As in the continuum we obtain two equivalent descriptions of
non--commu\-ta\-tive space--time; either we use Weyl--operators or
functions on $(a\mathbb{Z})^d$ with a deformed product.

\section{Non--commutative field theory on the lattice}
\label{sec:lat-dft}

With the definitions of the last Section it is straightforward to
construct field theories on a non--commutative lattice. In the
case of a scalar theory we use the map (\ref{eq:lat-map1}) to define
the operators $\hat{\phi}$. The action in terms of Weyl operators then
reads
\begin{equation}
  \label{eq:lat-scalar-action}
  S[\hat{\phi}]=\trace\lrx{\frac{1}{2}\sum_\mu\lrx{\hat{D}_\mu\hat{\phi}\,\hat{D}_\mu^\dagger-\hat{\phi}}^2
    +\frac{1}{2}\hat{\phi}^2+\frac{1}{4}\hat{\phi}^4}\,,
\end{equation}
where $\hat{D}_\mu$ is the lattice shift operator defined in equation
(\ref{eq:lat-shift})\,. Using the inverse map (\ref{eq:lat-map2})
leads to the lattice action of the field $\phi(x)$
\begin{equation}
  \label{eq:lat-phi-action}
  \hspace{0.12cm}S[\phi]\!=\!\sum_x\lrx{\frac{1}{2}\sum_\mu\lrx{\phi(x\!+\!a\hat{\mu})\!-\!\phi(x)}^2
  +\frac{m^2}{2}\phi^2(x)+\frac{\lambda}{4}\phi(x)\!\star\!\phi(x)\!\star\!\phi(x)\!\star\!\phi(x)}\,,
\end{equation}
where $\hat{\mu}$ is the unit vector in the direction $\mu$.  Since
also the lattice version of the star--product is invariant under
cyclic permutation, only the interaction term differs from the
commutative case, as in the continuum.

To construct a non--commutative Yang--Mills theory on the lattice we
need the non--commutative counterpart of the link variables
$U_\mu(x)$. In order to preserve star--gauge invariance the link
variables have to be star--unitary. This can be formally achieved by
taking the parallel transporter defined in
(\ref{eq:SGO-parallel-transporter}) between the lattice sites $x$ and
$x+a\hat{\mu}$
\begin{equation}
  \label{eq:lat-parallel-transporter}
  U_\mu(x)=\text{P}\exp_\star\lrx{i\int_x^{x+a\hat{\mu}}d\xi A_\mu(\xi)}\,,
\end{equation}
where $A_\mu$ is a non--compact $U(n)$ gauge field and $\exp_\star$
indicates that in the expansion of the exponential function the
star--product has to be used. $U_\mu$ is star--unitary and transforms
under star--gauge transformations as
\begin{equation}
  \label{eq:lat-star-gauge-trans}
  U_\mu(x)\to G(x)\star U_\mu(x)\star G^\dagger(x+a\hat{\mu})\,,
\end{equation}
where $G(x)$ is again a star--unitary matrix field introduced in
equation (\ref{eq:gau-star-unitary}).  The corresponding Weyl
operators are given by
\begin{equation}
  \label{eq:lat-weyl-gauge}
  \hat{U}_\mu=\sum_{x}\Delta(x)\otimes U_\mu(x)\,.
\end{equation}
The star-unitarity of the link variables $U_\mu$ is now equivalent
to unitarity of the Weyl operators. 

The action of a non--commutative Yang--Mills theory can again be
constructed with the use of the lattice shift operators
(\ref{eq:lat-shift}) or with the star--product
\begin{equation}
  \label{eq:lat-ymnc-action}
  \begin{split}
    S&=-\frac{1}{g^2}\sum_{\mu\neq\nu}\trace\,\tr_N\lry{\hat{U}_\mu\lrx{\hat{D}_\mu\hat{U}_\nu\hat{D}_\mu^\dagger}
      \lrx{\hat{D}_\nu\hat{U}_\mu^\dagger\hat{D}_\nu^\dagger}\hat{U}^\dagger_\nu}\\
    &=-\frac{1}{g^2}\sum_x\sum_{\mu\neq\nu}\tr_N\lry{U_\mu(x)\star U_\nu(x+a\hat{\mu})\star
      U^\dagger_\mu(x+a\hat{\nu})\star U_\nu^\dagger(x)}\,,
  \end{split}
\end{equation}
where $\tr_N$ is the trace in the fundamental representation of the
unitary group $U(N)$.\\

As in the continuum we find in coordinate space a simple recipe to
construct a field theory on a non--commutative plane: replace all
usual products in the commutative theory by star--products. In this
spirit we now define star--gauge invariant observables in non--commutative
lattice gauge theory, by replacing the products in the commutative Wilson
lines $\mathcal{U}(x,C_v)$ with star--products
\begin{equation}
  \label{eq:lat-wilson--loops}
  \mathcal{U}(x,C_v)=U(x,\mu_1)\star U(x+\hat{\mu}_1,\mu_2)\star\dots\,.
\end{equation}
Here $C_v$ is an arbitrary lattice contour with the displacement vector $v$
between the two end point of the contour.  In the commutative
case the trace over these lines is gauge invariant if the lines are
closed, either on the plane or over the boundary. This implies $v=0$
or $v_\mu=\Sigma_{\mu \nu}n_\nu$, respectively, where $n_\nu\in\mathbb{Z}$ indicates how
often the contour winds around the torus in the $\nu$-th direction.

On the non--commutative torus we can construct star--gauge invariant
observables for any separation vector $v$ by multiplication by plane waves
\begin{equation}
  \label{eq:lat-star-gauge-obs}
  \mathcal{O}(C_v)=\sum_x\tr_N\,\mathcal{U}(x,C_v)\star e^{ik_\mu x_\mu}\,,
\end{equation}
where the relation between the momentum $k$, carried by the contour, and the 
separation vector $v$ on the torus is given by
\begin{equation}
  \label{eq:lat-k-v}
  v_\mu=\Theta_{\mu\nu}k_\nu+\Sigma_{\mu \nu}n_\nu\,.
\end{equation}
The fact that open loops carry a momentum proportional to the
separation vector $v$ again displays the UV/IR mixing.

\section{Matrix model formulation}
\label{sec:finite}

In the last Section we constructed a scalar and a pure gauge action on
a non--commutative lattice in terms of Weyl operators and in terms of
functions with a deformed product, i.e.\ the star--product. The latter
allows for an intuitive extension from commutative field theories to
their non--commutative counterparts and it may seem to be suitable for
a direct simulation.  However, from a practical point of view this
formulation rises enormous problems for Monte Carlo simulations.

Already in the scalar case we have to implement the lattice version of
the star--product, which involves a sum over the complete lattice for
every product. This increases the needed computer time to an almost
unreachable amount. In addition to this, in gauge theory we have to
construct star--gauge invariant link variables $U_\mu$, which makes the
simulation even more expensive.  To avoid these problems we will use a
finite dimensional representation of the operator description.\\

Here we are leaving the general discussion and restrict ourselves to
the case that we studied numerically. This is a 2d non--commutative
space or subspace with the period matrix given by 
\begin{equation}
  \label{eq:lat-periods}
  \Sigma_{\mu\nu}=Na\delta_{\mu\nu}\,,
\end{equation}
where $a$ is the lattice spacing on a hyper--cubic lattice
and $N$ is the number of lattice sites in each direction. Then the
coordinate operator (\ref{eq:lat-coordinate}) and the dimensionless
non-commutativity tensor (\ref{eq:lat-theta}) are given by
\begin{equation}
  \label{eq:lat-coordinate-theta}
  \hat{Z}_\mu=e^{2\pi i \hat{x}_\mu/aN}\andx \tilde{\Theta}_{\mu\nu}=\frac{2\pi\Theta_{\mu\nu}}{a^2N^2}\,.
\end{equation}
The quantization of momentum and the non--commutativity parameter
$\theta$ read
\begin{equation}
  \label{eq:momenta-Theta}
  p_\mu=\frac{2\pi m_\mu}{aN}\andx \theta=\frac{1}{\pi}a^2N\,.
\end{equation}
The algebra (\ref{eq:lat-comm1}) then simplifies to
\begin{equation}
  \label{eq:lat-comm2}
    \hat{Z}_\mu\hat{Z}_\nu=e^{-4\pi i\epsilon_{\mu\nu}/N}\hat{Z}_\nu\hat{Z}_\mu\,;\qquad
    \hat{D}_\mu\hat{Z}_\nu\hat{D}_\mu^\dagger=e^{2\pi i \delta_{\mu\nu}/N}\hat{Z}_\nu\,,
\end{equation}
where $\epsilon_{\mu\nu}$ is the totally antisymmetric tensor.

The key step is now to replace the operators $\hat{Z}_\mu$ and
$\hat{D}_\mu$ in the algebra (\ref{eq:lat-comm1}) by $N\times N$
matrices, satisfying the relations (\ref{eq:lat-comm2}). For odd
values of $N$ there is a simple choice to achieve this, given by the
so--called {\em twist eaters} $\Gamma_\mu$. In two dimensions this
amounts to
\begin{equation}
  \label{eq:lat-gamma}
  \begin{split}
  \hat{D_1}=\Gamma_1=\delta_{i+1,j}\!\mod N &\andx \hat{D_2}=\Gamma_2=\mathcal{Z}_{12}^{*\,i-1}\delta_{i,j}\,\\
  \hat{Z}_1=\lrx{\Gamma_2}^{(N+1)/2}&\andx \hat{Z}_2=\lrx{\Gamma_1^\dagger}^{(N+1)/2}\,,
  \end{split}
\end{equation}
where we introduced the {\em twist} $\mathcal{Z}_{\mu\nu}$ in a way that will
be useful in the next Section. Inserting (\ref{eq:lat-gamma}) in
(\ref{eq:lat-comm2}) fixes the twist to
\begin{equation}
  \label{eq:lat-twist}
  \mathcal{Z}_{12}=\mathcal{Z}^*_{21}=\exp\lrx{\pi i (N+1)/N}\,. 
\end{equation}
The equations (\ref{eq:lat-gamma}) are of course only one
possible choice. For a general construction see for example
Refs.\ \cite{Ambjorn:1999ts,Ambjorn:2000cs,vanBaal:1986na}.

With these definitions an explicit form of the star--product
(\ref{eq:lat-star-product}) is provided and the operator $\Delta(x)$
becomes a $N\times N$ matrix
\begin{equation}
  \label{eq:lat-delta-explicit}
  \Delta(x)=\sum_{m_1,m_2=1}^N \hat{Z}_1^{m_1}\hat{Z}_2^{m_2}
  e^{-2\pi i m_1m_2/N}e^{-2\pi i m_\mu x_\mu/N}\,.
\end{equation}
The map between the functions $f(x)$ and the $N\times N$ matrices $\hat{f}$ in
terms of $\Delta(x)$ then reads
\begin{equation}
  \label{eq:lat-mapx}
  \hat{f}=\frac{1}{N^2}\sum_x f(x)\Delta(x)\,;\qquad f(x)=\frac{1}{N}\trace\lrx{\hat{f}\Delta(x)}\,,
\end{equation}
where the trace now refers to the $N\times N$ matrices. If we insert
$\Delta(x)$ explicitly in the left equation of (\ref{eq:lat-mapx}) we
obtain the matrices $J({m})$. Since the Fourier transform of $f(x)$ is
given by
\begin{equation}
  \label{eq:lat-J1}
  \tilde{f}({{m}}) = \frac{1}{N}\sum_{{x}}f(x)e^{-2\pi i m_\mu x_\mu/N}\,,
\end{equation}
the matrices $J({m})$ read
\begin{equation}
  \label{eq:lat-J2}
  J({m})=\hat{Z}_1^{m_1}\hat{Z}_2^{m_2}e^{-2\pi i m_1m_2/N}
\end{equation}
and we obtain a map between Weyl operators and the Fourier transform
$\tilde{f}({m})$.
\begin{equation}
  \label{eq:lat-J3}
  \hat{f}=\frac{1}{N}\sum_{{m}}\tilde{f}({m})J({m})\andx \tilde{f}({m})
  =\frac{1}{N}\trace\lrx{\hat{f}\,J^\dagger({m})}\,.
\end{equation}

This completes the construction of a finite dimensional representation
of field theory on a non--commutative lattice. In Chapter
\ref{cha:TEK} and \ref{phi} we will construct explicitly the action of
2d Yang--Mills and 3d scalar field theory (with two non--commuting
dimensions) in this representation.



\chapter{Numerical studies of non--commutative gauge theory}
\label{cha:TEK}

In this Chapter we present the results of our studies of 2d
non--commutative $U(1)$ gauge theory. As already discussed in the
previous Chapter the lattice version of non--commutative gauge theory
in coordinate space (\ref{eq:lat-ymnc-action}) is not immediately
suitable for computer simulations. Therefore we make use of the finite
dimensional representation introduced at the end of the last Section.
The model we will arrive at is the {\em twisted Eguchi--Kawai model}
(TEK).  We start with a discussion of the origins of this model.

\section{The twisted Eguchi--Kawai model }
\label{sec:TEK}

\subsection{History of the TEK}
\label{sec:TEK-hist}

In 1982 Eguchi and Kawai conjectured that standard (commutative)
$U(N)$ and $SU(N)$  lattice gauge theory {\em in the large $N$ limit}
is equivalent to their dimensional reduction to $d=0$ (one point)
\cite{Eguchi:1982nm}. The link variables are replaced by
$U_{\mu}(x)\to U_\mu$, and the partition function of the Eguchi--Kawai
model (EK) simplifies to
\begin{equation}
  \label{eq:tek-ek-action}
  \begin{split}
    Z&=\int\prod_{\mu=1}^d dU_\mu \,e^{-S_{\text{EK}}[U]}\\
    S_{\text{EK}}[U] &= -\beta\sum_{\mu\neq\nu}\text{Tr}\lrx{U_\mu\, U_\nu\, U^\dagger_\mu\, U_\nu^\dagger}\,,
  \end{split}
\end{equation}
where $\mu,\nu=1\dots d$. Eguchi and Kawai proved that in the large
$N$ limit both models have the same Schwinger--Dyson equations --- and
therefore the same Wilson loops --- if the global $U(1)^d$ symmetry of
the phases is not spontaneously broken. In $d=2$ this symmetry is
unbroken and the equivalence holds exactly, but in $d>2$ this is not
the case anymore.  There the equivalence holds only at
strong coupling \cite{Bhanot:1982sh}.\\

A way out of this restriction to two dimensions is using {\em twisted
boundary conditions} rather than periodic ones
\cite{Gonzalez-Arroyo:1983ub}.  It is known that the behavior of the
partition function at weak coupling differs significantly between
%
%
twisted and periodic bounda\-ries. Therefore one can expect that the
$U(1)^d$ symmetry breaking at weak coupling could be cured.

To obtain the twisted Eguchi--Kawai model one applies the
transformation
\begin{equation}
  \label{eq:tek-twist1}
  U_{\mu}(x)\to U'_{\mu}(x)=\mathcal{Z}_{\mu}U_{\mu}(x)\,,\qquad \mathcal{Z}_{\mu}\in \mathbb{Z}_N
\end{equation}
to the partition function with the standard Wilson gauge action \cite{Wilson:1974sk}
\begin{equation}
  \label{eq:tek-wilson-part}
  \begin{split}
  Z&=\int\mathcal{D}U\exp(-S[U])\\
  S[U]&=-\beta\sum_{x,\mu\neq\nu}\text{Tr}
  \lrx{U_{\mu}(x)U_{\nu}(x+a\hat{\mu})U^\dagger_{\mu}(x+a\hat{\nu})U_{\nu}(x)^\dagger}\,.
  \end{split}
\end{equation}
The $\mathcal{Z}_\mu$'s can in general depend on space--time, but for
the purpose here it is enough to keep just the dependence on the
orientation.  The integration measure is invariant under this
transformation and the transformed action reads
\begin{equation}
  \label{eq:tek-wil-twist}
  S[U]=-\beta\sum_{x,\mu\neq\nu}\mathcal{Z}_{\mu\nu}
    \text{Tr}\lrx{U_{\mu}(x)U_{\nu}(x+a\hat{\mu})U^\dagger_{\mu}(x+a\hat{\nu})U_{\nu}(x)^\dagger}\,.
\end{equation}
The twist $\mathcal{Z_{\mu\nu}}$ is the product of the factors
$\mathcal{Z}_{\mu}$ at the boundary of the plaquette. It can be
characterized by an integer valued anti--symmetric $d\times d$ matrix $n_{\mu\nu}$
\begin{equation}
  \label{eq:tek-gen-twist}
  \mathcal{Z}_{\mu\nu}=e^{2\pi i n_{\mu\nu}/N}
  =\mathcal{Z}^*_{\nu\mu}\qquad\text{with}\qquad n_{\mu\nu}\in\mathbb{Z}\,.
\end{equation}
Neglecting the space--time dependence of the link variables $U_\mu$
then defines the TEK action
\begin{equation}
  \label{eq:tek-tek-action}
    S_{\text{TEK}}[U] = -\beta\sum_{\mu\neq\nu}\mathcal{Z}_{\mu\nu}
    \text{Tr}\lrx{U_\mu\, U_\nu\, U^\dagger_\mu\, U_\nu^\dagger}\,,
\end{equation}
and a general Wilson loop spanned by $I\times J$ plaquettes is given by
\begin{equation}
  \label{eq:tek-wilsonloop}
    W_{\mu\nu}(I\times J) = \mathcal{Z}_{\mu\nu}^{IJ}\text{Tr}
  \lrx{U_\mu^I\, U_\nu^J\, U^{\dagger I}_\mu\, U_\nu^{\dagger J}} \,.
\end{equation}
In Ref.\ \cite{Gonzalez-Arroyo:1983ac} it is shown that with
this action the equivalence between the TEK and commutative $U(N)$
and $SU(N)$ lattice gauge theory also holds in $d>2$.

Here we are concerned with the 2d TEK. It seemed for a long time
that adding a twist is not highly motivated in $d=2$, since there 
even the ordinary EK model coincides with lattice gauge theory
in the planar large $N$ limit, which was solved analytically
by Gross and Witten \cite{Gross:1980he}.\\

However, the situation changed suddenly due to a new interpretation of
the TEK as an equivalent description of non--commutative gauge theory
\cite{Aoki:1999vr}. This equivalence was established by embedding the
(dynamically generated) coordinates and momenta of the reduced model
into matrices.  These matrices can be mapped on functions, where the
trace turns into an integral and the star products arise, so that one
arrives at action (\ref{eq:gau-yang-mills2}).  The construction of non--commutative
$U(n)$ gauge theories (for certain $n \in \{ 1,2, \dots \}$) works out
in this way at $N = \infty$. 
\footnote{Note that $N$ is here the size of the matrices $U_\mu$,
  whereas $n$ refers to the gauge group $U(n)$.}
At finite $N$ the conditions cannot be
matched at the boundaries.

\subsection{TEK at finite $N$}
\label{sec:TEK-nc}

An interpretation of the TEK at finite $N$ occurred when Refs.\ 
\cite{Ambjorn:1999ts,Ambjorn:2000nb,Ambjorn:2000cs} pointed out
that the twisted reduced model can be mapped exactly on a
non--commutative $U(n)$ lattice gauge theory 
\footnote{Remember that $SU(n)$ gauge theories so far cannot be
  constructed on a non--commutative plane and therefore this
  equivalence exists only for $U(n)$ gauge theories.}
in general dimension and
rank of the gauge group. Here we will describe the case of $d=2$ and the
star--unitary gauge group $U(1)$.

To show the equivalence we will use the finite dimensional
representation of a non--commutative geometry, developed in Section
\ref{sec:finite}.  The starting point is the action of 2d
non--commutative lattice gauge theory in terms of operators
(\ref{eq:lat-ymnc-action})
\begin{equation}
  \label{eq:TEK-action-special}
  S=-\beta\,\trace\,\lry{\hat{U}_1\lrx{\hat{D}_1\hat{U}_2\hat{D}_1^\dagger}
    \lrx{\hat{D}_2\hat{U}_1^\dagger\hat{D}_2^\dagger}\hat{U}^\dagger_2}\,.
\end{equation}
Inserting the finite dimensional representation
(\ref{eq:lat-delta-explicit}) of the general shift operator
$\hat{D}_\mu$ leads to
\begin{equation}
  \label{eq:TEK-action-special1}
  S=-\beta\,\trace\,\lry{\hat{U}_1\lrx{\Gamma_1\hat{U}_2\Gamma_1^\dagger}
    \lrx{\Gamma_2\hat{U}_1^\dagger\Gamma_2^\dagger}\hat{U}^\dagger_2}\,.
\end{equation}
The twist eaters $\Gamma_\mu$ obey the Weyl--'t Hooft commutation relation
\begin{equation}
  \label{eq:TEK-twisteater}
  \Gamma_1\Gamma_2=\mathcal{Z}_{12}^*\Gamma_2\Gamma_1\,.
\end{equation}
The substitution $\hat{U}_\mu\Gamma_\mu=V_\mu$ in equation
(\ref{eq:TEK-action-special1}) along with the commutation relation
(\ref{eq:TEK-twisteater}) for the inner $\Gamma$'s leads to the TEK
action (\ref{eq:tek-tek-action}). 

In this representation the twist reads
$\mathcal{Z}_{12}=\mathcal{Z}^*_{21}=\exp\lrx{\pi i (N+1)/N}$.
Comparing this with the twist needed for the TEK, where the exponent
is given by $2\pi i /N$ multiplied by some integer, we find that
$(N+1)/2\in \mathbb{Z}$ has to be fulfilled. Therefore we have to use
{\em odd} values of $N$.\\

Note that this interpretation of the TEK differs significantly from
the original one. While in the original interpretation the space--time
degrees of freedom become irrelevant in the large $N$ limit, in the
interpretation here the space--time degrees of freedom are {\em
  exactly} mapped onto matrix degrees of freedom already at finite
$N$.

\subsection{Continuum limits}
\label{sec:TEK-cont}

The continuum limit $a\to0$ also requires the limit $N\to\infty$, where 
the order of these limits is important for the resulting continuum theory.

\subsubsection{The planar limit}

In the planar limit one first sends $N\to\infty$, keeping the lattice
spacing $a$ fixed, followed by the continuum limit $a\to0$ along with
$\beta \rightarrow \infty$ in a particular way dictated by the
coupling constant renormalization. Equation (\ref{eq:momenta-Theta})
implies that in this limit $\theta\to\infty$ and only planar diagrams
remain. In $d=2$ there exists an exact solution of planar $U(\infty)$
and $SU(\infty)$ \cite{Gross:1980he}: rectangular Wilson loops follow
the area law
\begin{equation}
  \label{eq:tek-GWarea}
  w(I\times J) = \exp(-\kappa(\beta)IJ)\,,
\end{equation}
where $w(I\times J)$ is the expectation value of the Wilson loops.
The string tension $\kappa(\beta)$ is given by
\begin{equation}
  \label{eq:tek-kappa}
  \kappa (\beta ) = \left\{
    \begin{array}{cccc}
      - \ln \beta &&& \beta < \frac{1}{2} \\
      - \ln \Big( 1 - \frac{1}{4 \beta} \Big) &&& \beta \geq \frac{1}{2}
\end{array} \right. \quad .
\end{equation}
Equation (\ref{eq:tek-GWarea}) shows how the bare coupling has to be tuned 
as a function of $a$ when one takes the continuum limit. In this case 
the scaling is exact if one identifies the lattice spacing $a$ as
\begin{equation}
  \label{eq:tek-lattice-spacing}
  a=\sqrt{\kappa(\beta)}\,.
\end{equation}

The equivalence between commutative $U(N)$ and $SU(N)$ theory and the
TEK (and in two dimensions also EK) states that
\begin{equation}
  \label{eq:tek-eqivalence}
  \lim_{N\to\infty}\frac{1}{N}\langle W_{12}(I\times J)\rangle=w(I\times J)\,.
\end{equation}
This agreement was already shown in Ref.\ 
\cite{Eguchi:1982nm,Fabricius:1983au} and later on in Ref.\ 
\cite{Nakajima:1998vj} were finite $N$ effects where studied.

\subsubsection{The double scaling limit}

In the double scaling limit one takes the thermodynamic limit
$N\to\infty$ and the continuum limit $a\to0$ at the same time keeping
$N a^2=\pi \theta$ fixed. Here we used the relation
(\ref{eq:momenta-Theta}) between $\theta$ and the lattice spacing.
This leads to a finite non--commutativity parameter $\theta$ in the
continuum limit and one recovers non--commutative 2d $U(1)$ theory in
$\mathbb{R}^d$.

This is the continuum limit we studied. $\beta$ has to be scaled as a
function of the lattice spacing $a$ in exactly the same way as in the
planar theory. From equation (\ref{eq:tek-kappa}) we read off
$\beta\propto a^{-2}$ at large $\beta$.  Hence in $d=2$ we are going
to search for a {\em double scaling limit} keeping the ratio $N /
\beta $ constant.  The question of renormalizability of 2d
non--commutative gauge theory at finite $\theta$ can be answered by
studying this large $N$ limit of the TEK.  If the observables converge
to finite values, we can conclude that 2d lattice non--commutative
gauge theory does have a finite continuum limit. Our results will be
presented in the next Section.


\section{2d non--commutative $U(1)$ theory}
\label{tek}

\subsection{The model}
\label{tek-model}

We study the double scaling limit in two dimensional non--commutative
$U(1)$ lattice theory. The explicit action in terms of dimensionless
quantities reads
\begin{equation}
  \label{eq:tek-action-explicit}
      S_{\text{TEK}}[U] = -N\beta\sum_{\mu\neq\nu}\mathcal{Z}_{\mu\nu}
    \text{Tr}\lrx{U_\mu\, U_\nu\, U^\dagger_\mu\, U_\nu^\dagger}\,,
\end{equation}
with the twist given by (\ref{eq:lat-twist}). If not stated differently
we set $N/\beta=32$.
For details of the simulation see Appendix \ref{sec:numerics-1}.\\

Our main interest in this model is whether a finite continuum limit
exists or not.  Thus we study the double scaling limit --- as
described in the previous Section --- of various observables. If we
find a large $N$ scaling of these observables (at fixed $N/\beta$)
this corresponds to a finite continuum limit.

The observables we measure are constructed from Wilson loops spanned
by $I\times J$ plaquettes
\begin{equation}
  \label{eq:Ntek-wilsonloops}
  W_{\mu\nu}(I\times J) = \mathcal{Z}_{\mu\nu}^{IJ}\text{Tr}
  \lrx{U_\mu^I\, U_\nu^J\, U^{\dagger I}_\mu\, U_\nu^{\dagger J}} 
\end{equation}
and from {\it open} Polyakov lines,
\begin{equation}
  \label{eq:Ntek-polone}
  P_\mu(I)=\text{Tr}\lrx{U_\mu^I}\qquad\text{and}\qquad P_{-\mu}(I)=\text{Tr}\lrx{U_{-\mu}^{\dagger I}}\,,
\end{equation}
where $I$ is the number of multiplied link variables and ${-\mu}$
indicates the backward direction.

In Section \ref{sec:lat-dft} we showed that one can construct
star--gauge invariant observables from general open lines. Here we
restrict ourselves to straight lines which are {\em not} necessarily
closed by the boundary.

\subsection{Wilson loops and area law}
\label{sec:tek-area}

First we study the expectation values of Wilson loops. The expectation
value of open Polyakov lines is vanishing. In fact the expectation
value of any open contour vanishes due to the $U(1)^d$ symmetry of the
TEK. But $n$--point functions of these contours are sensible
observables.

In the EK model it was observed that square shaped Wilson loops
converge faster to the known exact result than other rectangles with
the same area \cite{Nakajima:1998vj}.  Hence we also focus on square
shaped Wilson loops, which corresponds to $I=J$ in the definition
(\ref{eq:Ntek-wilsonloops}). We define the normalized Wilson loop
\begin{equation}
  \label{eq:Ntek-norm-wilsonloop}
  W(I)=\frac{1}{N}\langle W_{12}(I\times I)\rangle\in\mathbb{C}\,.
\end{equation}
Due to the presence of the twist, there is no invariance under the
transformation $U_1\to U_2$, $U_2\to U_1$. As a consequence
$W_{\mu\nu}$ is in general complex and $W_{12}=W_{21}^*$, hence the
Wilson loop depends on the orientation. The real part represents the
average over both orientations.
\begin{figure}[htbp]
  \centering
  \includegraphics[width=.85\linewidth]{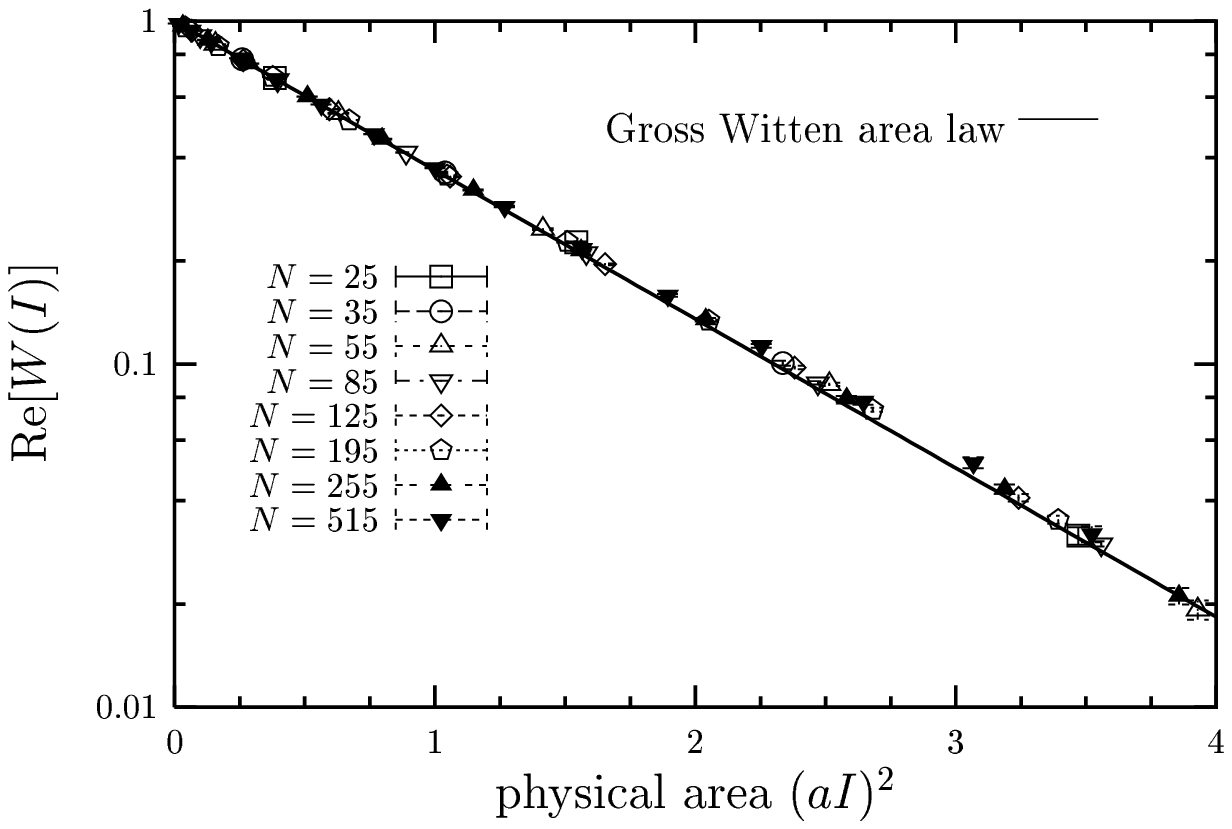}

  \vspace{.8cm}
  \includegraphics[width=.84\linewidth]{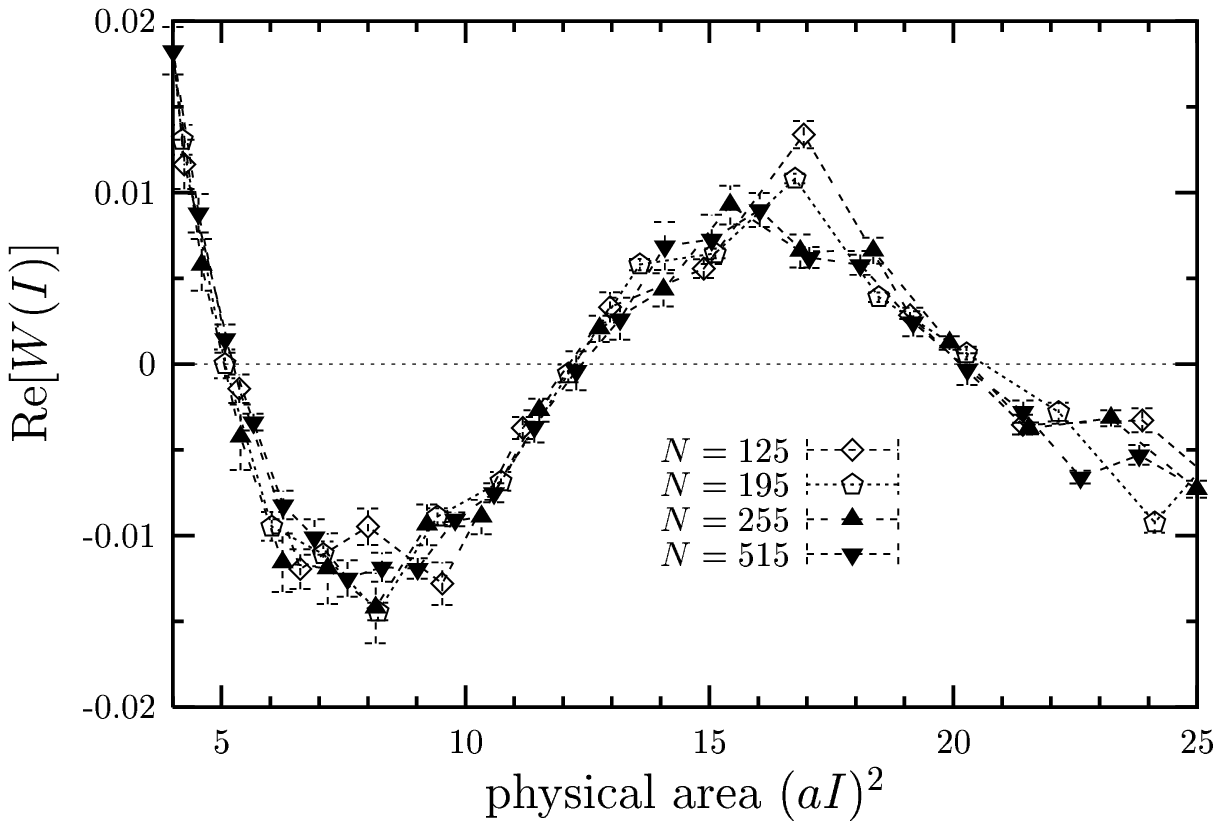}
  \caption{The real part of square shaped Wilson loops defined in Eq.\ (\ref{eq:Ntek-wilsonloops})
    against the physical area.  The plot at the top shows the Wilson
    loops for small areas, where the data follow the Gross--Witten
    area law (solid line). For large areas the real part of the Wilson loops deviate
    from this law and oscillates around zero instead.}
  \label{fig:wilsonloops1}
\end{figure}
\begin{figure}[htbp]
  \centering
  \includegraphics[width=.85\linewidth]{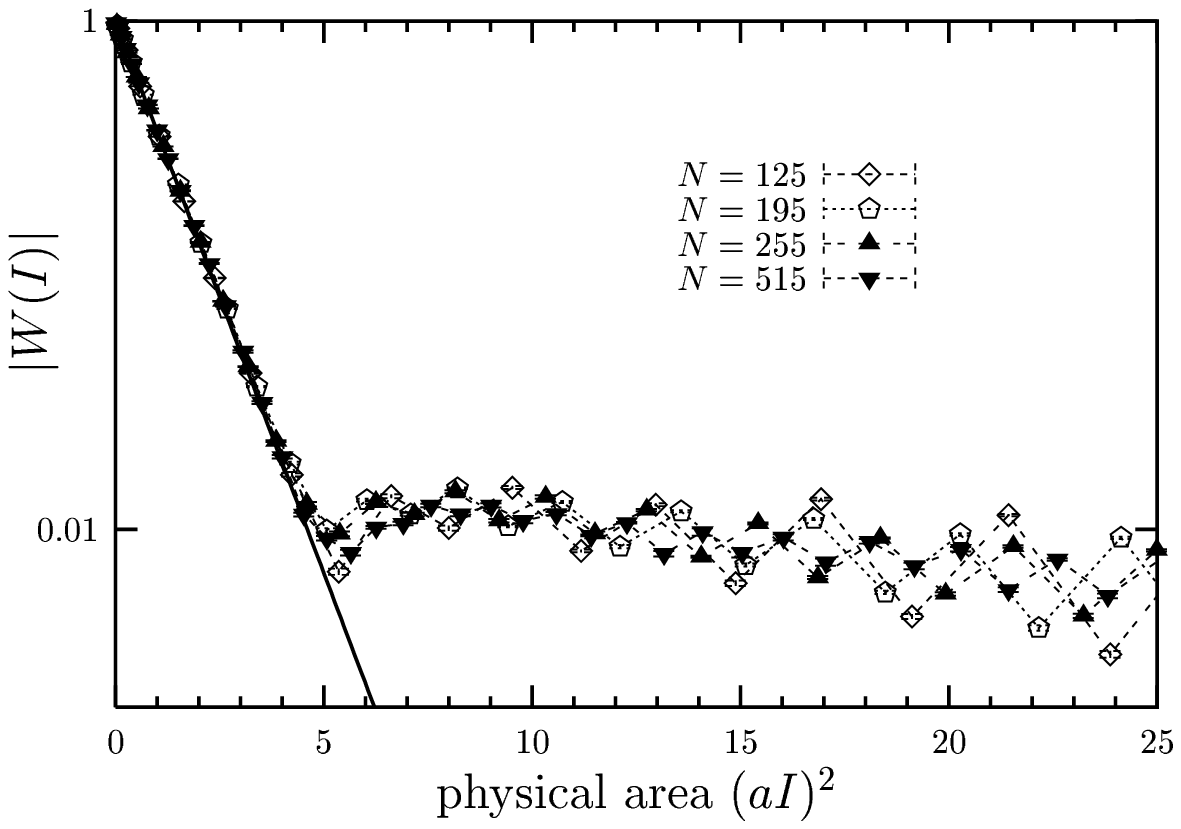}
  
  \vspace{.5cm}
  \includegraphics[width=.85\linewidth]{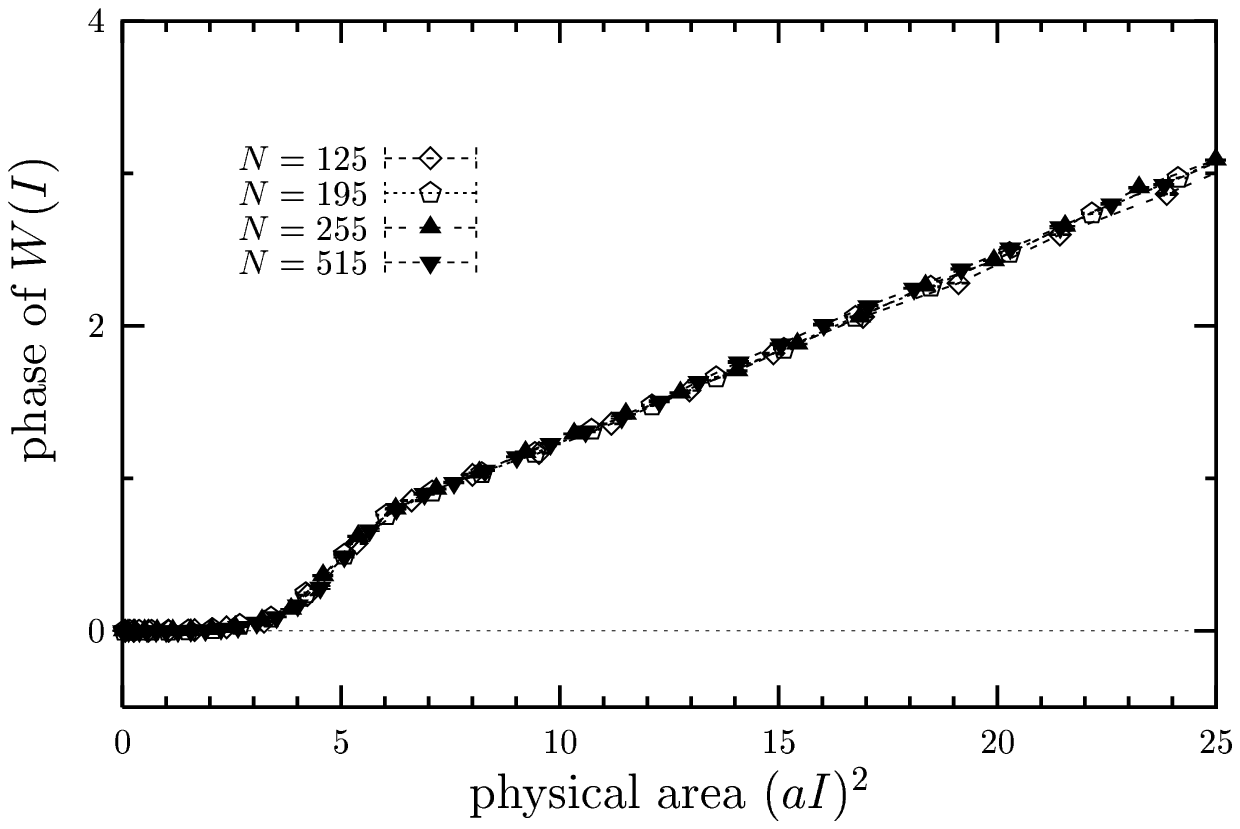}
  \caption{Wilson loops against the physical area in polar coordinates. The solid
    line in the plot at the top shows the Gross--Witten area law. The
    phase of the Wilson loops increases linearly in the area beyond
    the area law regime (bottom).}
  \label{fig:wilsonloops2}
\end{figure}
Figure \ref{fig:wilsonloops1} shows the real part of the Wilson loops
$\text{Re}[W(I)]$ as a function of the physical loop area.

Large $N$ scaling is clearly confirmed. At small up to moderate areas
the Wilson loop follows the area law of planar lattice gauge theory
given in Eq.\ (\ref{eq:tek-GWarea}), but at large areas it deviates
and the real part oscillates around zero instead. This observation is 
consistent over a wide range of $N$. Remarkably, not even the absolute
value decays monotonously; at large areas it seems to fluctuate
around an approximately constant value. This is illustrated in Figure
\ref{fig:wilsonloops2} (top).

\begin{figure}[htbp]
  \centering \includegraphics[width=.90\linewidth]{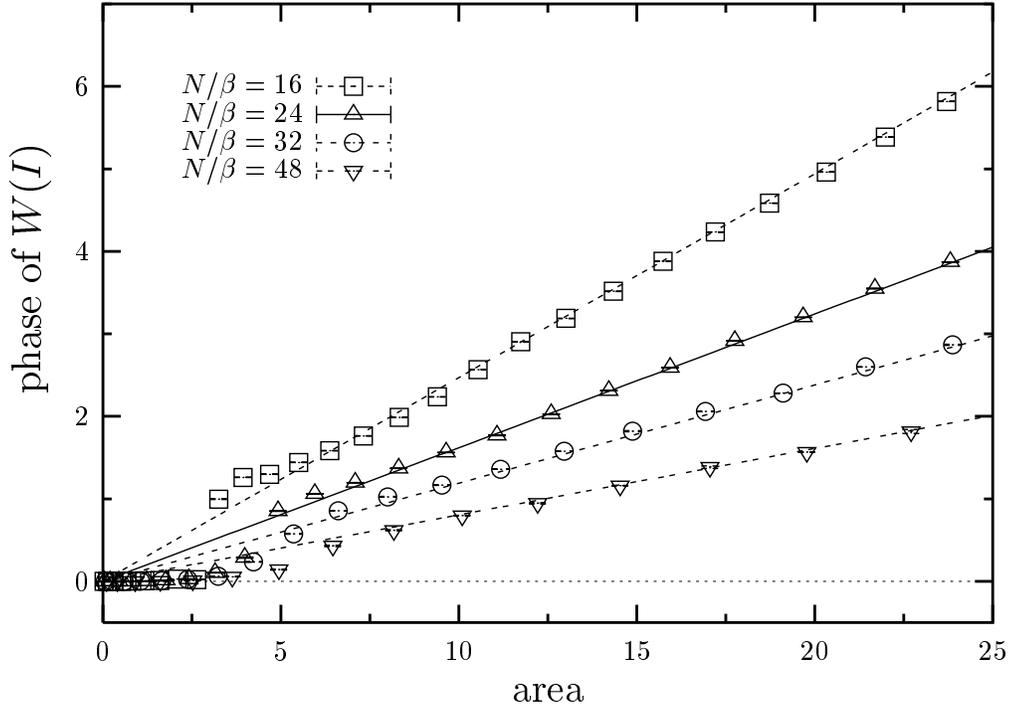}
  \caption{The phase of Wilson loops against the physical area for different
  values of $\theta$. The phase increases linear in the area beyond the 
  Gross--Witten regime for all $\theta$.}
  \label{fig:wilson-theta}
\end{figure}

Figure \ref{fig:wilsonloops2} (below) shows that beyond the
Gross--Witten regime, the phase increases linearly in the area.
Additional measurements at $N/\beta=16,24,48$, corresponding to
different values of $\theta$, are shown in Figure
\ref{fig:wilson-theta}. They reveal that the phase $\Phi$ of the
Wilson loop $W(I)$ is, in fact, given the simple relation
\begin{equation}
  \label{eq:Ntek-phase}
  \Phi=\frac{\text{area}}{\theta}\,.
\end{equation}
where the area is given by $(aI)^2$.  This relation holds to a high
accuracy at large areas $(aI)^2>O(\theta)$, i.e.\ beyond the
Gross--Witten regime.  Relation (\ref{eq:Ntek-phase}) has been
confirmed also for other (non--square) rectangular Wilson loops, which
shows that the effect does not depend on the shape of the Wilson
loops.
\footnote{Note, however, that the expectation values of Wilson loops
  with the same area but with different shapes have in general
  different absolute values in the double scaling limit.}
Indeed the formula (\ref{eq:Ntek-phase}) agrees with a Aharonov--Bohm
effect in the presence of a constant magnetic field $B=1/\theta$
across the plane.  This is reminiscent of the description of
non--commutative gauge theory by Seiberg and Witten
\cite{Seiberg:1999vs}. This interpretation is also known in solid
state physics. Our numerical results seem to support a picture of this
kind.

The observed large area behavior of the Wilson loops confirms that the
continuum limit of non--commutative gauge theory is different from any
ordinary gauge theory, hence we have found a new universality class.
Since the non--locality in this model is of order $\sqrt{\theta}$, one
might naively think that in the large area regime,
$\text{area}\gg\sqrt{\theta}$, the effect of non--commutativity is
invisible. The fact that we the observe contrary can be understood as
a manifestation of non--perturbative UV/IR mixing.

In the two dimensional model there is no perturbative UV/IR mixing,
since there are no UV divergent diagrams in the commutative case. Here
we found this mixing at a completely non--perturbative level.\\

In the double scaling limit of the untwisted Eguchi--Kawai model, the expectation
values of the Wilson loops is real, and it remains positive even at
large areas \cite{Nakajima:1998vj}.
This means that the two models, twisted and untwisted, yield 
qualitatively different double scaling limits, although they become
identical in the planar large $N$ limit.

\subsection{2--point functions}
\label{sec:tek-two-2}

\begin{figure}[htbp]
  \centering
  \includegraphics[width=.85\linewidth]{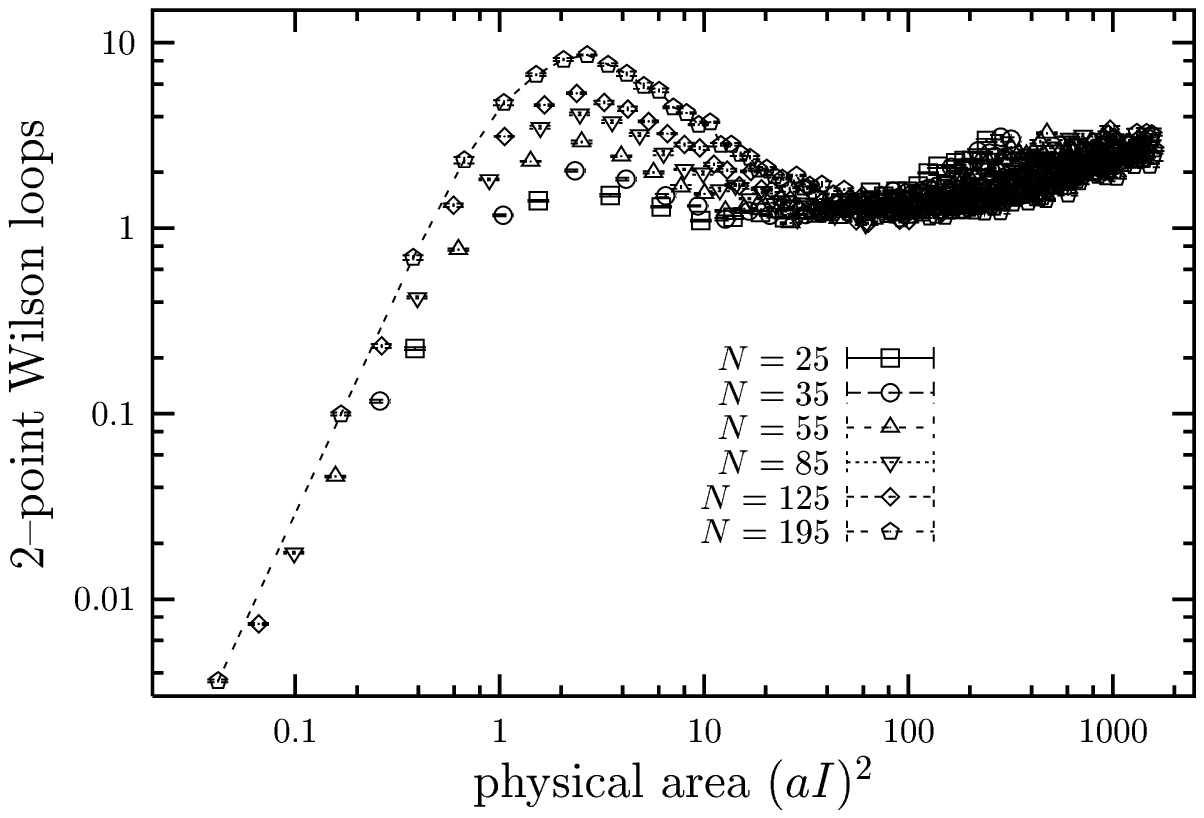}\\

  \vspace{2.5cm}
  \includegraphics[width=.85\linewidth]{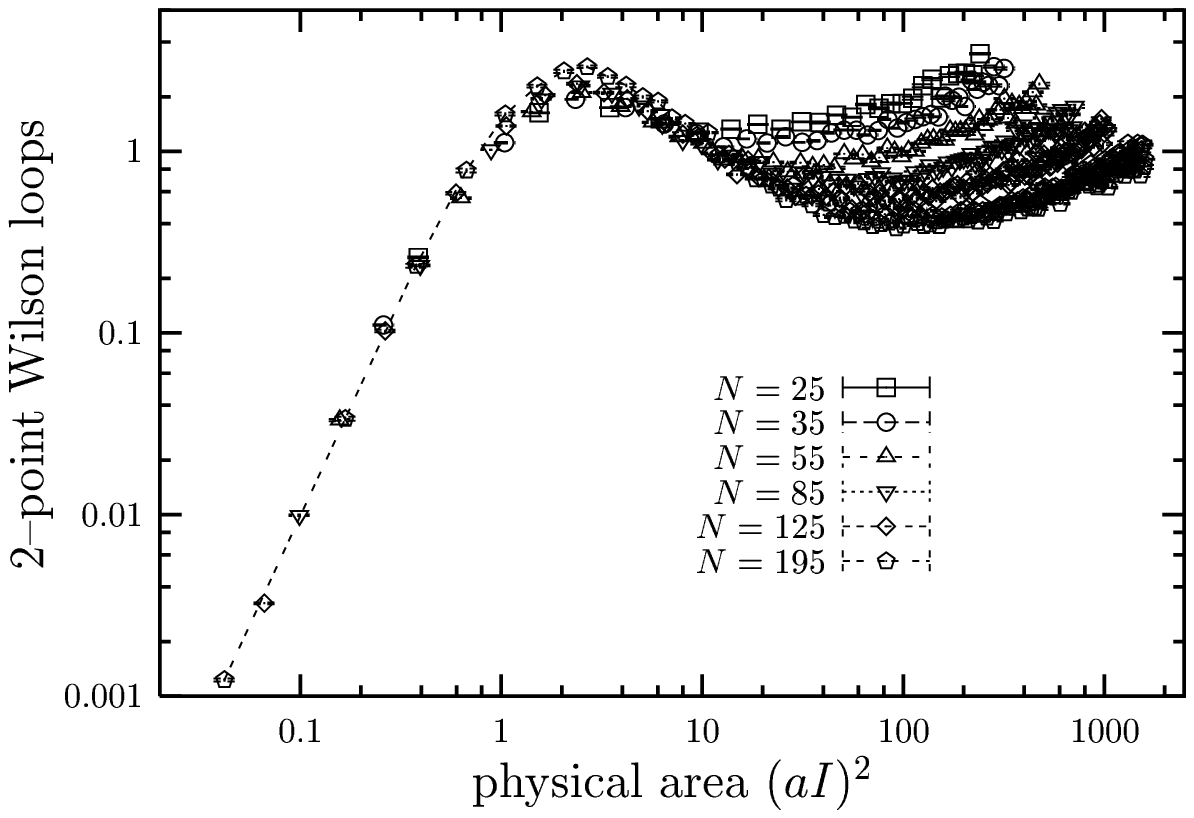}
  \caption{Two--point functions of square shaped Wilson loops (\ref{eq:Ntek-wiltwo}) against the physical area.
    The raw data at the top do not show any scaling behavior, but with
    a wave function renormalization $G^W_2\to\beta^{-0.6}G^W_2$ there
    is a clear scaling regime (bottom).}
  \label{fig:wiltwo}
\end{figure}

Here we consider 2--point functions.  Figure \ref{fig:wiltwo} (top)
shows the connected Wilson loop 2--point function
\begin{equation}
  \label{eq:Ntek-wiltwo}
  G^W_2(I)=\langle W_{12}(I\times I)W_{21}(I\times I)\rangle
  -\langle W_{12}(I\times I)\rangle\langle W_{21}(I\times I)\rangle \in \mathbb{R}\,,
\end{equation}
again plotted against the physical area, for $I=1 \dots N$.  In
contrast to the Wilson loop itself, $G^W_2(I)$ is a real quantity,
since both orientations of the Wilson loop are involved. 

In Figure \ref{fig:wiltwo} (below) we include a wave function
renormalization factor
\begin{equation}
  \label{wave1}
  G_{2}^{(W)} \to \beta^{-0.6} G_{2}^{(W)} \,.
\end{equation}
The exponent $-0.6$ was found to be optimal for $G_{2}^{(W)}$ to scale.
Indeed it leads to a neat large $N$ scaling over more than two 
orders of magnitude in the physical area.\\

\noindent Next we consider the Polyakov line
\begin{equation}
\label{eq:tek-pol}
P_{\mu}(I) = {\rm Tr} \Big( U_{\mu}^{I} \Big) \,,
\end{equation}
which is also $U(N)$ invariant and therefore has an interpretation as
a star gauge invariant observable in non--commutative gauge theory.
Their momentum $\vec{p}$ is related to the separation vector $\vec{v}$
between the two ends of the line.  In general the relation is given by
$v_\mu = \Theta_{\mu\nu} p_{\nu}$ modulo the periodicity of the torus,
as discussed in Section (\ref{sec:lat-dft}). In the present case, the
Polyakov line $P_{\mu}(I)$ corresponds to a momentum mode with
$p_{\nu} = 2 \pi \ell / (Na)$, where the integer $\ell$ is given by
$I/2$ and by $(I+N)/2$ for even and odd $I$, respectively.  In the
following, we plot our results against the physical distance $aI$ for
even $I = 2,\dots, N-1$.

\begin{figure}[htbp]
  \centering
  \includegraphics[width=.84\linewidth]{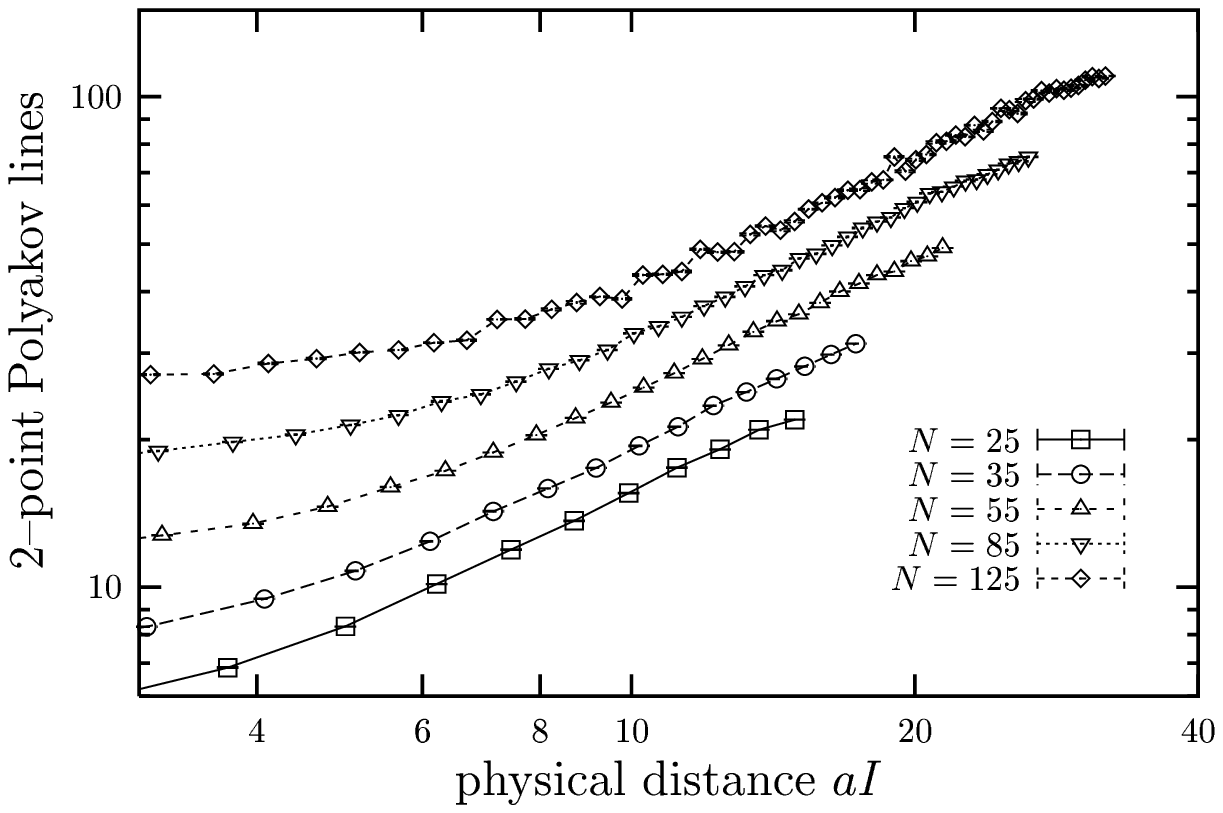}\\

  \vspace{2.5cm}
  \includegraphics[width=.83\linewidth]{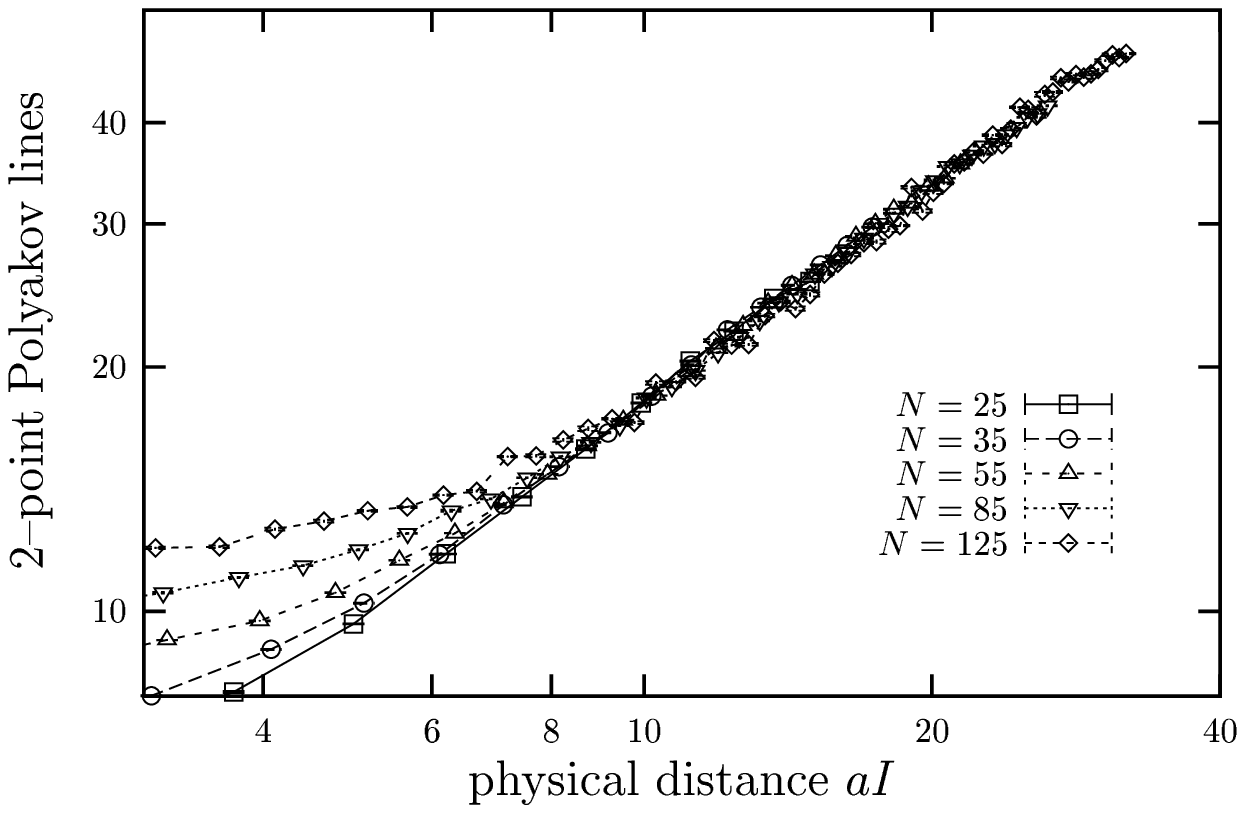}
  \caption{Two--point functions of open Polyakov (\ref{eq:Ntek-poltwo}) against the physical area.
    The plot at the top shows that the raw data do not scale in $N$.
    Applying a wave function renormalization
    $G^P_2\to\beta^{-0.6}G^P_2$ leads to large $N$ scaling in a
    significant interval.}
  \label{fig:pol2}
\end{figure}

The phase symmetry \footnote{In the terminology of non--commutative gauge theory,
this corresponds to momentum conservation.}
makes $\langle P_{\mu}(I) \rangle$ vanish, 
but the connected $n$-point functions ($n>1$) of Polyakov lines are
sensible observables.
In Figure \ref{fig:pol2} we show the 2--point function
\footnote{The choice of the direction $\mu$ is irrelevant. In practice 
we average over both possibilities in order to enhance the statistics.}
\begin{equation}
  \label{eq:Ntek-poltwo}
  G^P_2(I)=\langle P_{\mu}(I)P_{-\mu}(I)\rangle \in \mathbb{R}\,.
\end{equation}
Note that there is no disconnected part in $G_{2}^{(P)}$.
Again we insert the wave function renormalization which was optimal for
the Wilson 2--point function
\begin{equation} 
  \label{renorm}
  G_{2}^{(P)} \to \beta^{-0.6 } \, G_{2}^{(P)} \,.
\end{equation}
As a function of the physical length $aI$, the result is consistent
with large $N$ scaling, as well as a {\em universal}
wave function renormalization.
A similar wave function renormalization was also observed in the
EK model \cite{Nakajima:1998vj}, where the optimal factor in
relation (\ref{renorm}) is modified to $\beta^{-0.65}$.\\

The large $N$ scaling of the Wilson loops as well as the large $N$
scaling of 2--point functions of Wilson loops and Polyakov lines,
described in this Chapter, correspond to a finite continuum limit in 2d
non--commutative $U(1)$ gauge theory. This observation therefore
demonstrates the renormalizability of 2d non--commutative $U(1)$
theory.



\chapter{Numerical studies of the $\lambda\phi^4$ model}
\label{phi}

The second model we investigated is the 3d non--commutative
$\lambda\phi^4$ theory. In Section \ref{scalar} we described the
effects of the UV/IR mixing effect based on results of one loop
calculations. We studied this model non--perturbatively and in this
Chapter we present our results.  For details of the simulations we
refer to Appendix \ref{sec:numerics-2}.

\section{Dimensionally reduced model}
\label{sec:phi-drm}

Since we are in odd dimensions we cannot apply directly the
construction of non--commutative field theories as described in
Chapter \ref{cha:NCFT} and \ref{cha:lattice}. There the
anti--symmetric non--commutativity tensor $\Theta$ had to be
invertible. This restricts the dimension to be even.

In addition to this rather technical problem related to odd
dimensions, a non--commutative time could give rise to unitarity
problems \cite{Seiberg:2000gc,Bozkaya:2002at}.  Ref.\ 
\cite{Gomis:2000zz} showed that field theories on a
non--commutative space satisfy the generalized unitarity relations. If
the time is also non--commutative this is not the case.  Therefore we
exclude the time direction from non--commutativity.
\footnote{In four dimensions the problem is often avoided by taking two
  commuting and two non--commuting directions.}

Then the non--commutativity tensor $\Theta$ is two dimensional
and acts only in the 2d non--commutative subspace. The star--product
defined in (\ref{eq:star-product}) then reads
\begin{equation}
  \label{eq:phi-str-product}
  \phi_1(x)\star \phi_2(x) \defi \phi_1(x) \exp\lrx{\frac{i}{2}\overset{\leftarrow}{\pd_i}\Theta_{ij}
    \overset{\rightarrow}{\pd_j}}\phi_2(x)\with i,j=1,2\,,
\end{equation}
and the lattice version of the star--product is analogous. Here $\phi(x)$
means $\phi(x_1,x_2,t)$, where $x_1,x_2$ satisfy the commutation
relation (\ref{eq:NC-comm-star}) and $t$ commutes with all
coordinates.

The corresponding Weyl operators then also depend on the time
$\hat{\phi}(t)$ and the lattice action of this version of
non--commutative $\lambda\phi^4$ theory reads, in analogy to
equation (\ref{eq:lat-scalar-action}),
\begin{equation}
  \label{eq:phi-action}
  \begin{split}
    S[\hat{\phi}]=N\trace\sum_{t=1}^T\biggl[\frac{1}{2}\sum_{i}&
      \lrx{\hat{D}_i\,\hat{\phi}(t)\,\hat{D}_i^\dagger-\hat{\phi}(t)}^2\\
      \frac{1}{2}&\lrx{\hat{\phi}(t+1)-\hat{\phi}(t)}^2
      +\frac{m^2}{2}\hat{\phi}(t)^2+\frac{\lambda}{4}\hat{\phi}(t)^4\biggl]\,,
  \end{split}
\end{equation}
where $i=1,2$. There are now two kinetic terms: the first one uses the
shift operators (\ref{eq:lat-shift}) to perform spatial translations
in units of the lattice spacing $a$. The second kinetic term is the
square of the standard discrete derivative in time direction.

As in the TEK model we use here the finite dimensional representation
(\ref{eq:lat-gamma}). In this representation the Hermitian operators
$\hat{\phi}(t)$ turn into Hermitian matrices and the shift operators
are replaced by the twist eaters $\Gamma_i$.

Effectively we are mapping here a non--commutative $\lambda\phi^4$
lattice theory, defined on a three dimensional $N^2T$ lattice, to a
one dimensional lattice with $T$ sites.  
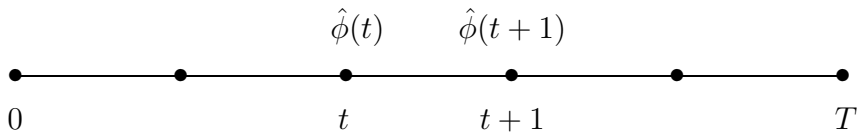
\begin{figure}[htbp]
  \begin{center}
    \setlength{\unitlength}{1mm}
    \begin{picture}(150,20)(-1,-1)
      \put(20,7){\line(1,0){110}}
      \put(19,6){$\bullet$}
      \put(41,6){$\bullet$}
      \put(63,6){$\bullet$}
      \put(85,6){$\bullet$}
      \put(107,6){$\bullet$}
      \put(129,6){$\bullet$}
      \put(19,0){$0$}
      \put(63,0){$t$}
      \put(81.8,0){$t+1$} 
      \put(129,0){$T$}    
      \put(62,12){$\hat{\phi}(t)$}
      \put(79,12){$\hat{\phi}(t+1)$}  
    \end{picture}
    \caption{The dimensionally reduced lattice. The fields $\phi(x_1,x_2,t)$ are mapped into
      Hermitian $N\times N$ matrices $\hat{\phi}(t)$.}
    \label{fig:phi-reduced-lattice}
  \end{center}
\end{figure}

 
%
On each site
there is a Hermitian $N\times N$ matrix $\hat{\phi}(t)$ representing
the Hermitian operators, see Figure \ref{fig:phi-reduced-lattice}.  We
use periodic boundary conditions $\hat{\phi}(T)=\hat{\phi}(0)$. In
our simulations we always set
$T=N$.\\

The dimensionless parameters $m^2$ and $\lambda$ in the action
(\ref{eq:phi-action}) can be identified with physical parameters
\begin{equation}
  \label{eq:phi-physical-parameters}
  \begin{split}
    m^2&=a^2m^2_\text{phys} =\frac{\pi\,\theta }{N}\,m^2_\text{phys}\\
    \lambda&=a \lambda_\text{phys}= \sqrt{\frac{\pi\,\theta }{N}} \,\lambda_\text{phys}\,.
  \end{split}
\end{equation}
Here we used the relation between the non--commutativity parameter
$\theta$ and the lattice spacing $\theta=\frac{1}{\pi}a^2N$
(\ref{eq:momenta-Theta}). With this identification the double scaling
limit described in Subsection \ref{sec:TEK-cont} leads to the
non--commutative $\lambda\phi^4$ model on $\mathbb{R}^3$. In this procedure
the limits $N\to\infty$ and $a\to0$ are taken such that $a^2N$ is kept
constant, leading to a finite non--commutativity parameter.

Here we will not study the continuum limit and the question of
renormalizability. Instead we study the phase diagram and the UV/IR
mixing effects in the regularized theory.\\

In Section \ref{sec:sc-uvir} we discussed this issue in $d=4$.  Since
the results depend on the dimension, we repeat the considerations of
Section \ref{sec:sc-uvir} for the case of three dimensions.

We start with the non--planar one loop contribution to the irreducible
two--point function in equation (\ref{eq:salar-poles}). In $d=3$ this
term reads
\begin{equation}
  \label{eq:phi-pert1}
    \Gamma^{(1)}_{\text{np}}(p)=\frac{\sqrt{m}}{{6(2\pi)^{3/2}}}
  \lrx{\frac{4}{\Lambda^2}+\theta^2 \vec{p}^{\;2}}^{-\frac{1}{4}}
  K_{\frac{1}{2}}\lrx{m\sqrt{\frac{4}{\Lambda^2}+\theta^2\vec{p}^{\;2}}}\,,
\end{equation}
where $\Lambda$ is again a momentum cut--off and $p=(p_0,\vec{p}\,)$.
Since we chose the time as commutative, only the spatial components
$\vec{p}$ of the momentum $p$ appear here.  Introducing the effective
cut--off $\Lambda_{\text{eff}}$ as in equation (\ref{eq:NC-cutoff})
\begin{equation}
    \label{eq:NC-cutoff-3d}
    \Lambda^2_{\text{eff}}=\frac{1}{\frac{1}{\Lambda^2}+\theta^2\vec{p}^{\,2}}\,,
\end{equation}
and evaluating the Bessel function $K_{1/2}$ leads the one loop
corrected two--point function
\begin{equation}
  \label{eq:phi-pert2}
    \Gamma(p)=p^2+M^2_\text{eff}+\xi\,\lambda\, 
    \Lambda_{\text{eff}} \;e^{-\frac{m}{\Lambda_{\text{eff}}}}+O(\lambda^2)\,,
\end{equation}
with $\xi=\frac{1}{6(2\pi)^{3/2}}$. As in Ref.\ \cite{Minwalla:1999px}
we absorbed the planar contribution into the effective mass
$M_\text{eff}$. After removing the cut--off $\Lambda$ in equation
(\ref{eq:phi-pert2}) we obtain the leading IR divergence
\begin{equation}
  \label{eq:phi-pert3}
    \Gamma(p)=p^2+M^2_\text{eff}+\xi\frac{\lambda}{\theta |\vec{p}\,|}\,.
\end{equation}
Again the two--point function is singular at zero momentum. However,
here $\Gamma(p)$ is not a function of $p^2$, which leads to a
different IR behavior of the theory.

\section{The phase diagram}
\label{sec:phi-pd}

As we discussed in Section \ref{gubser-sondhi} the phase diagram of
the non--commutative 3d $\lambda\phi^4$ theory is expected to differ
significantly from the phase diagram in the commutative case. In this
Section we present our Monte Carlo results for this phase diagram.

\subsection{The order parameter}
\label{sec:phi-order}

When studying a phase diagram one first has to identify a suitable
{\em order parameter} that indicates the symmetry breaking. In the
model here we expect an {\em Ising type phase}, where the discrete
symmetry $\phi(x)\to-\phi(x)$ is broken spontaneously. In addition we
expect --- for sufficiently large coupling $\lambda$ --- a {\em
  striped phase}, where the translation symmetry is broken
spontaneously (see Section \ref{gubser-sondhi}).  Therefore we need an
order parameter that is sensitive to both variants of symmetry
breaking, to distinguish the two types of ordered phases.

The momentum dependent quantity
\begin{equation}
  \label{eq:phi-order-1}
  \bar{M}(\vec{m})=\frac{1}{NT}\left|\sum_t\tilde{\phi}(\vec{m},t)\right|\,,
\end{equation}
turned out to be a good choice. The vector $\vec{m}$ is the integer
representation of the momenta introduced in equation
(\ref{eq:NCT-momenta1}). Here $\tilde{\phi}(\vec{m},t)$ is the spatial
Fourier transform of the field $\phi(\vec{x},t)$, where only the
non--commutating coordinates are transformed. The expectation value of
$\bar{M}(\vec{m})$ is zero in the disordered phase, where both
symmetries under consideration are unbroken. In an Ising type or
uniform phase only the expectation value of $\bar{M}(\vec{0})$ is
non--zero, since
\begin{equation}
  \label{eq:phi-order-uniform}
  \left\langle \bar{M}(\vec{0})\right\rangle
  =\frac{1}{NT}\left\langle\left|\sum_t\tilde{\phi}(\vec{m}=\vec{0},t)\right|\right\rangle
  =\left\langle\frac{1}{N^2T}\left|\sum_{t,\vec{x}}\phi(\vec{x},t)\right|\right\rangle
\end{equation}
is the standard order parameter of the spontaneous breakdown of the
$Z_2$ symmetry.  A non--vanishing order parameter at $\vec{m}\neq\vec{0}$ indicates
a spontaneous breakdown of the translation symmetry and therefore it
implies the striped or non--uniformly ordered phase.

Since we are simulating the dimensionally reduced model in terms of
the matrices $\hat{\phi}(t)$ we have to express this order parameter
by these matrices. To this end we use the map defined in equation
(\ref{eq:lat-J3}) and we obtain
\begin{equation}
  \label{eq:phi-order-trans}
  \bar{M}(\vec{m})=\frac{1}{NT}\left|\sum_t\tilde{\phi}(\vec{m},t)\right|
  =\frac{1}{NT}\left|\trace\sum_t\hat{\phi}(t)J^\dagger(\vec{m})\right|\,.
\end{equation}

This order parameter detects the disordered as well as the uniformly
ordered phase. In the striped phase there will arise problems when the
pattern for different configurations are rotated. For example if there
are stripes for one configuration parallel to the $x_1$ and for
another configuration parallel to the $x_2$ there would be two
different non--vanishing order parameters. This effect is related to
the rotation symmetry, which is also broken spontaneously in the
striped phase.
\footnote{Actually rotation symmetry is explicitly broken on the
  lattice.  However, in the commutative case this symmetry is reduced
  to an invariance under rotations of $\pi/2$, which can be broken
  spontaneously in the non--commutative lattice theory.}
To avoid this problem we defined the rotation invariant order
parameter as the expectation value of
\begin{equation}
  \label{eq:phi-order-final}
  M(k)=\max_{|\vec{m}|=k} \bar{M}(\vec{m})\,.
\end{equation}
This order parameter depends only on the absolute value of the momentum
and therefore the above mentioned problem does not occur. Based on
this order parameter we explored the phase diagram of the 3d
$\lambda\phi^4$ model.\\

\subsection{Numerical results}
\label{sec:phi-phase}

\begin{figure}[htbp]
  \centering
  \includegraphics[width=.85\linewidth]{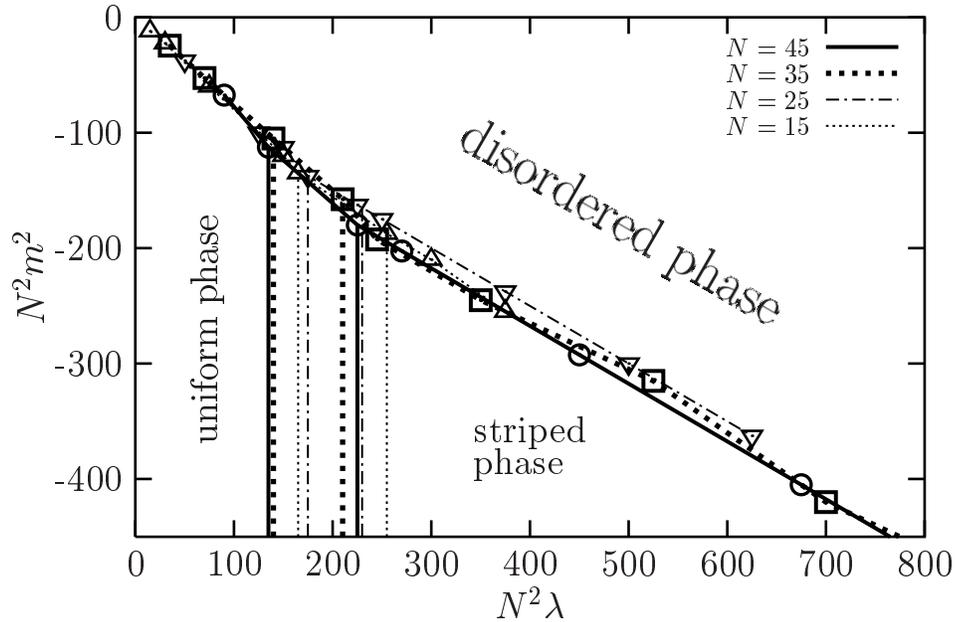}
  \caption{The phase diagram of the non--commutative 3d $\lambda\phi^4$ in the
    $m^2$ -- $\lambda$ plane.}
  \label{fig:phi-phase-diagram}
\end{figure}

With the algorithm described in Appendix \ref{sec:numerics-2} we
generated configurations at various values of $N$, $\lambda$ and $m^2$
and measured the order parameter (\ref{eq:phi-order-final}). The
result is the phase diagram plotted in Figure
\ref{fig:phi-phase-diagram}. The points connected by lines display
the phase transition between disordered phase and the ordered regime.
The ordered regime clearly splits into a uniform and in a striped phase,
where the transition region between these two phases is marked by two
vertical lines for each value of $N$. For each $N$ the left line
represents the largest value of $\lambda$ at which we are still in
the uniform phase, and the right lines show the smallest value of
$\lambda$, where we are already in the striped phase. Both transitions 
stabilize in $N$ if we multiply the axes by $N^2$.

To illustrate the different phases we mapped the matrices
$\hat{\phi}(t)$ back to coordinate space ($\phi(\vec{x},t)$) using the
map (\ref{eq:lat-delta-explicit}).  We chose configurations in the
four areas that can be distinguished.
\begin{figure}[htbp]
  \centering
  \hspace{-.3cm}\subfigure[{\hspace{-.05cm}$N^2m^2\!=-\!22.5$}]{\epsfig{figure=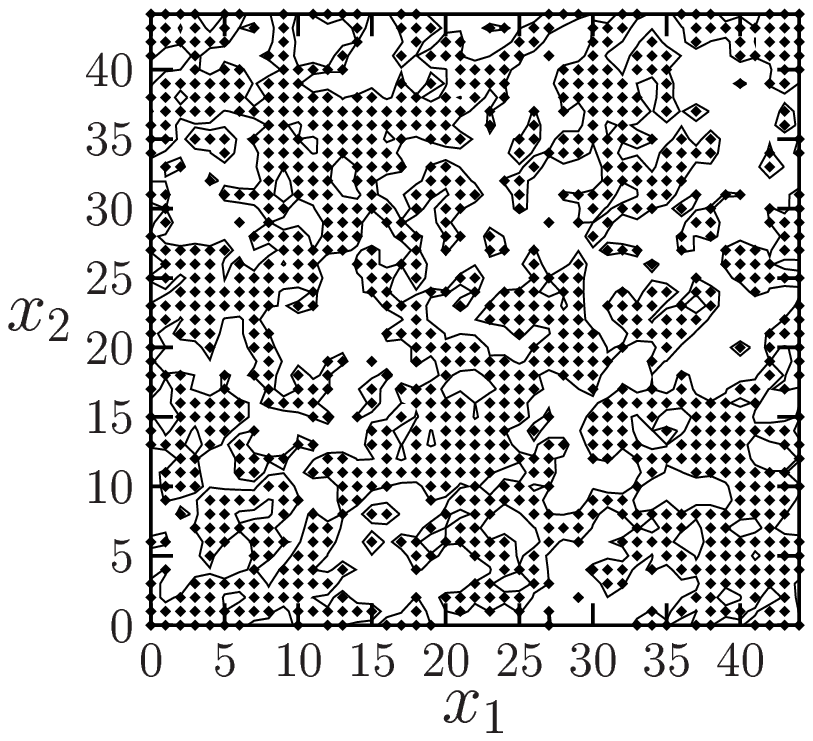,width=.23\linewidth}}%
  \hspace{.3cm}\subfigure[{$N^2m^2\!=\!-225$}]{\epsfig{figure=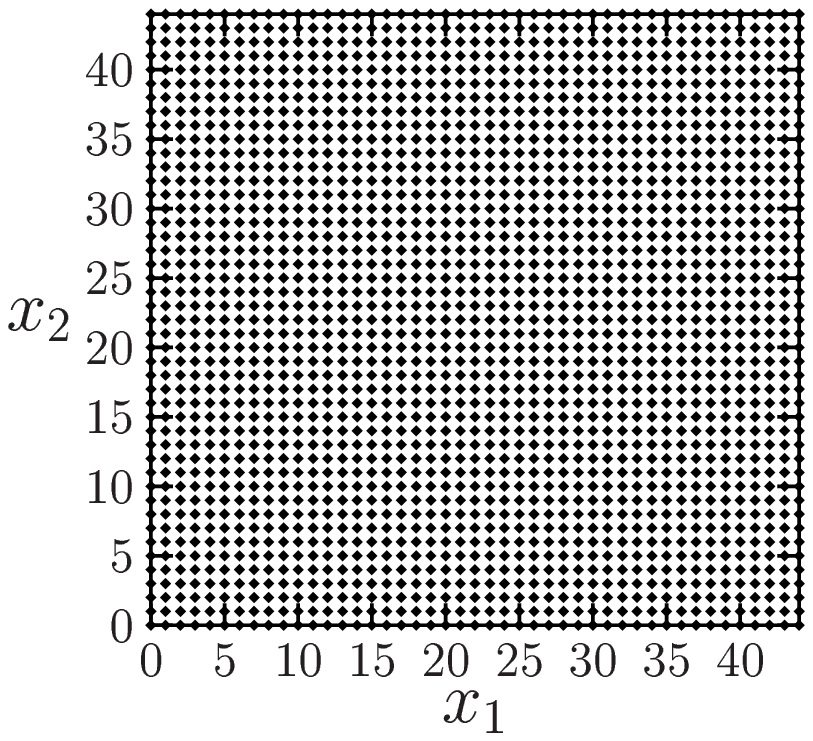,width=.23\linewidth}}%
  \hspace{.3cm}\subfigure[$N^2m^2\!=\!-90$]{\epsfig{figure=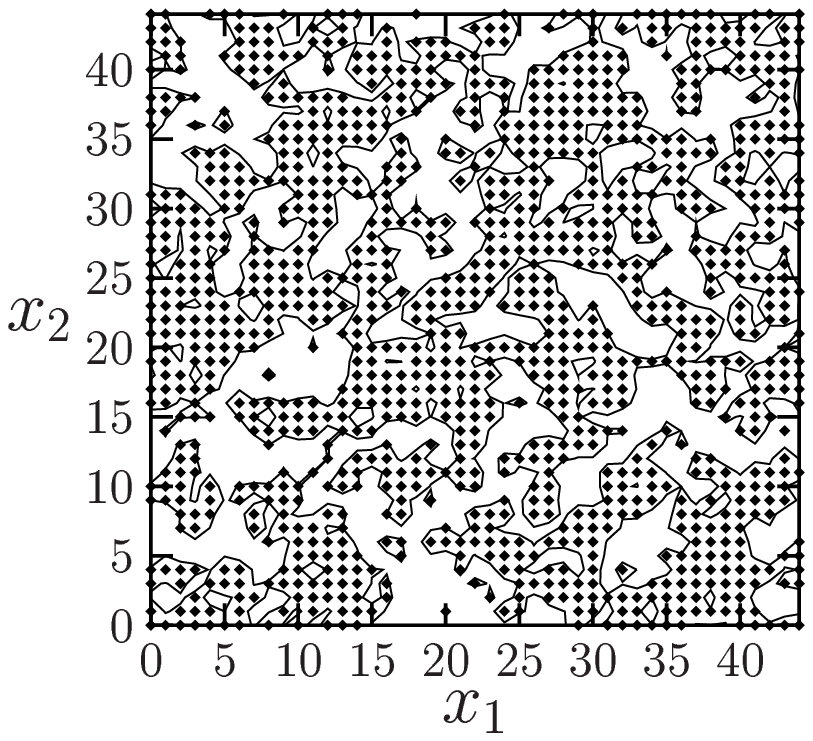,width=.23\linewidth}}%
  \hspace{.3cm}\subfigure[$N^2m^2\!=\!-400$]{\epsfig{figure=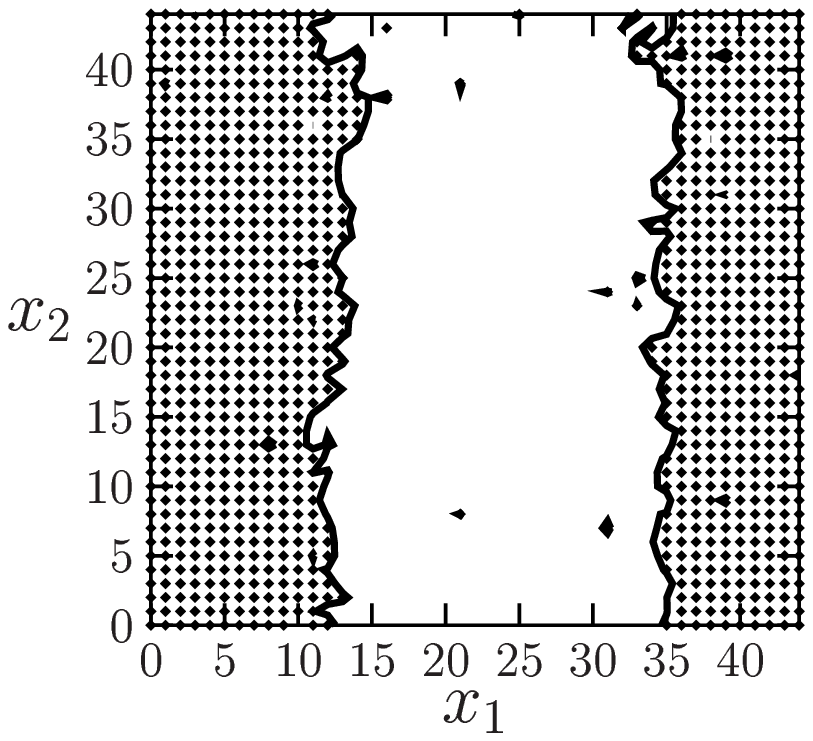,width=.23\linewidth}}
  \caption{Snapshots of single configurations 
    $\phi(\vec{x},t)$ at some time $t$, at $N=45$, $N^2\lambda=90$
    (a,b) and $N^2\lambda=450$ (c,d).}
  \label{fig:stripe}
\end{figure}
Some example snapshots of $\phi(\vec{x},t)$ are shown in Figure
\ref{fig:stripe}.  The dotted areas indicate $\phi(\vec{x},t)>0$ and
in the blank areas $\phi(\vec{x},t)$ is negative.  Figure
\ref{fig:stripe}(a) and \ref{fig:stripe}(c) show $\phi(\vec{x},t)$ in
the disordered phase. The positive and
negative areas are spread all over as one expects in the disordered
phase. Also in the uniformly ordered phase we find the expected
behavior; $\phi(\vec{x},t)$ is either positive for all $\vec{x}$ or
negative for all $\vec{x}$ (Figure \ref{fig:stripe}(b)).

Figure \ref{fig:stripe}(d) shows a typical pattern with two stripes.
In the range of the parameters plotted in the phase diagram
\ref{fig:phi-phase-diagram} we always found two stripes (in the
non--uniformly ordered phase). They were either parallel to the $x_1$ or to the
$x_2$ axis. We will discuss the number of stripes separately in the
next Subsection.  In the rest of this Subsection we discuss the
measurements that
the phase diagram \ref{fig:phi-phase-diagram} is based on.\\

To localize the phase transitions we started simulations in the
disordered phase. After we reached equilibrium we measured $M(k)$.
Then we decreased slowly $m^2$ towards the expected phase transition.
\begin{figure}[htbp]
  \centering
  \includegraphics[width=.45\linewidth]{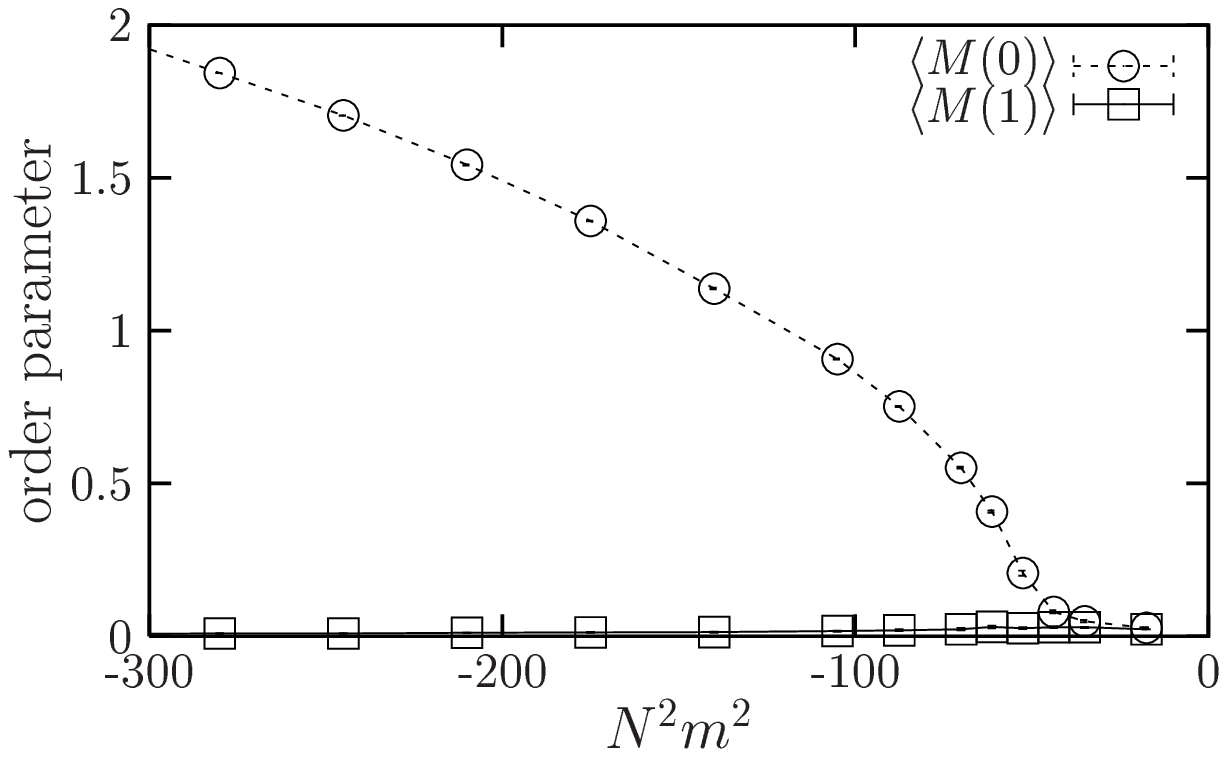} %
  \hspace{.5cm}\includegraphics[width=.46\linewidth]{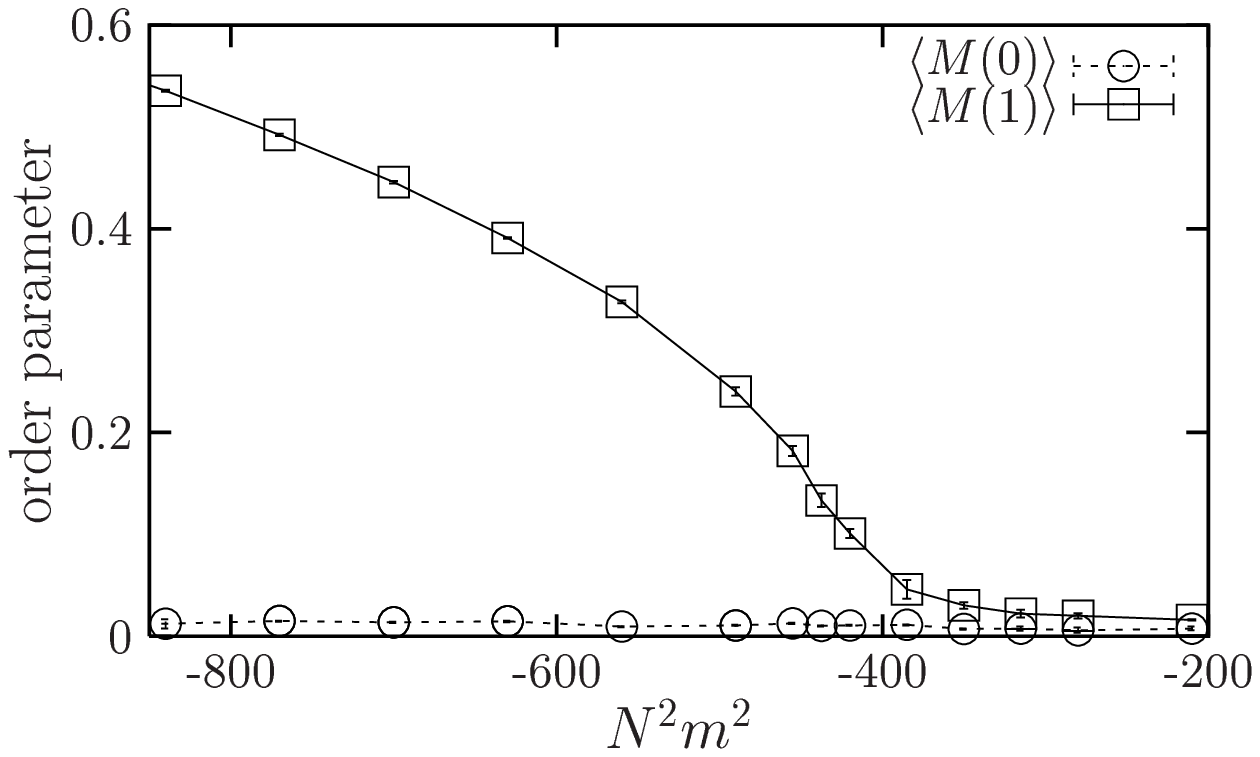}
  \caption{The momentum dependent order parameter $\langle M(k)\rangle$ 
    against $N^2m^2$ at $N=35$. On the left we fixed at
    $N^2\lambda=70$, which leads to the uniform phase and on the right
    we are at $N^2\lambda=350$, leading to the striped phase.}
  \label{fig:phi-determine1}
\end{figure}
Figure \ref{fig:phi-determine1} shows two example measurements at
$N=35$. We plotted in both cases the order parameter at $k=0$ and
$k=1$ against $N^2m^2$.  On the left we started in the disordered
phase at $N^2\lambda=70$. Below a critical value of $m^2$, we
see a clearly non--vanishing $\langle M(k)\rangle$ at $k=0$, while
$\langle M(1)\rangle$ is zero for all values of $m^2$.  This indicates
that we are entering the uniform phase.  On the right we started in
the disordered phase at $N^2\lambda=350$.  Lowering $m^2$ leads to a
non--zero order parameter at $k=1$ and the standard order
parameter $\langle M(0)\rangle$ is zero.  The fact that $\langle
M(1)\rangle\neq0$ implies that the system is in the striped phase.

Measurements of this kind allowed us to separate the uniform phase from
the striped phase. %
\footnote{A natural method the determine the transition line between
  uniform and striped phase is to start in the uniform phase and
  increase $\lambda$ or vice versa. However, it turned out that the
  initial patterns remain stable when the transition region is
  crossed.}
Whenever we see a dominant order parameter with $k\neq0$ we are in the
striped phase. Unfortunately we were not able to determine an accurate
transition line between these phases.  We obtained a transition
region, in which it was not possible to identify the order of the
system. Depending on the starting conditions (see Appendix
\ref{numerics}) we found indication for both phases. We come back to
this point in Section \ref{sec:phi-stripe-rev}.

From plots like shown in Figure \ref{fig:phi-determine1} one can also
identify the transition line between disordered phase and ordered
regime, at least roughly. Whenever any order parameter becomes
non--zero we hit the transition line. However, one can localize the
transition more accurately by looking at the connected part of the
two--point function of $M(k)$,
\begin{equation}
  \label{eq:phi-order-two}
  \langle M(k)^2\rangle_c=\langle M(k)^2\rangle-\langle M(k)\rangle^2\,.
\end{equation}
From statistical mechanics it is known
that this two--point function has a peak at the phase transition. 
\begin{figure}[htbp]
  \centering
  \includegraphics[width=.45\linewidth]{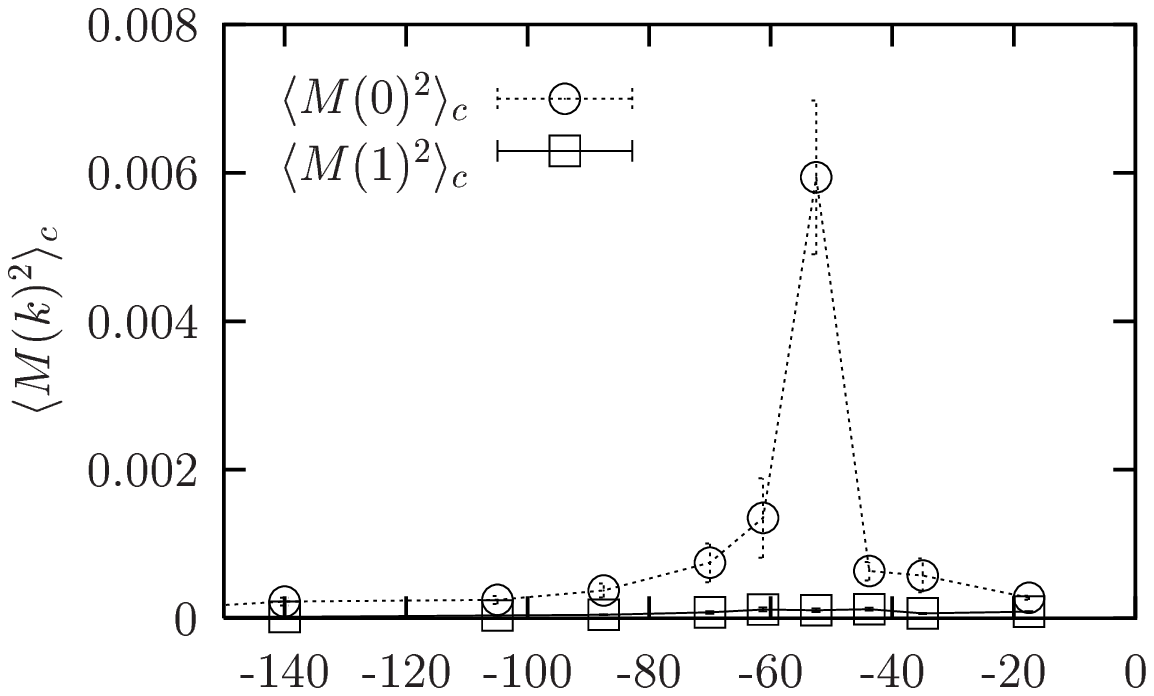} 
  \hspace{.5cm}\includegraphics[width=.45\linewidth]{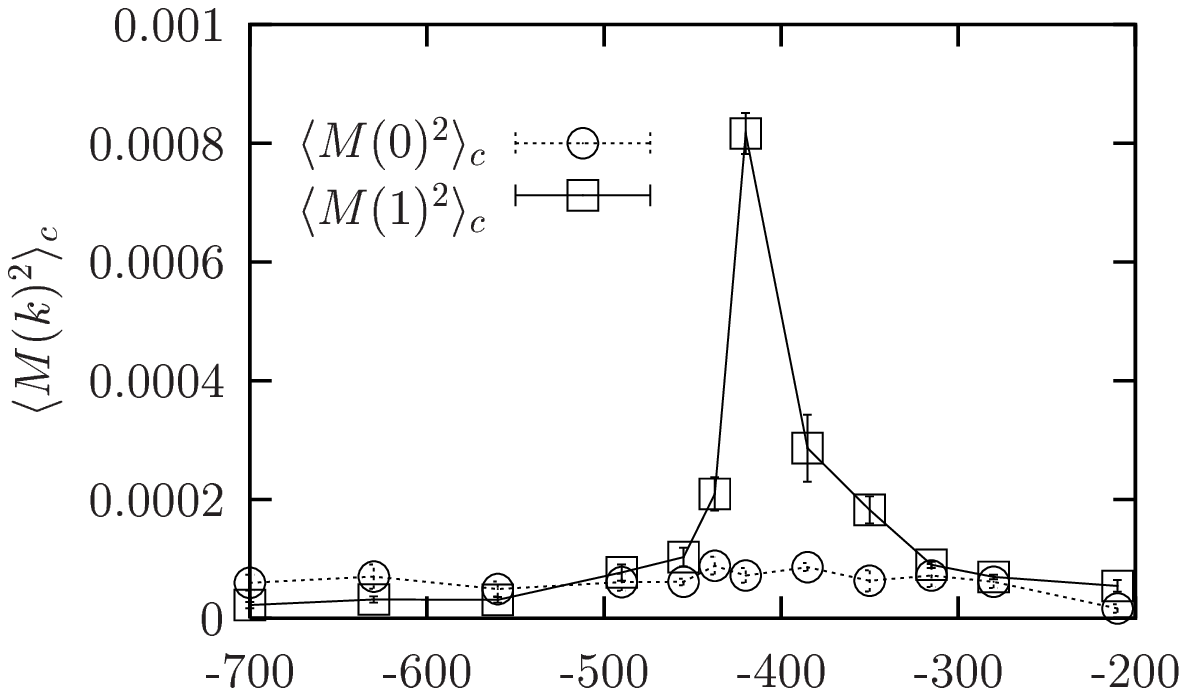}
  \caption{The connected two--point function of $M(k)$ defined in equation (\ref{eq:phi-order-two}) 
    against $N^2m^2$ at $N=35$. On the left at $N^2\lambda=70$ and on
    the right at $N^2\lambda=350$.}
  \label{fig:phi-determine3}
\end{figure}
Figure \ref{fig:phi-determine3} shows two examples of these
measurements, where we used the same configurations as we did in
Figure \ref{fig:phi-determine1}. This allowed us the determine the
transition line between disordered phase and ordered regime to a high
accuracy.

\subsection{The striped phase}
\label{sec:phi-stripe}

In Figure \ref{fig:phi-determine1} we have seen that the two types of
ordered phases can be distinguished by the momentum $k$ of the
non--vanishing order parameter $\langle M(k)\rangle$. We showed the
two possibilities $k=0,1$. It was left as an open question how
$\langle M(k)\rangle$ behaves for other values of $k$. This we want to
discuss here.

\begin{figure}[htbp]
  \centering
  \includegraphics[width=.45\linewidth]{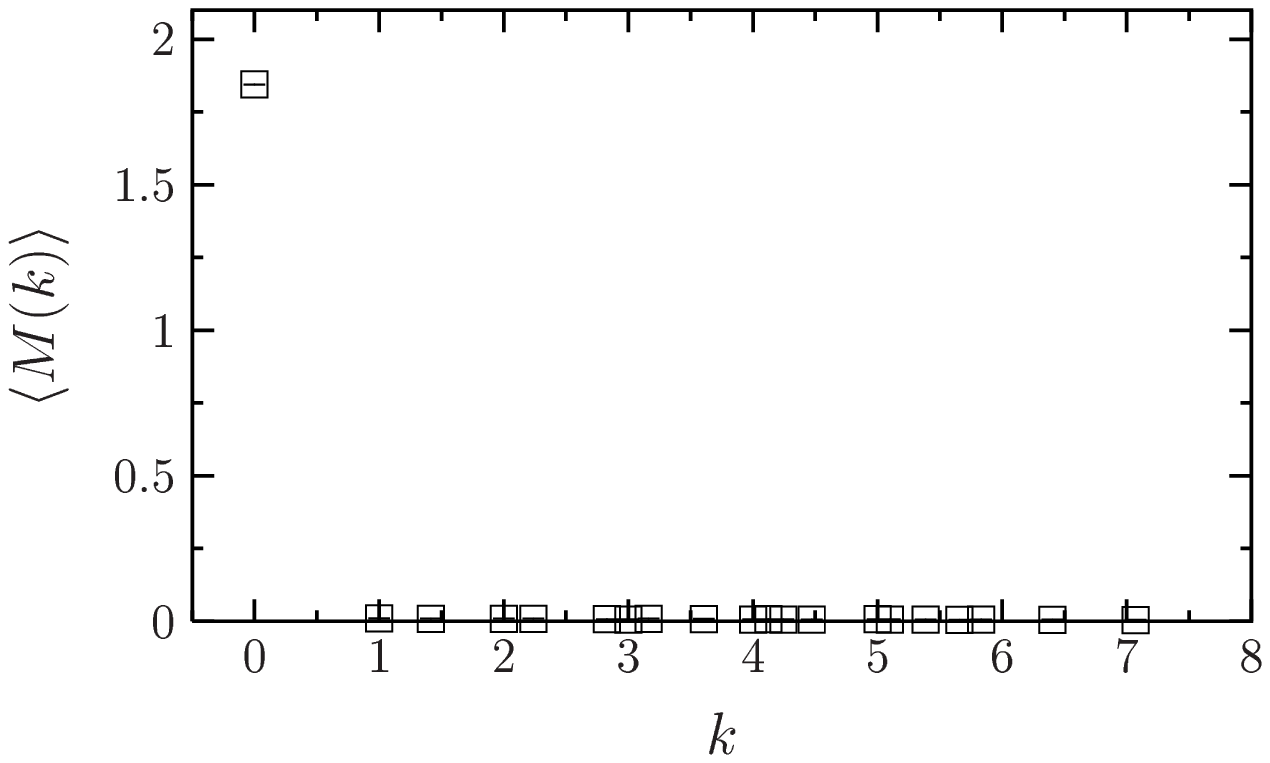} %
  \hspace{.5cm}\includegraphics[width=.45\linewidth]{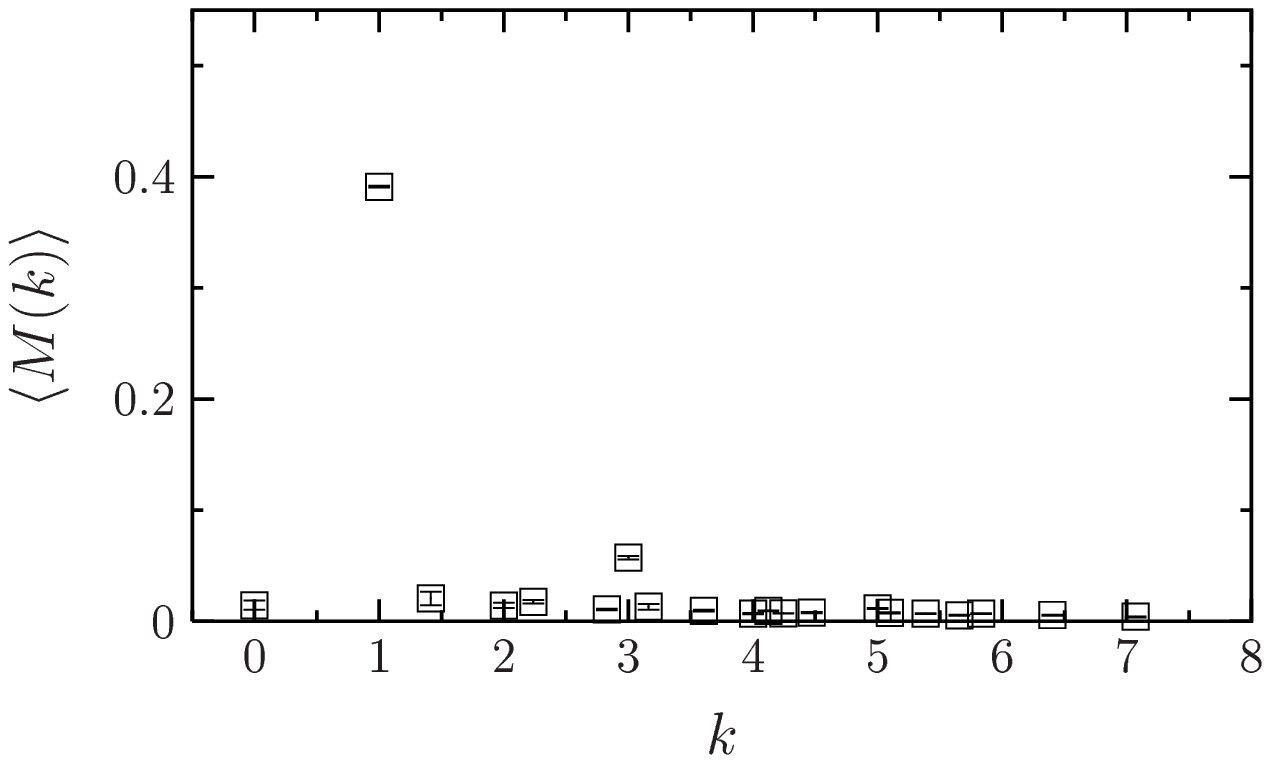}
  \caption{The momentum dependent order parameter $\langle M(k)\rangle$ 
    against $k$ at $N=35$. On the left in the uniform phase at
    $N^2\lambda=70$ and $N^2m^2=-210$, and on the right in the striped phase at
    $N^2\lambda=350$ and $N^2m^2=-400$.}
  \label{fig:phi-striped-phase1}
\end{figure}

In Figure \ref{fig:phi-striped-phase1} we plotted two examples of the
order parameter against the momentum $k$, where in both cases we are
clearly in the ordered regime.  On the left the system is in the
uniformly ordered phase. Only the standard order parameter is
non--zero, for all other momenta the order parameter is zero. This
defines the uniform phase.

The right plot in Figure \ref{fig:phi-striped-phase1} shows the system
in the striped phase. There we see again a non--zero order parameter
at $k=1$. In addition we also see a signal for $\langle M(3)\rangle$.
This may appear as an indication for an underlying multi--stripe structure,
which is not visible when we plot the sign of $\phi(\vec{x},t)$, as we
did in Figure \ref{fig:stripe}. In fact this is not the case. To
illustrate this we define the function
\begin{equation}
  \label{eq:phi-illustrate-stripes}
  \Phi(x_1)=\frac{1}{NT}\sum_{x_2,t}\phi(x_1,x_2,t)\,.
\end{equation}
In this definition we assume that the stripes are parallel to the
$x_2$ axis. 
The function $\Phi(x_1)$ averages $\phi(\vec{x},t)$ over $x_2$
and $t$ at $x_1$. If there is an additional structure 
\footnote{Of course the method will not find all possible patterns, but
  we also looked for additional structures in the other direction,
  which were not present.}
\begin{figure}[htbp]
  \centering
  \includegraphics[width=.45\linewidth]{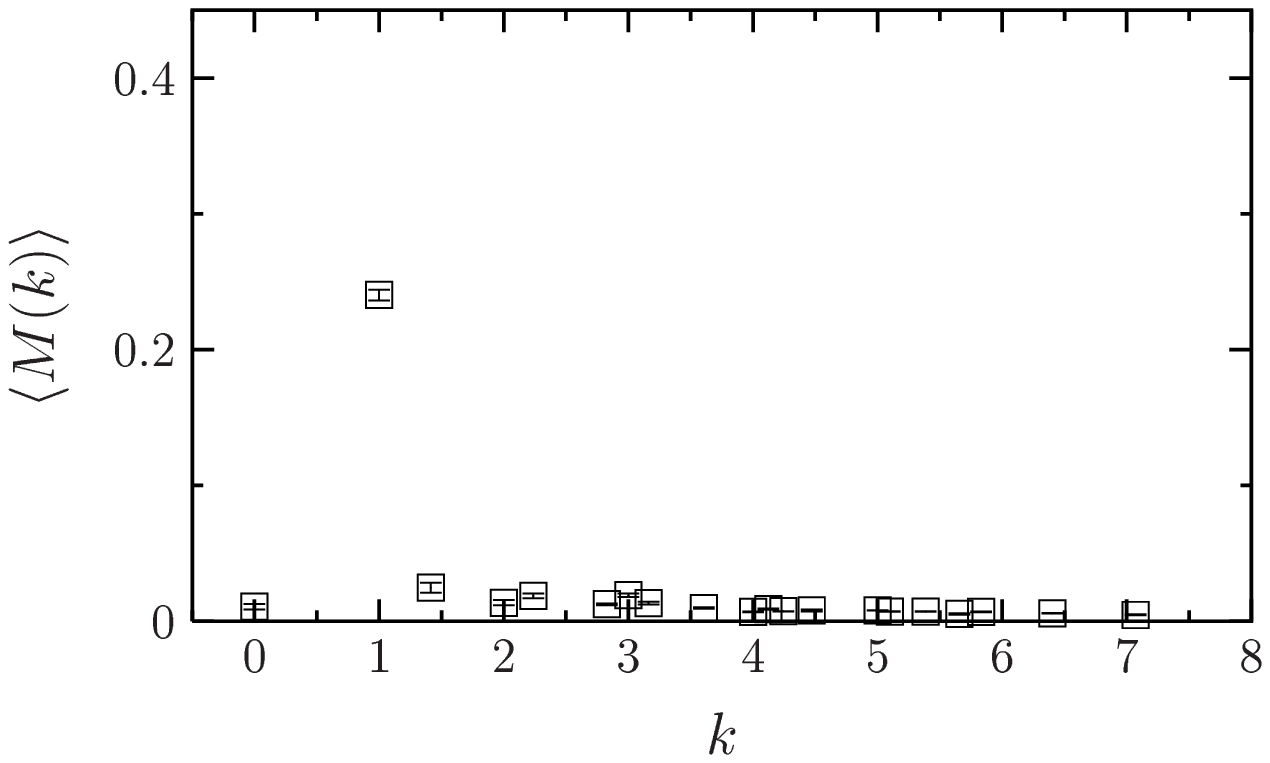} %
  \hspace{.5cm}\includegraphics[width=.45\linewidth]{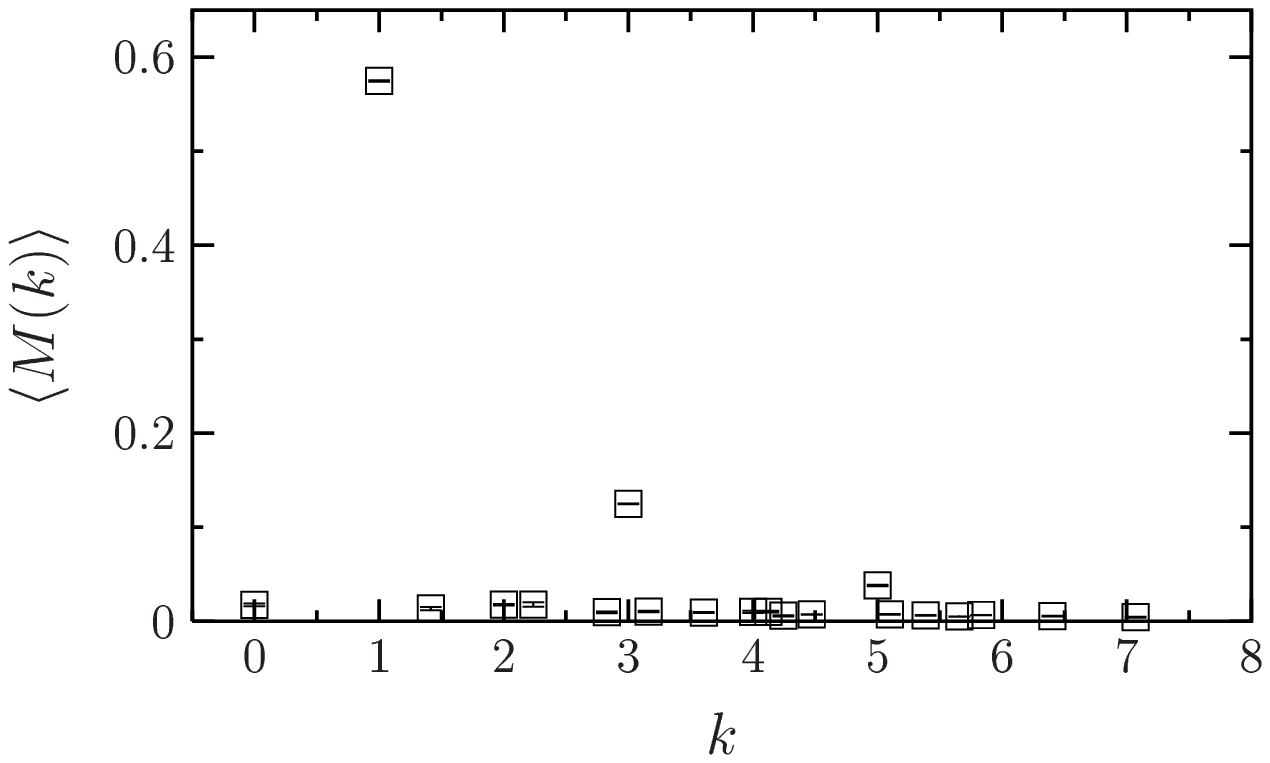}\\
  \hspace{-0.1cm}\includegraphics[width=.475\linewidth]{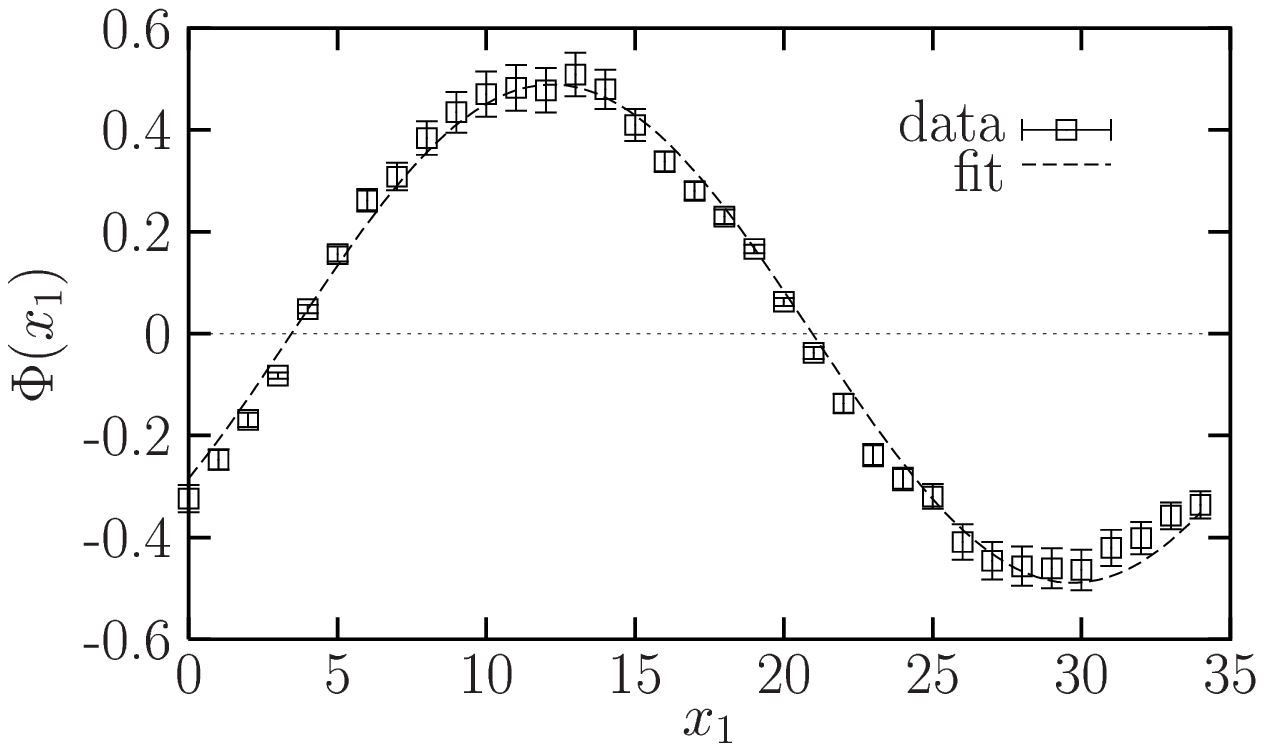}%
  \hspace{.3cm}\includegraphics[width=.475\linewidth]{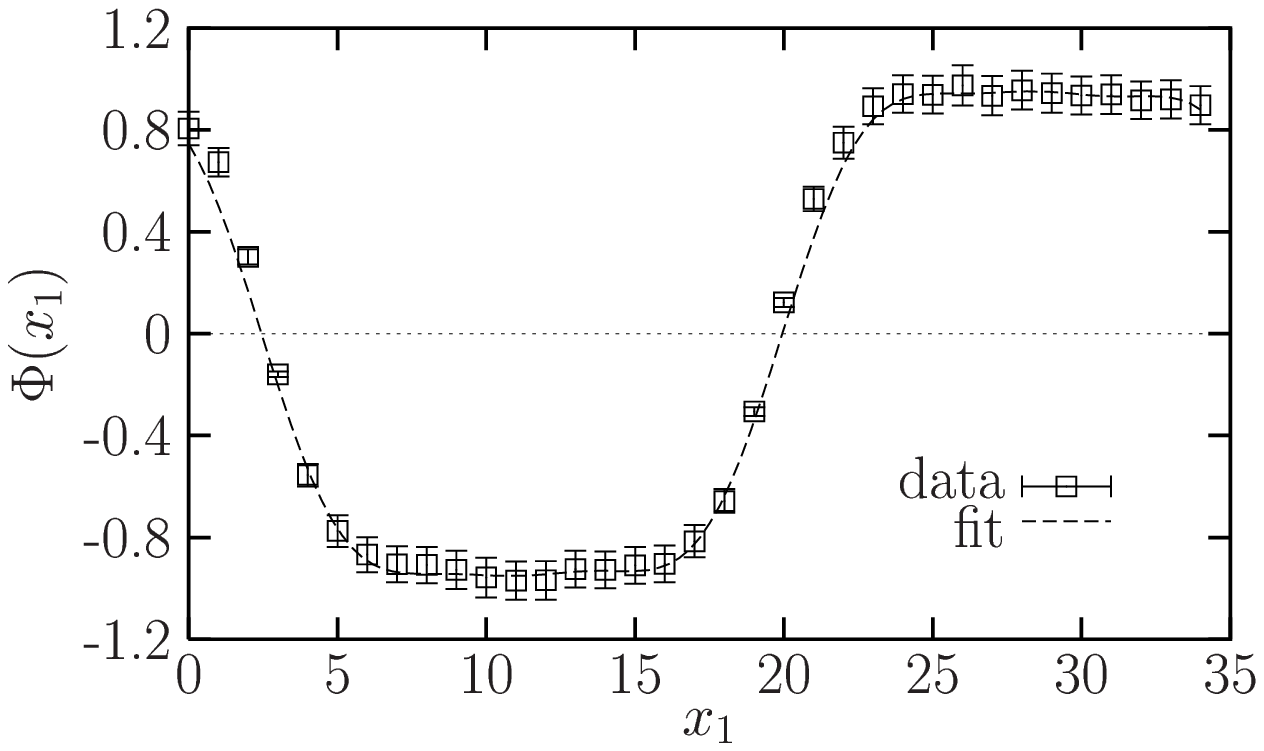}
  \caption{The momentum dependent order parameter $\langle M(k)\rangle$ 
    against $k$ at $N=35$ and $\Phi(x_1)$ in the striped phase. On the
    left close to the phase transition at $N^2\lambda=350$ and
    $N^2m^2=-250$ and on the right at large negative values of $m^2$
    in the striped phase at $N^2\lambda=350$ and $N^2m^2=-620$.}
  \label{fig:phi-striped-phase2}
\end{figure}
this should be seen in the behavior of $\Phi(x_1)$. To examine this we
first computed the order parameter from configurations close to the
phase transition and from configurations at large negative values of
$m^2$. The results are shown in Figure \ref{fig:phi-striped-phase2}
(top). On the left the system is close to the phase transition and
only $\langle M(1)\rangle$ is non--zero. On the right, at large
negative values of $m^2$, we see that also $\langle M(3)\rangle$ and
$\langle M(5)\rangle$ are different from zero. It is evident that when
$m^2$ is further decreased, $\langle M(k)\rangle$ will also be
non--vanishing for larger (odd) momenta $k$.

We computed $\Phi(x_1)$ on one configurations in each of these
regions, to study how the additional momentum modes influence this
function. The results are plotted in Figure
\ref{fig:phi-striped-phase2} (bottom). If there is only one
non--vanishing mode of the field this corresponds to a pure {\tt sine}
behavior. This is clearly confirmed by the plot on the left. The
dashed line is a fit of the data to a {\tt sine} function.
\footnote{The function $\Phi(x_1)$ is an average over $x_2$ and $t$.
  The errors displayed in Figure \ref{fig:phi-striped-phase2} are
  computed with respect to this average with the binning method (see
  Appendix \ref{sec:monte}).}
On the right $\Phi(x_1)$ is plotted in the case of various
non--vanishing momentum modes. Here we see that these additional modes
do {\em not} indicate a more complicated structure. It simply displays
that deeply in the striped phase $\Phi(x_1)$ is no more a pure {\tt
  sine}, but the Fourier series involves higher momenta. As a fit
function in this plot we used the first three odd terms in such an
expansion, according to the three non--vanishing modes in Figure
\ref{fig:phi-striped-phase2} (right, top), which fits the data
perfectly. This result means for the structure of the field
$\phi(\vec{x},t)$ that at large negative values of $m^2$ the number of
stripes does not change, but the field is here restricted to only
two values. At $m^2\ll 0$ the kinetic term is negligible and the 
anti--ferromagnetic coupling dominates the system.

As we mentioned above we obtained --- in the range of the parameters
shown in the phase diagram \ref{fig:phi-phase-diagram} --- never more
than two stripes. Analytically one expects not only two stripes, but
also more complicated patterns \cite{Gubser:2000cd}. 
To be in agreement with the perturbative conjecture 
these complex patterns should also occur on the lattice.
The multi--stripe
patterns are expected either on larger lattices or at larger
non--commutativity parameter $\theta$ or at larger coupling $\lambda$.
\begin{figure}[htbp]
  \centering
  \hspace{-.3cm}\subfigure[{$\lambda=10$}]{\epsfig{figure=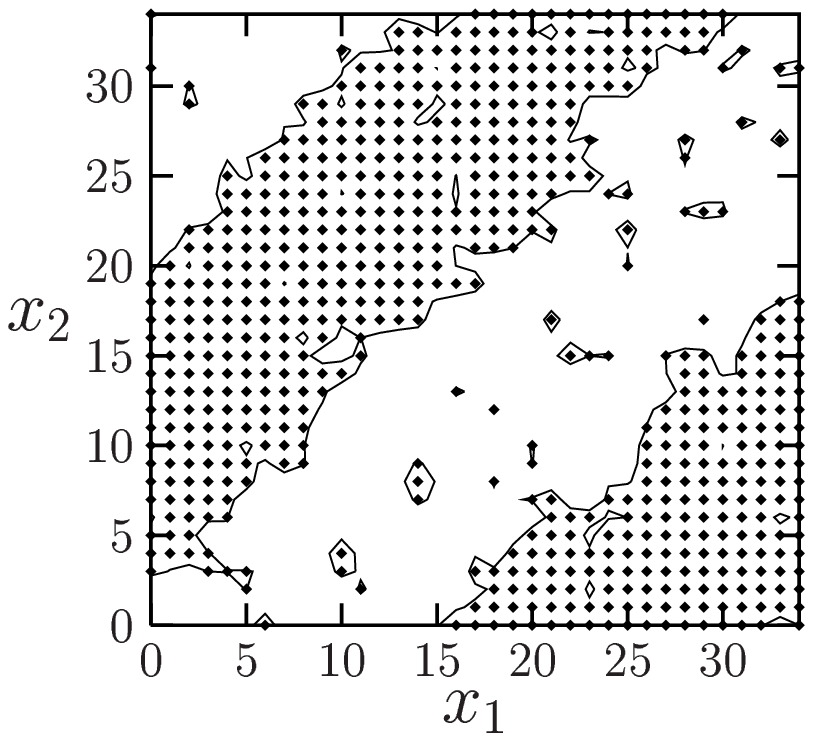,width=.23\linewidth}}%
  \hspace{.3cm}\subfigure[{$\lambda=10$}]{\epsfig{figure=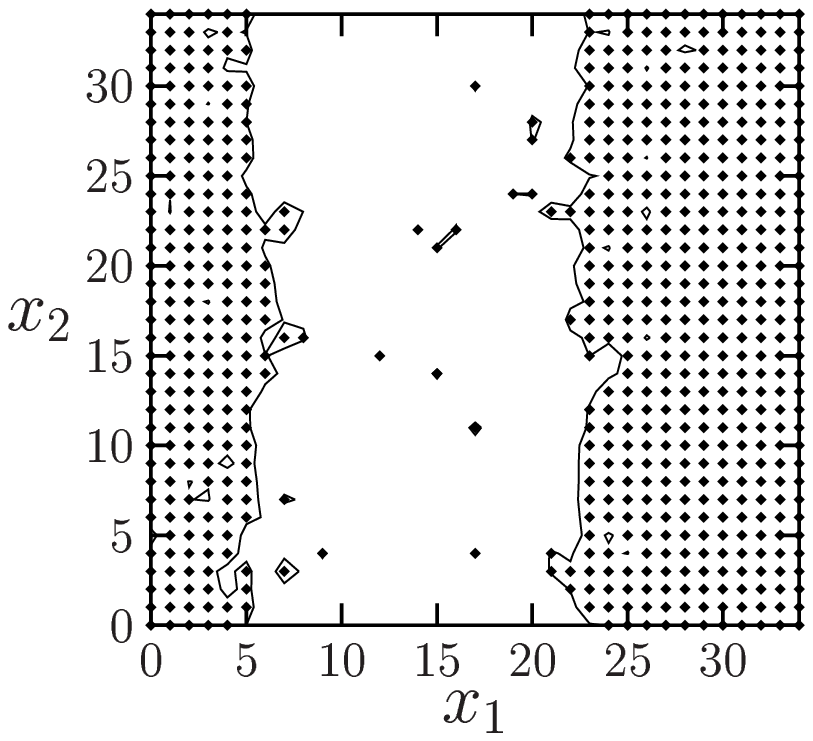,width=.23\linewidth}}%
  \hspace{.3cm}\subfigure[$\lambda=100$]{\epsfig{figure=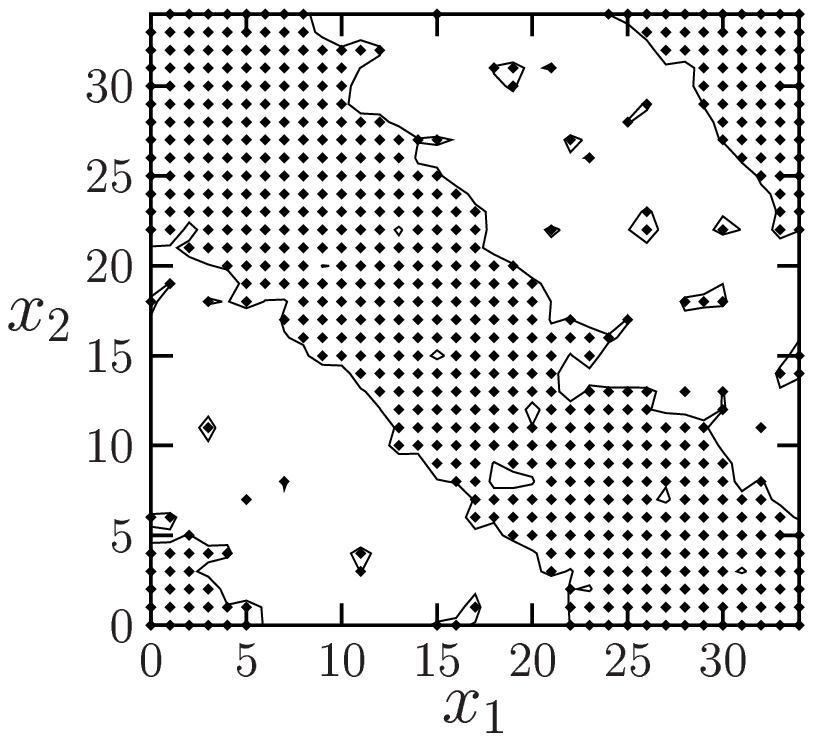,width=.23\linewidth}}%
  \hspace{.3cm}\subfigure[$\lambda=100$]{\epsfig{figure=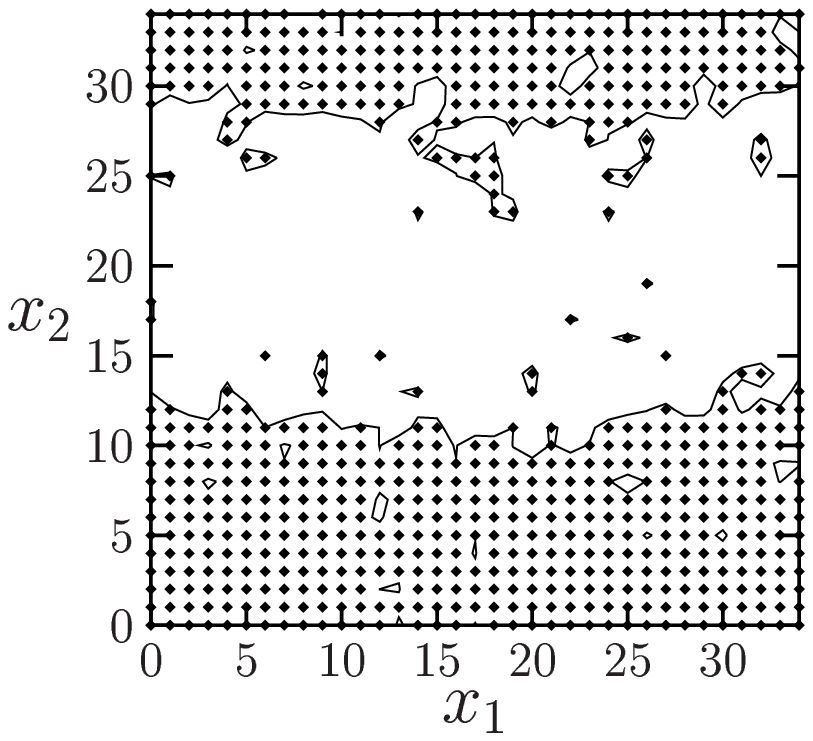,width=.23\linewidth}}
  \caption{Snapshots of single configurations 
    $\phi(\vec{x},t)$ at a certain time $t$, at $N=35$, $\lambda=10$
    and $m^2=-6$ (a,b) and at $\lambda=100$ and $m^2=-40$. The
    different patterns at each values of the parameters correspond to
    different starting configuration.}
  \label{fig:stripe-multi}
\end{figure}
In fact for increased $\lambda$ we obtain also diagonal stripes, which
support the picture of Gubser and Sondhi. However, only in the
continuum limit, where the number of stripes should diverge, this
picture can be ultimately verified.  Some example snapshots at $N=35$
are displayed in Figure \ref{fig:stripe-multi}. The different patterns
shown here correspond to different starting configurations. We
postpone the discussion of the vacuum until we have considered the
dispersion relation.

\section{Correlation functions}
\label{sec:phi-obs}

In this Section we present our results concerning the correlation
functions.  The main interest here is to study the dispersion
relation, but also the behavior of two--point functions in position
space. 

\subsection{Spatial correlators}
\label{sec:phi-cor}

There are several predictions about the behavior of the correlation
functions in coordinate space in the 4d $\lambda\phi^4$ theory. It is
expected from one loop perturbation theory that the decay of this
correlator at small $\lambda$ is not exponential
\cite{Minwalla:1999px}. For larger $\lambda$ the exponential behavior
is restored but now depending on the coupling instead of the mass.
This refers to the disordered phase. In the non--uniformly ordered phase an
oscillation of the correlator is expected.  Here we are studying the
behavior of these correlators in $d=3$.

\begin{figure}[htbp]
  \centering
  \includegraphics[width=.47\linewidth]{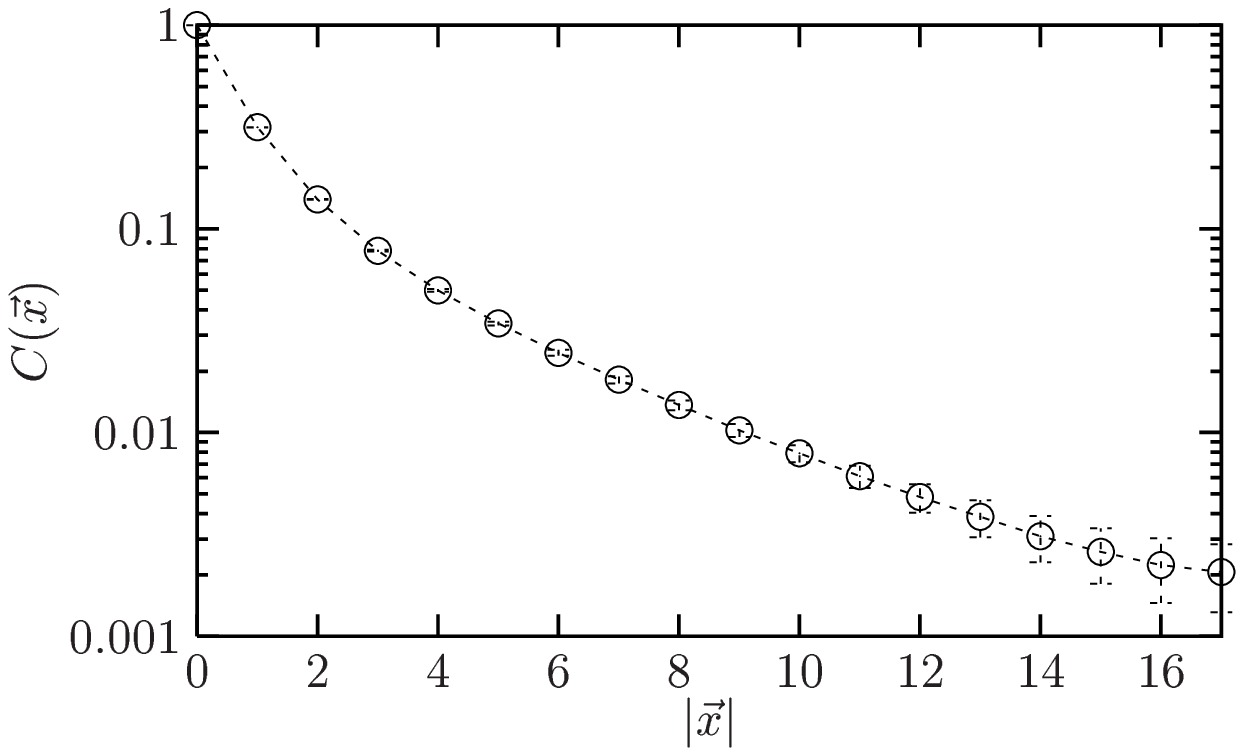} %
  \hspace{.3cm}\includegraphics[width=.47\linewidth]{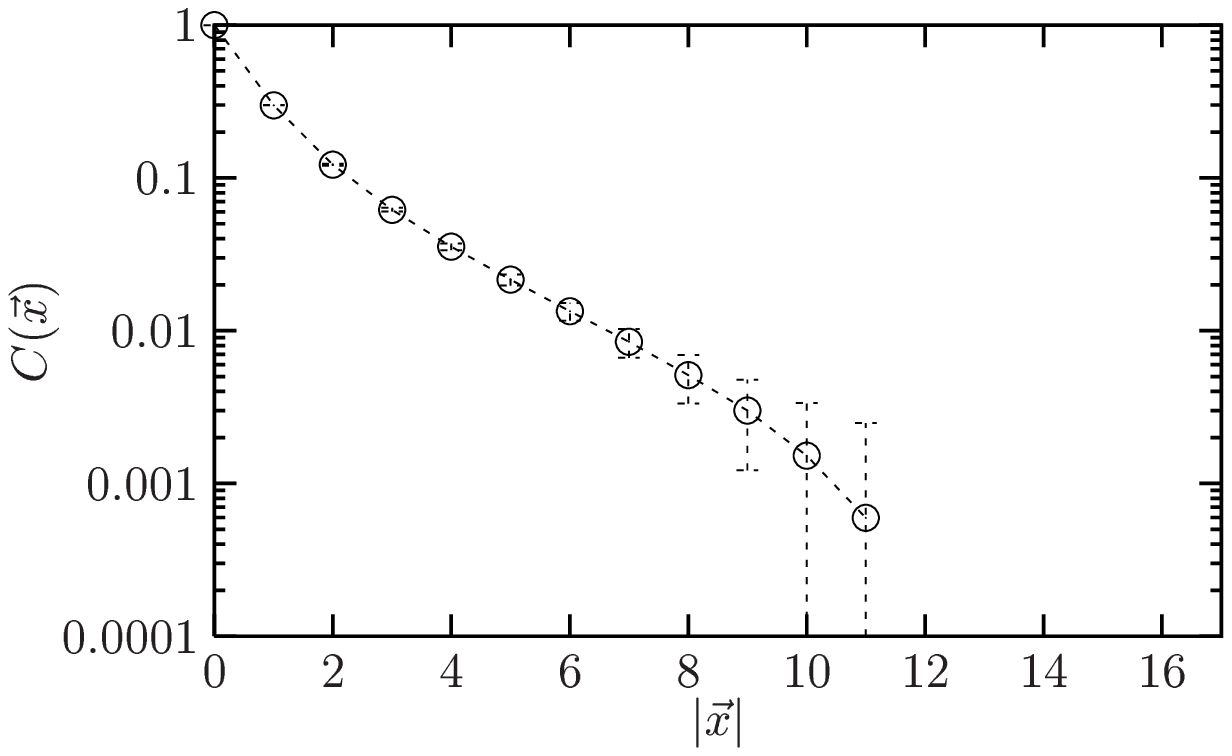} %
  \caption{The correlator (\ref{eq:phi-spat-corr}) against $|\vec{x}|$ in the disordered
    phase at $N=35$.  On the left the selfcoupling amounts to
    $\lambda=0.06$ and $m^2=-0.015$. On the right the correlator is
    plotted at $\lambda=0.6$ and $m^2=-0.15$.}
  \label{fig:phi-spat-corr1} 
\end{figure}

We are dealing here with one commuting and two non--commuting
coordinates, namely the spatial coordinates.  Since only the
correlators in the non--commuting directions are expected to have an
exotic behavior, we focus on spatial correlators.  To be explicit, we
study the correlation function
\footnote{In practice we use the twist eaters (\ref{eq:lat-gamma}),
  which are a finite dimensional representation of the shift operators
  (\ref{eq:lat-shift}) to compute this quantity. We discuss this issue
  here and in the following in coordinate space since then the
  quantities are intuitively readable. }
\begin{equation}
  \label{eq:phi-spat-corr}
  C(\vec{x})=\frac{1}{N^2T}\sum_{\vec{y},t}\langle\phi(\vec{y},t)\star\phi(\vec{y}+\vec{x},t)\rangle\,.
\end{equation}
In Figure \ref{fig:phi-spat-corr1} the correlator
(\ref{eq:phi-spat-corr}) in the disordered phase at $N=35$ is shown.
We measured the $C(\vec{x})$ in one direction $x_i$ keeping $x$ fixed in
the other direction. In Figure \ref{fig:phi-spat-corr1} we averaged
over both directions. On the left we plotted the correlator
(\ref{eq:phi-spat-corr}) at $\lambda=0.06$ and on the right at
$\lambda=0.6$. The decay is clearly not exponential at these values of
the coupling.

\begin{figure}[htbp]
  \centering
  \hspace{-.5cm}\includegraphics[width=.47\linewidth]{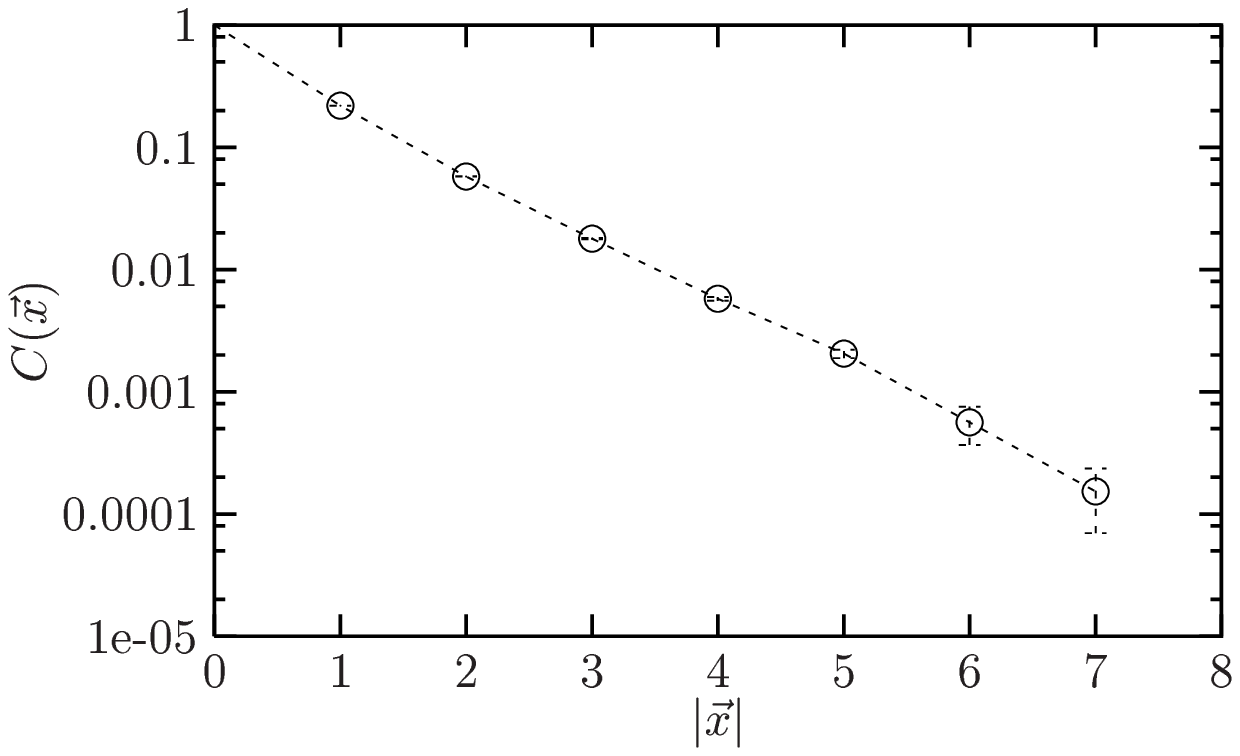} %
  \hspace{.3cm}\includegraphics[width=.445\linewidth]{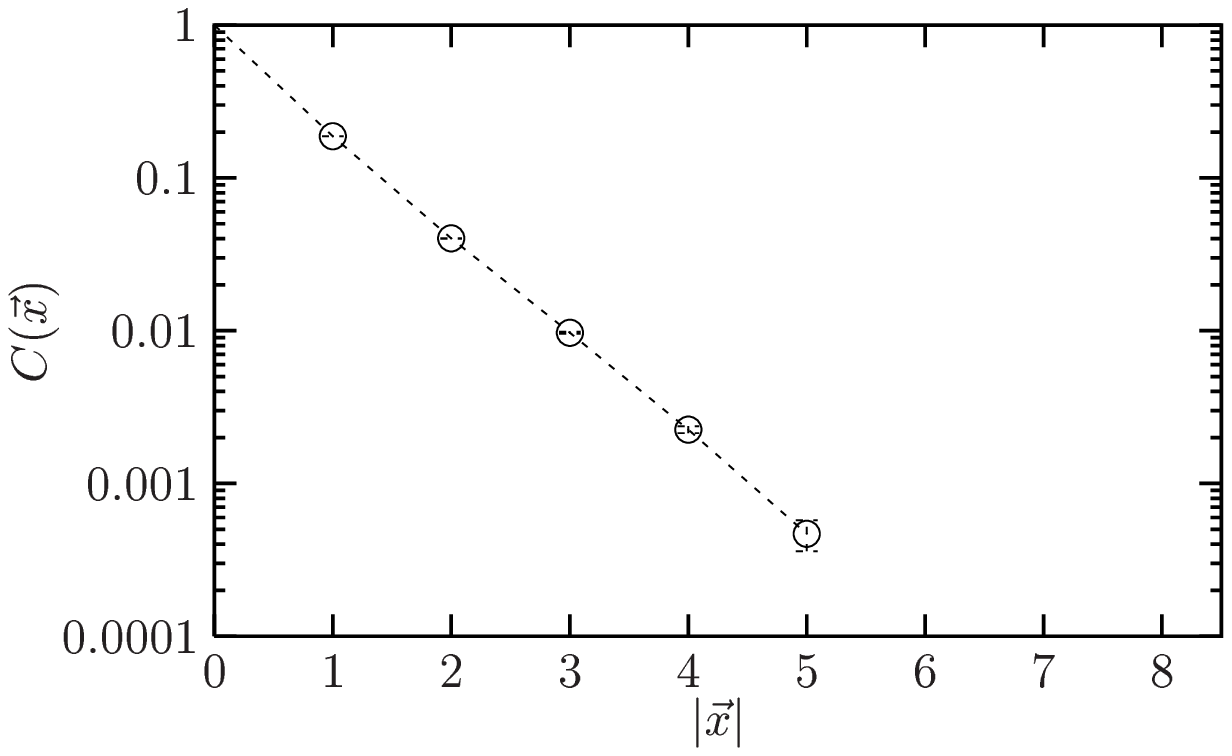} %
  \caption{The correlator (\ref{eq:phi-spat-corr}) against $|\vec{x}|$ in the disordered
    phase at $N=35$. On the left we set $\lambda=2$ and $m^2=-0.3$ and on the right
    to $\lambda=10$ and $m^2=-2$.}
  \label{fig:phi-spat-corr2} 
\end{figure}

In Figure \ref{fig:phi-spat-corr2} we show results at increased coupling.
At moderately enlarged $\lambda$ the decay appears almost exponential (on the left),
and at very large coupling the exponential decay is restored (on the
right). The non--exponential decay of the spatial correlators was a
result obtained from perturbation theory in four dimension. Here we
observed this behavior in the 3d model.\\

In the striped phase we have to take more care about the direction in
which we measure the correlation function (\ref{eq:phi-spat-corr}),
since here we have patterns leading to a dependence of the correlator
on the direction. Therefore an analysis of the stripe pattern has to
be done first. This is achieved by evaluating the order parameter
$\bar{M}(\vec{m})$ defined in equation (\ref{eq:phi-order-1}). This
order parameter depends also on the orientation of the condensed
momentum mode and it is therefore not suitable for detecting the
disordered -- striped phase transition. However, here it
is a perfect observable to detect the patterns. The momentum that
maximizes $\bar{M}(\vec{m})$ dominates the pattern in the striped
phase. For example, a maximum at $\vec{m}=(1,0)$ indicates two stripes
parallel to the $x_2$ axis, and a maximum at $\vec{m}=(1,1)$ or at
$\vec{m}=(1,-1)$ indicates diagonal stripes.

We measured the correlation function parallel and vertical
to the stripes separately. For completeness we plotted in Figure
\begin{figure}[htbp]
  \centering
  \includegraphics[width=.48\linewidth]{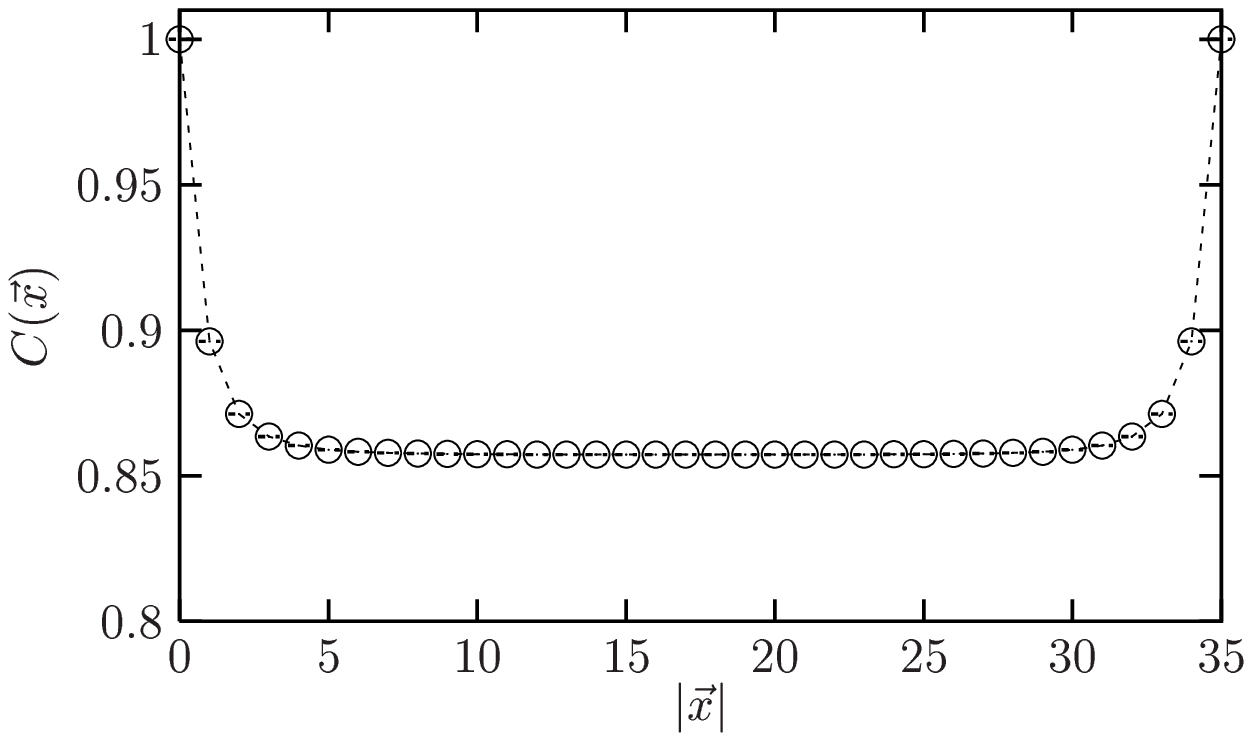} %
  \hspace{.3cm}\includegraphics[width=.48\linewidth]{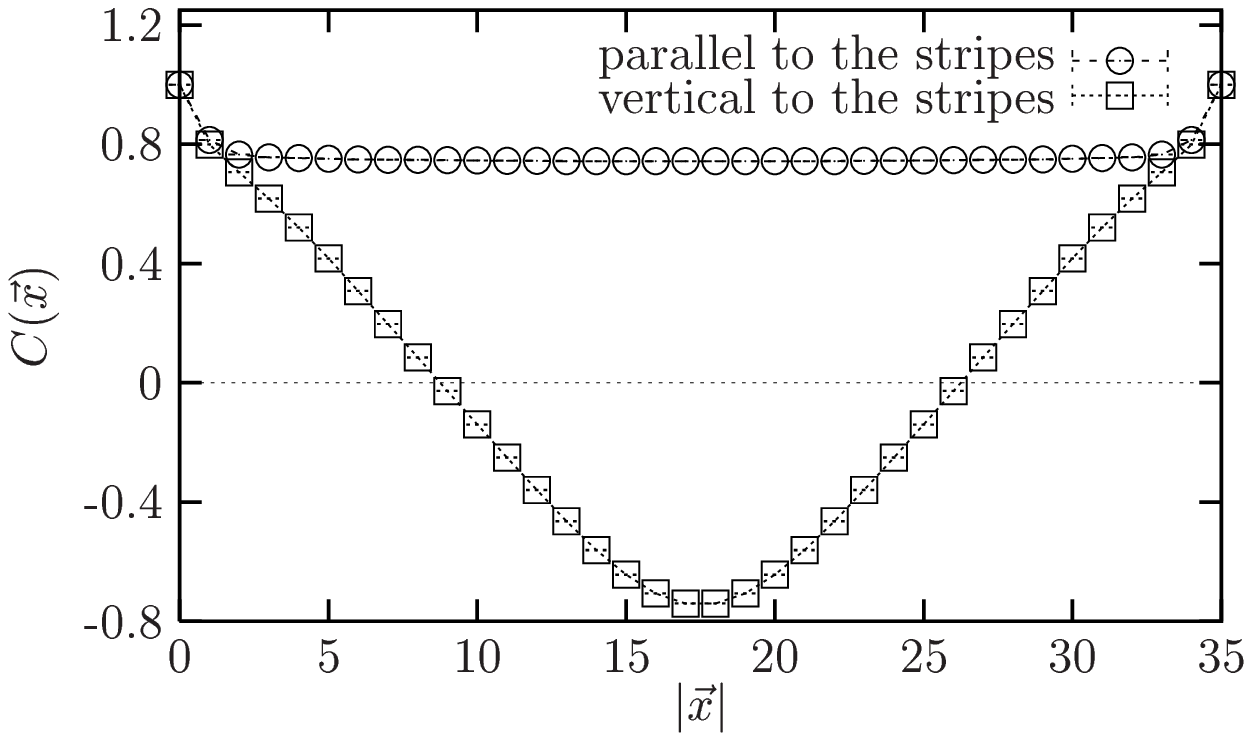} %
  \caption{The correlator (\ref{eq:phi-spat-corr}) against $|\vec{x}|$ in the ordered
    regime at $N=35$. On the left at $\lambda=0.06$ and $m^2=-0.1$ in the uniform phase and on
    the right at $\lambda=0.6$ and $m^2=-0.7$ in the non--uniform phase.}
  \label{fig:phi-spat-corr3} 
\end{figure}
\ref{fig:phi-spat-corr3} on the left also an example of $C(\vec{x})$
in the uniform phase.  In this phase we see a strong correlation as
expected in a uniform phase. However, in the striped phase (on the
right) at $\lambda=0.6$ the spatial correlator behaves differently in
the two directions. At this value of $\lambda$ we obtained two stripes
parallel to one of the axis. Therefore we find in one direction
(parallel to stripes) still a strong correlation. In the direction
vertical to the stripes we see a strong anti--correlation.  Since
at this value of the coupling we have only two
stripes, the correlator does not oscillate.\\

\begin{figure}[htbp]
  \centering
  \includegraphics[width=.48\linewidth]{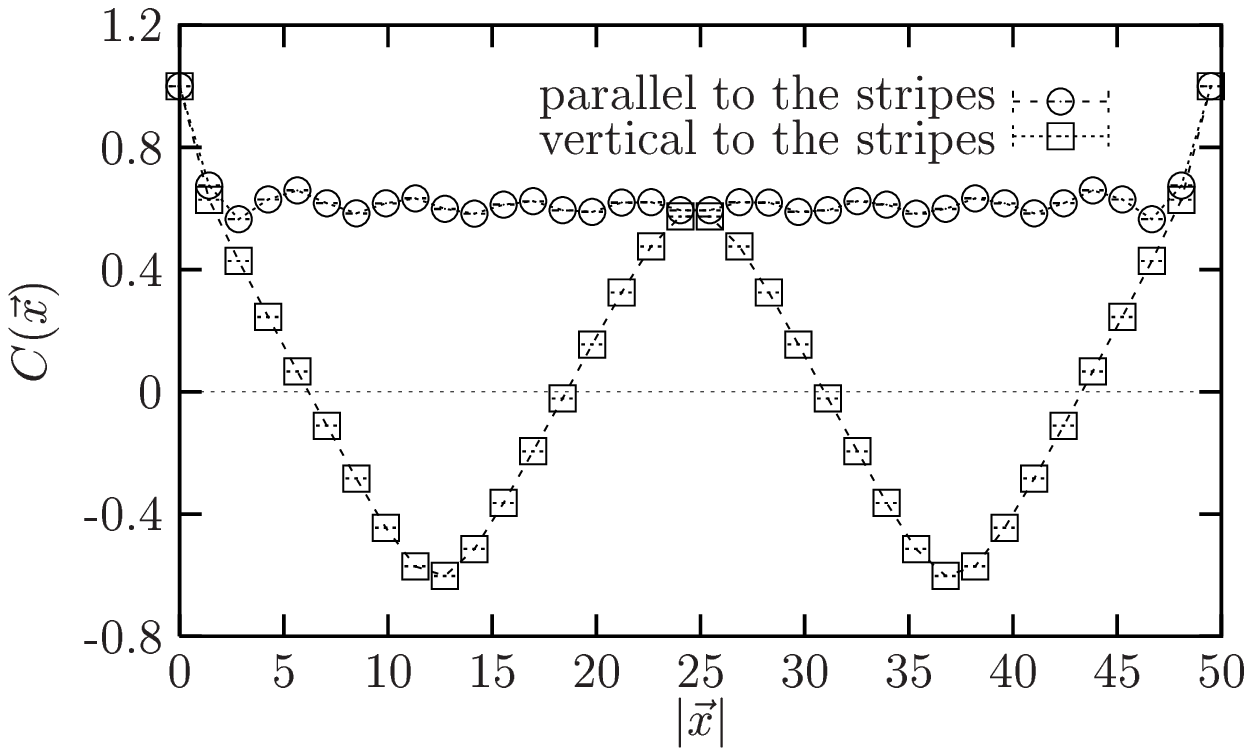} %
  \hspace{.3cm}\includegraphics[width=.48\linewidth]{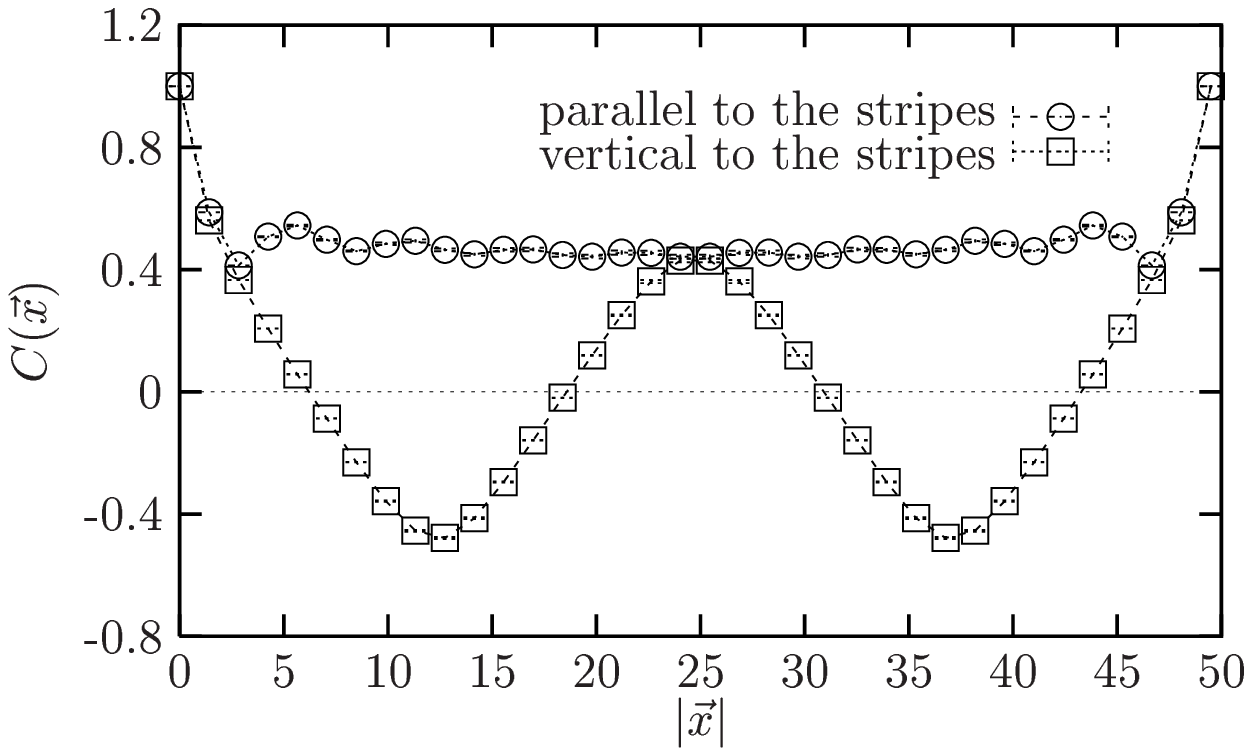} %
  \caption{The correlator (\ref{eq:phi-spat-corr}) against $|\vec{x}|$ in the striped
    phase at $N=35$. On the left at $\lambda=10$ and $m^2=-4$ and on the right at
    $\lambda=100$ and $m^2=-40$. The maximum of the order parameter is here at $\vec{m} = (1,1)$.}
  \label{fig:phi-spat-corr4} 
\end{figure}

The situation changes if we increase the coupling further. In Figure
\ref{fig:phi-spat-corr4} $C(\vec{x})$ is plotted at $\lambda=10$ and
$100$. In the last Section we showed already snapshots of
$\phi(\vec{x},t)$ at these values of the coupling.  We saw there that
depending on the starting configuration we obtain qualitatively
different patterns. This leads to a different behavior of the spatial
correlator (\ref{eq:phi-spat-corr}). Again we refer to Section
\ref{sec:phi-stripe-rev} for a discussion. Here we show results for
the different striped phases separately.

At these values of the coupling we find the maximum of
$\bar{M}(\vec{m})$ at $\vec{m}=(1,0)$ or at $\vec{m}=(1,1)$
corresponding to two stripes parallel to the $x_2$--axis or two
diagonal stripes, respectively. In the first case the correlator
$C(\vec{x})$ behaves in the same way as in Figure
\ref{fig:phi-spat-corr3} on the right.
In the case of diagonal stripes we obtain a clear oscillation of the
correlation function vertical to the stripes as shown in Figure
\ref{fig:phi-spat-corr4}. Parallel to the stripes $C(\vec{x})$ is
strongly correlated. Note that here the oscillations do not indicate
more than two stripes.

\subsection{Dispersion relation}
\label{sec:phi-disp}

In Section \ref{standard} we discussed the broken Lorentz symmetry and
the corresponding deformation of the dispersion relation in
non--commutative $\lambda\phi^4$ theory, in the framework of
perturbation theory.

In a Lorentz invariant theory the energy squared is linear in
$\vec{p}^{\,2}$ and the dispersion relation is given by
\begin{equation}
  \label{eq:phi-Lorentz-dispersion}
  E(\vec{p})^2=\vec{p}^{\,2}+M_\text{eff}^2\,,
\end{equation}
where $M_\text{eff}$ is the effective mass.  In the non--commutative
case equation (\ref{eq:phi-pert2}) implies a dispersion relation of
the form
\begin{equation}
  \label{eq:phi-deformed-dispersion}
  E(\vec{p})^2=\vec{p}^{\,2}+M_\text{eff}^2+\xi\frac{\lambda}{|\theta\vec{p}\,|}\,,
\end{equation}
where the last term is the leading one loop IR divergence.

Here we study the dispersion relation non--perturbatively.  Since we
do not consider renormalization aspects in this model, we are only
interested in the momentum dependence of the energy. To this end we
considered the two--point correlation function in time direction
\begin{equation}
  \label{eq:phi-gm}
  G(\vec{m},\tau)=\frac{1}{N^2T}\sum_t\left\langle\text{Re}\lrx{\tilde{\phi}^*(\vec{m},t)
      \tilde{\phi}(\vec{m},t+\tau)}\right\rangle\,,
\end{equation}
where $\tilde{\phi}(\vec{m},t)$ is again the spatial Fourier transform
that we already used in the definition of the order parameter
(\ref{eq:phi-order-1}). On an infinite lattice $G(\vec{m},\tau)$
decays exponentially in $\tau$.  Since we are on a finite lattice with
periodic boundary conditions $G(\vec{m},
\tau+T)=G(\vec{m},\tau)$, the correlator behaves like a
{\tt cosh} function,
\begin{equation}
  \label{eq:phi-gm-behavior}
  G(\vec{m},\tau) \propto \lrx{e^{-E(\vec{p})\,\tau}+e^{-E(\vec{p})(T-\tau)}}\,.
\end{equation}
Here $E(\vec{p})$ is the energy, where we rescaled the integer
representation of the momenta $\vec{m}$ to their physical value
$\vec{p}$
\begin{equation}
  \label{eq:phi-momenta}
  \vec{p}=\frac{2\pi}{N}\vec{m}\,.
\end{equation}
The correlator $G(\vec{m},\tau)$
allows us to determine the momentum dependence of the energy,
i.e.\ the dispersion relation, by studying its exponential decay.
This can be done either by fitting $G(\vec{m},\tau)$ to the function
(\ref{eq:phi-gm-behavior}) or by studying the ratio of two
subsequent values $-\log [G(\vec{m},\tau+1)/G(\vec{m},\tau)]$. 
\begin{figure}[htbp]
  \centering
  \includegraphics[width=.45\linewidth]{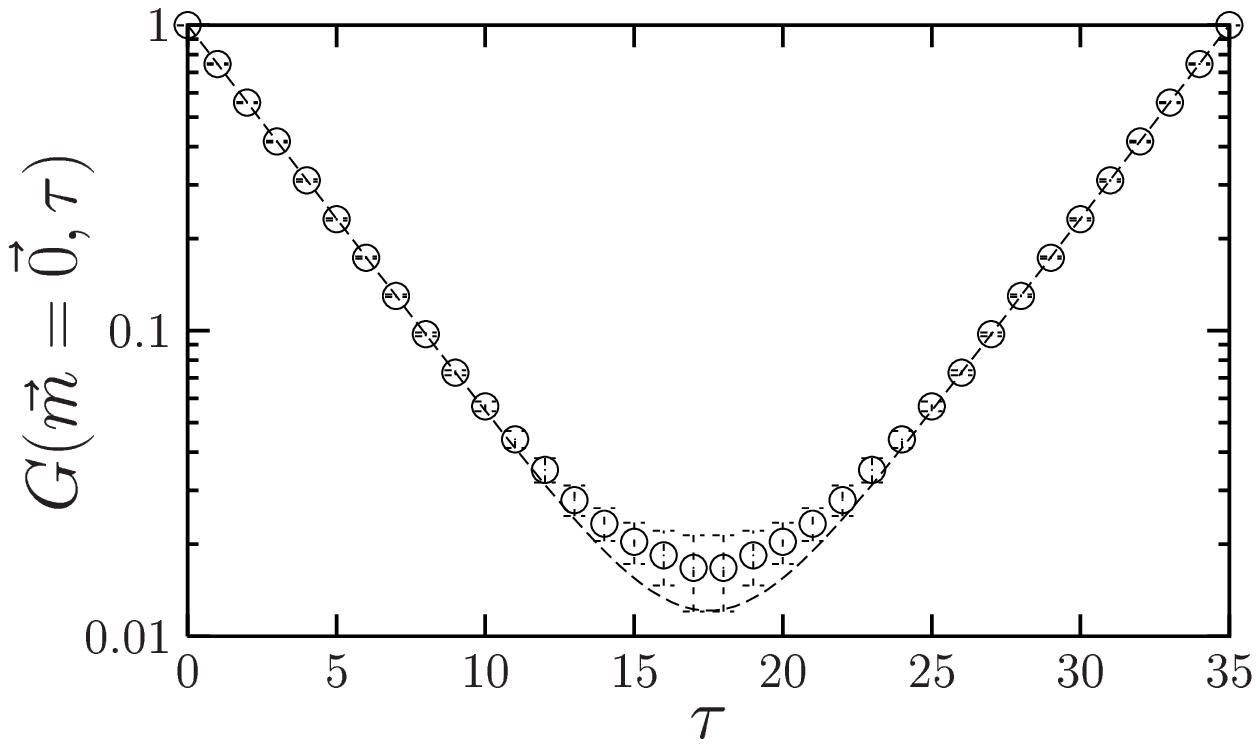} %
  \hspace{.5cm}\includegraphics[width=.45\linewidth]{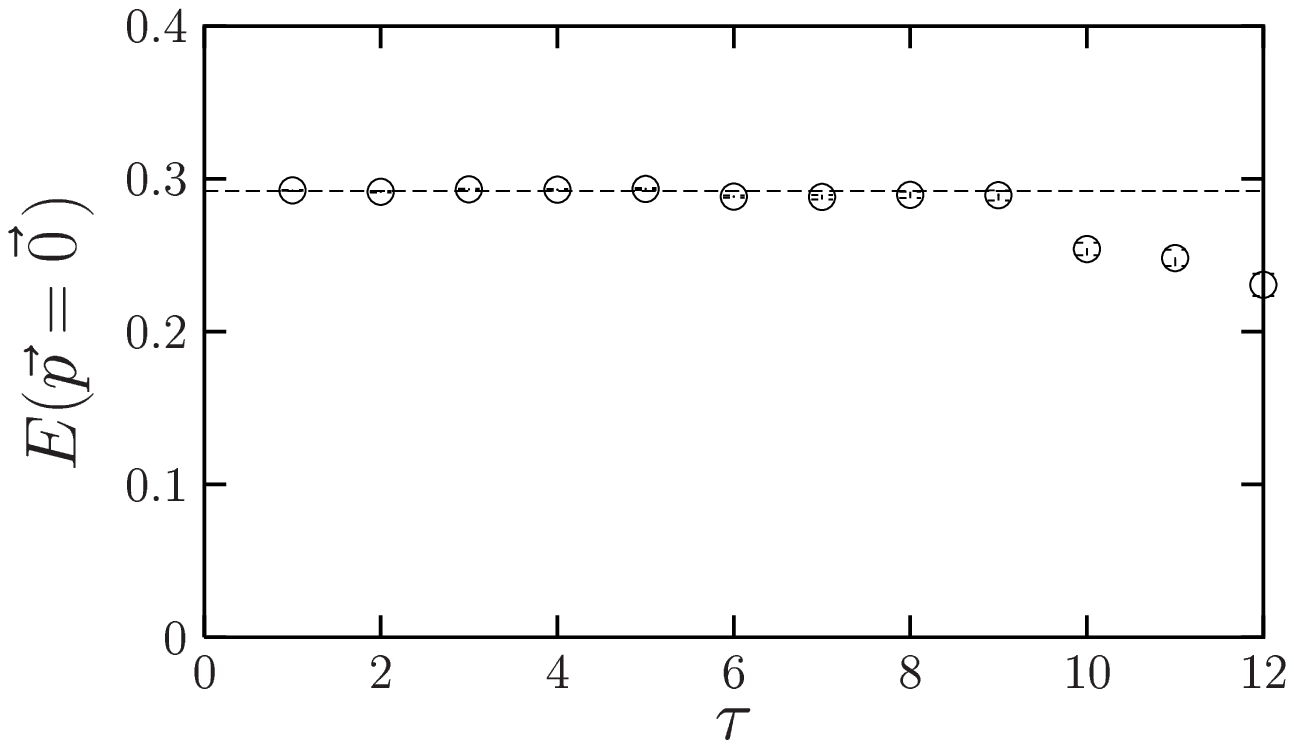}
  \caption{Determination of the energy $E(\vec{p})$. On the left the data are fitted to the function
    (\ref{eq:phi-gm-behavior}) and on the right we show $-\log
    (G(\vec{m},\tau+1)/G(\vec{m},\tau))$.}
  \label{fig:phi-gm-determine}
\end{figure}
With the first method we obtain $E(\vec{p})$ as a result of the fit, with
the second method the energy is determined by a plateau.  Results are
shown in Figure \ref{fig:phi-gm-determine}. In this example the
system is in the disordered phase and we computed
$G(\vec{m}=\vec{0},\tau)$. These measurements allow us to study the
dispersion relation.\\

\begin{figure}[htbp]
  \centering
  \includegraphics[width=.475\linewidth]{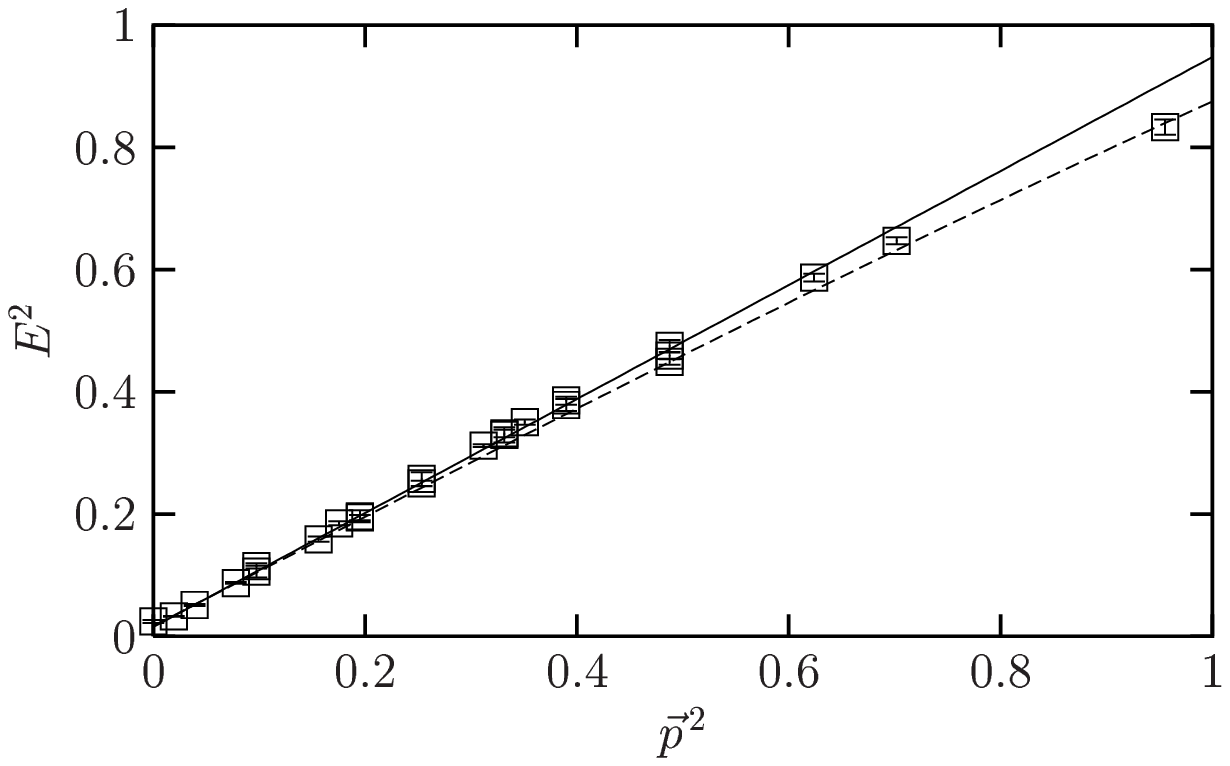} %
  \hspace{.25cm}\includegraphics[width=.495\linewidth]{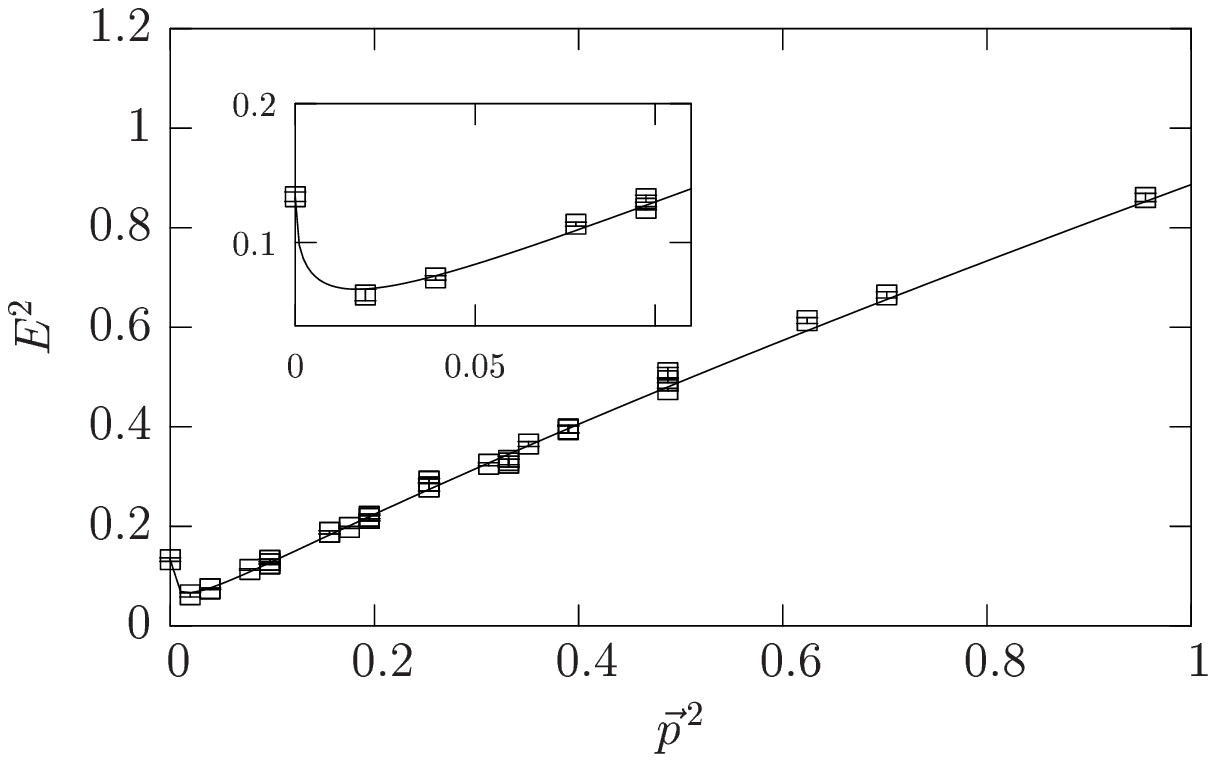}
  \caption{The energy $E^2(\vec{p})$ in the disordered phase at $N=45$, 
    on the left at $N^2\lambda=70$ and $N^2m^2=-17.5$, and on the right
    at $N^2\lambda=700$ and $N^2m^2=-280$.}
  \label{fig:phi-disp1}
\end{figure}

We evaluated configurations close to the disordered -- uniform
transition and close to disordered -- stripe transition. On all
configurations we measured $E(\vec{p})$ with the methods described above, for
various momenta $\vec{p}$. Two example results at $N=45$ are displayed
in Figure \ref{fig:phi-disp1}.
On the left $E^2$ is linear in $\vec{p}^{\,2}$ as one
expects in a Lorentz invariant theory. The solid line in this plot is
the result of a fit to the dispersion relation
(\ref{eq:phi-Lorentz-dispersion}), where we used the effective mass
$M_{\text{eff}}$ as the only free parameter. Since we are on a finite
lattice we see at larger momenta a deviation. Here the momentum
dependence of the energy is $E^2\propto(2\sin{|\vec{p}\,|/2})^2$ (if Lorentz
symmetry holds). A fit to this function is represented by the dashed
line in Figure \ref{fig:phi-disp1}. This line fits the data very well.
Since the energy minimum is at $|\vec{p}\,|=0$ at this value of
$\lambda$, we end up in the uniform phase when we decrease $m^2$.

The situation is different for larger $\lambda$. This is shown in
Figure \ref{fig:phi-disp1} on the right. In the vicinity of
$|\vec{p}\,|=0$ we see here a clear deviation from the dispersion
relation (\ref{eq:phi-Lorentz-dispersion}), at large momenta the
linear behavior is restored.  The increased energy at zero momentum is
in full agreement with the perturbatively predicted IR divergence.
Due to the discrete compactification (see Section \ref{sec:lat-dsp}) a
finite lattice spacing serves also as an IR cut--off, along with the
finite volume. Therefore we do not see a divergence. Since the
cut--off influences the IR behavior we included an ad hoc cut--off
$\kappa$, as a consequence of equation (\ref{eq:NC-cutoff}). We fitted
the data to the function
\begin{equation}
  \label{eq:phi-fit-function}
  E(p)^2=p^2+m^2+\frac{a_1}{p+\kappa}+a_2 (p+\kappa)\,.
\end{equation}
It turned out that the leading IR divergence in equation
(\ref{eq:phi-deformed-dispersion}) does not describe the low momentum
behavior sufficiently well. Therefore we added a linear term
\footnote{Adding more terms in the expansion does not change the
  results within the errors.}
in $p$ from the expansion of the exponential function in equation
(\ref{eq:phi-pert2}). The relative errors of the fit parameters are
shown in Table \ref{tab:phi-fit-error1}. The errors of the parameters
of the deformation are rather large, which is not surprising since we
have only one data point displaying this effect.

However, the minimum of the energy is here clearly at the smallest
non--vanishing momentum.  This minimum implies that for decreased
$m^2$ the system will be in the striped phase and the vacuum pattern
will have two stripes parallel to one of the axes. For $|\vec{p}\,|=0$
the energy increases again, which may indicate the IR divergence (at
$N\to\infty$).

Here we observe that at small $\lambda$, and equivalently for small
$\theta$, the effects of UV/IR mixing are strongly suppressed and we
restore the behavior of the commutative theory, including an Ising
type phase transition. At larger $\lambda$ the UV/IR mixing effects
become dominant and the phase transition changes its nature. This is
in qualitatively agreement with
the conjecture by Gubser and Sondhi \cite{Gubser:2000cd}.\\

\begin{figure}[htbp]
  \centering
  \hspace{-.11cm}\includegraphics[width=.48\linewidth]{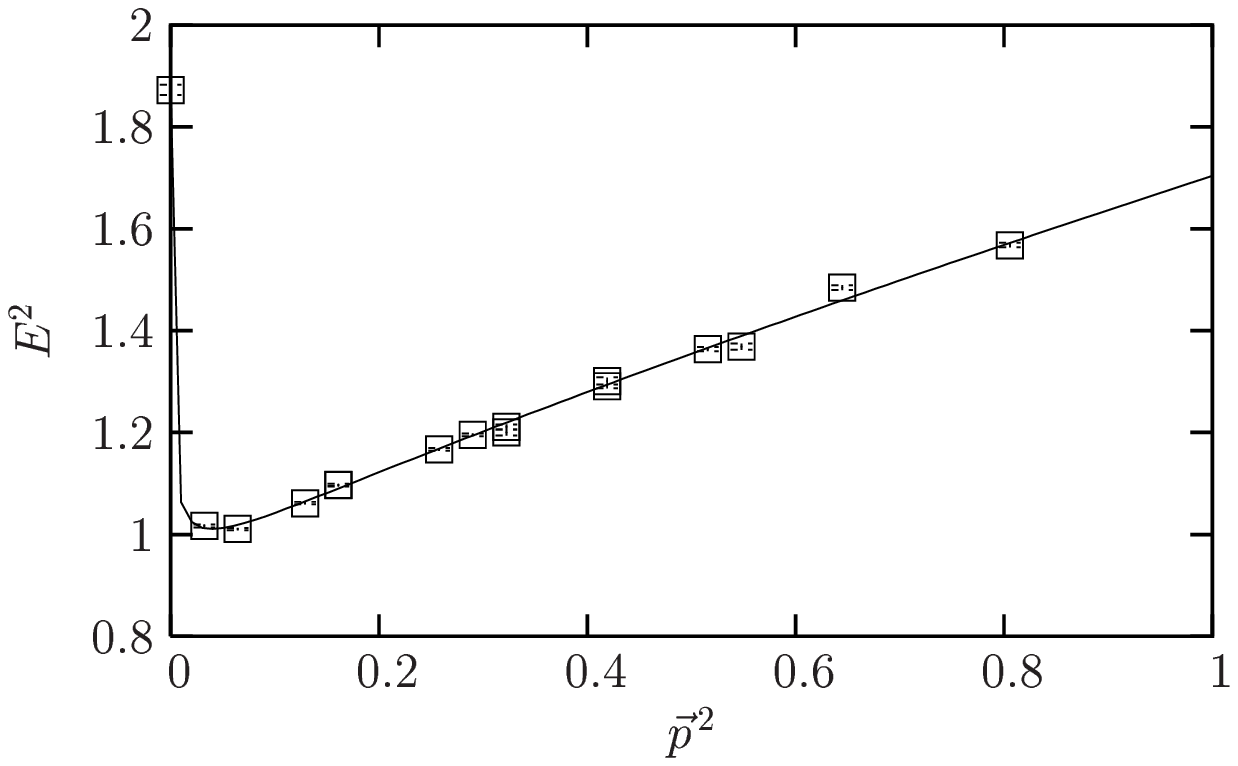} %
  \hspace{.4cm}\includegraphics[width=.48\linewidth]{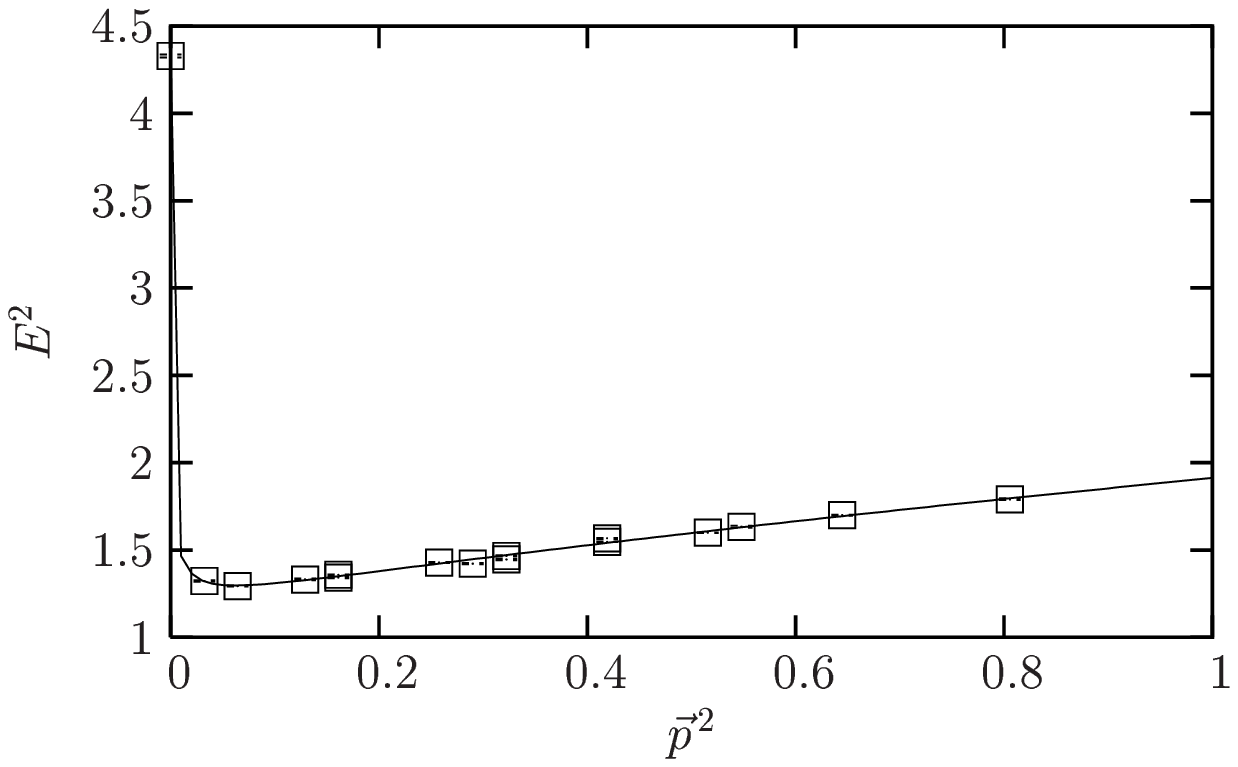}
  \caption{The energy $E^2(\vec{p})$ in the disordered phase at $N=35$. 
    On the left we are at $\lambda=10$ and $m^2=-1$, and on the right at
    $\lambda=100$ and $m^2=-10$. The energy at $\vec{p}=\vec{0}$
    increases with $\lambda$.}
  \label{fig:phi-disp2}
\end{figure}

\begin{figure}[htbp]
  \centering

  \hspace{.2cm}\includegraphics[width=.48\linewidth]{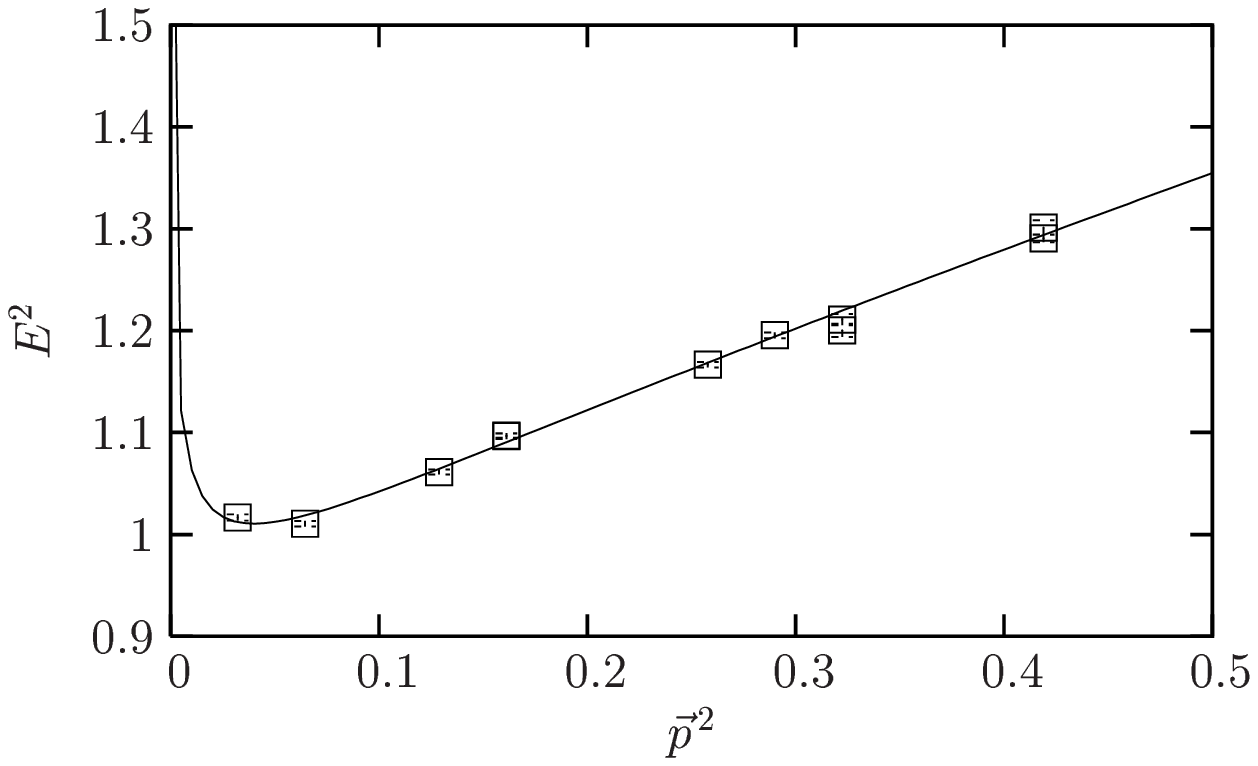} %
  \hspace{.2cm}\includegraphics[width=.48\linewidth]{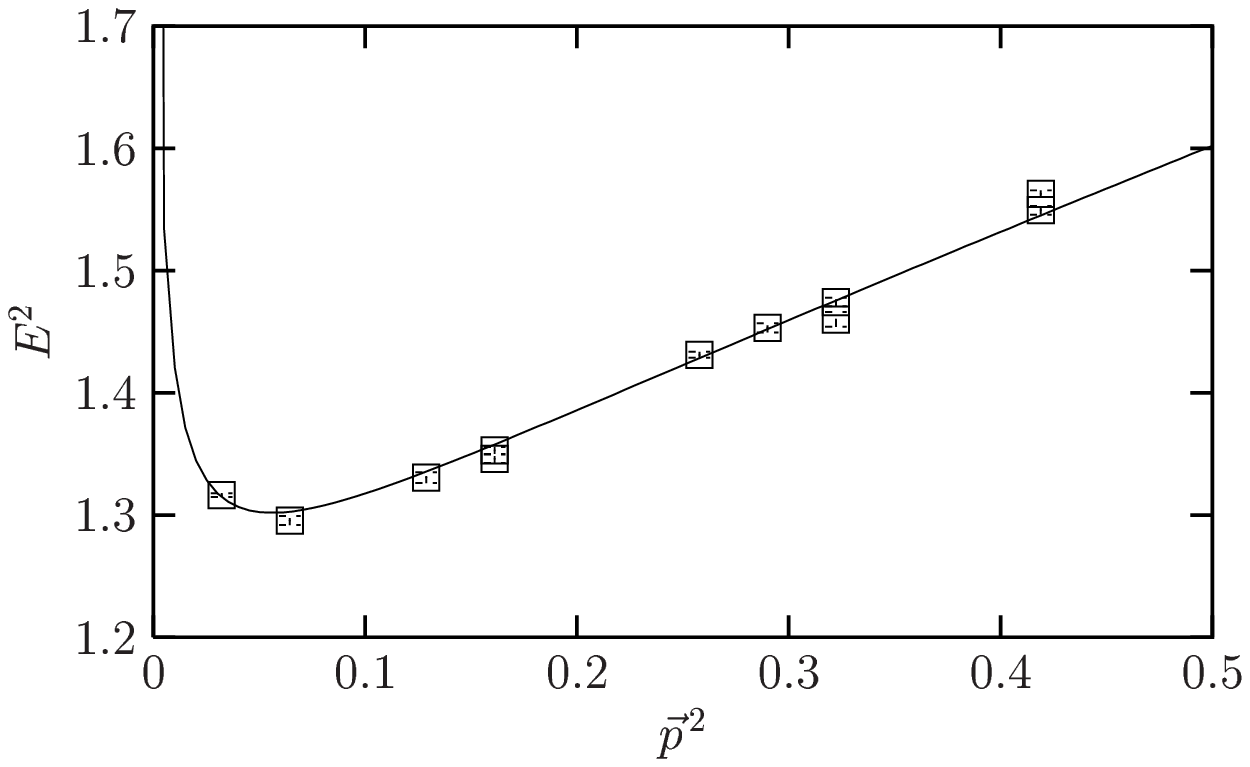}
  \caption{The energy $E^2(\vec{p})$ in the disordered phase at $N=35$. 
    On the left at $\lambda=10$ and $m^2=-1$, at $k=1$ and
    $k=\sqrt{2}$ the energy is equal within the statistical errors. On
    the right at $\lambda=100$ and $m^2=-10$, the minimum of the
    energy is shifted from $k=1$ to $k=\sqrt{2}$.}
  \label{fig:phi-disp3}
\end{figure}

For further increased $\lambda$ one expects the minimum at larger
momenta. In addition $E(0)$ should increase with $\lambda$.  Since at
a slightly larger coupling the expected effect did not show up, we
increased coupling drastically from $\lambda=0.6$ (the largest value
at $N=35$ that is plotted in the phase diagram
\ref{fig:phi-phase-diagram}) to $\lambda=10$ and $100$. The results
are shown in Figure \ref{fig:phi-disp2}.  We clearly see in these
plots that $E(0)$ increases with $\lambda$.  We fitted the data again
to the fit function (\ref{eq:phi-fit-function}). The results are
the solid lines in Figures \ref{fig:phi-disp2} and \ref{fig:phi-disp3}
and the relative errors of the fit parameters are shown in Table
\ref{tab:phi-fit-error1} in the last two lines. At these values of
$\lambda$ also the errors of the parameters describing the deformation
of the dispersion relation are under control. The results of the fits
are consistent with the {\em one loop} result of perturbation theory
(\ref{eq:phi-deformed-dispersion}) over a wide range of $\lambda$.
This is an unexpected result, since at these values of $\lambda$
effects from higher order perturbation theory are expected.  One might
conclude here that there are no qualitatively new IR singularities
from higher loop contributions.  However, this needs confirmation on
larger lattices.\\

To identify the minimum of the energy we plotted the dispersion
relation in a smaller range (Figure \ref{fig:phi-disp3}).  On the left,
at $\lambda=10$, the values of the energy at $k=|\vec{m}|=1$ and
$k=|\vec{m}|=\sqrt{2}$ are equal within the statistical errors. From
our data it is not possible to predict which of these momentum modes
will condense for decreased $m^2$. On the right in Figure
\ref{fig:phi-disp3}, at $\lambda=100$, the minimum of the energy is
clearly at the second smallest non--vanishing momentum, which
corresponds to $k=|\vec{m}|=\sqrt{2}$. Therefore the pattern in the
non--uniform phase will have two diagonal stripes.
\begin{table}[htbp]
  \centering
  \begin{tabular}{|c|c|c|c|c|}
    \hline\hline  \phantom{\LARGE A}parameter \phantom{\LARGE A}&$m^2$&$\kappa$&$a_1$&$a_2$\\\hline
    \phantom{\LARGE A} relative error in Figure \ref{fig:phi-disp1} right \phantom{\LARGE A}&0.05&0.25&0.2&0.2\\\hline
    \phantom{\LARGE A} relative error in Figure \ref{fig:phi-disp2} left \phantom{\LARGE A}&0.02&0.11&0.08&0.12\\\hline
    \phantom{\LARGE A} relative error in Figure \ref{fig:phi-disp2} right \phantom{\LARGE A}&0.02&0.15&0.13&0.15\\\hline
  \end{tabular}
  \caption{The relative errors of the fit parameters.}
  \label{tab:phi-fit-error1}
\end{table}

\section{The phase diagram revisited}
\label{sec:phi-stripe-rev}

As we have seen in Subsection \ref{sec:phi-disp}, the investigation of
the dispersion relation in the disordered phase provides information
about the ordered regime. The mode that drives the phase transition,
and therefore indicates the pattern of the ordered regime, is given by
the momentum that minimizes the energy.

This might give more insight into the transition region between uniform
and striped phase. The difference between the energy at
$\vec{m}=(0,0)$ and at $\vec{m}=(1,0)$ indicates the type of
ordering that will occur in the ordered regime. In the uniform phase
this difference is negative and in the striped phase positive. 
Equal energies indicate the phase transition.

Since this is a small effect we performed high statistic 
\footnote{To extract this information we used $20000$ configurations.}
simulations at $N=25$ in the range of $\lambda$, which is marked as 
transition region in Figure \ref{fig:phi-phase-diagram}. On these
configurations we measured the energy gap $E(0)-E(1)$.
\begin{figure}[htbp]
  \centering
  \includegraphics[width=.6\linewidth]{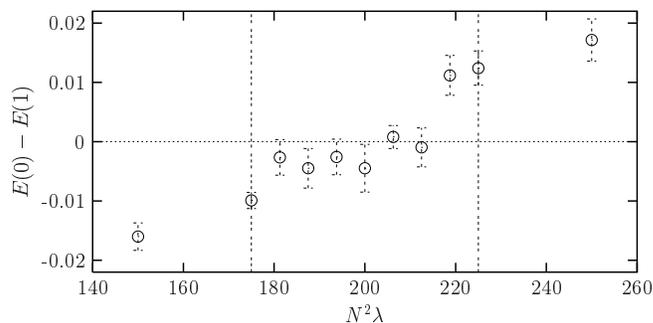} %
  \caption{The difference $E(0)-E(1)$ in the disordered phase at $N=25$ against
    $N^2\lambda$. The vertical lines mark the transition region
    between uniform and striped phase, which was determined by
    measurements of the order parameter.}
  \label{fig:phi-stripe-trans}
\end{figure}
The result is plotted in Figure \ref{fig:phi-stripe-trans}, where the
region between vertical lines represent the aforementioned transition
region. Outside this region and at its boundaries we clearly identify
the uniform phase (left) and the striped phase (right). Inside the
region the energy gap is zero within the errors with one exception
close to the striped phase. With this method we cannot identify a
transition line either. However, these measurements clarify why here
the pattern depends on the starting configurations. If $E(0)$ and
$E(1)$ are almost equal it is (numerically) not clear which mode
drives the phase transitions.  Since numerical studies always suffer
from the finite accuracy of computer numbers, it is hardly possible to
resolve these small effects.  Therefore already for two slightly
different configurations the system might end up in one case in the
uniform phase and in the other case in the striped phase,
both of which appear to be stable.\\

Regions where the minimum of the energy is difficult the identify are
of course not restricted to the case discussed here.  This occurs
whenever the difference between the minimum of the energy and the next
larger value of the energy is small compared to the absolute value of
the energies. This is certainly the case when the minimum of the
energy is changing with respect to the momentum. This obviously takes
place in Figure \ref{fig:phi-disp3} on the left. We also see this
behavior at large $\lambda$ as Figure \ref{fig:phi-disp3} on the right
shows. In the first case we face the same problem as in the uniform --
non--uniform transition area, which does not allow us to determine the
behavior in the ordered regime.

In the latter case the two lowest values of the energy are too close
to resolve the correct pattern in the striped phase with our
algorithm. However, from the analysis of the dispersion relation in the
disordered phase we know that at $N=35$ and $\lambda=100$ the striped
phase has $(1,1)$--patterns. The additional $(1,0)$--patterns in
Figure \ref{fig:stripe-multi} are only meta--stable.

It is evident that more complex pattern will occur at further
increased $\lambda$. However, in the limit of strong coupling the
kinetic term in the action (\ref{eq:phi-action}) becomes irrelevant.
In Monte Carlo studies this causes a dramatic increase of the
simulations steps that are needed to achieve the equilibrium (see
Appendix \ref{numerics}).\\

Finally we comment on the orders of the phase transitions.  We did not
study this topic systematically, but we have some indications of which
orders the transitions could be. 

The phase transition between disordered and ordered phase seems to be
most likely of second order. We assume that because we do not see any
indication of hysteresis at this transition. This holds for both, the
disordered -- uniform and the disordered -- non--uniform transition.
To study the hysteresis we performed simulations starting from the
disordered phase and by slowly decreasing $m^2$ we entered the ordered
regime. Once the system was clearly in the ordered regime we increased
$m^2$ slowly towards the disordered phase.  On these configurations we
measured the order parameter. In a first order phase transition the
order parameter would be different on both way. Since this hysteresis
did not show up we assume a second order phase transition.

We did not found a convincing indication of the order of the
transition from the uniform to the striped phase. Here we obtained a
transition region which prevents a prediction of the order.


\chapter{The 2d non--commutative scalar model}
\label{A-2d}

The occurrence of stripes in the ground state implies the spontaneous
breakdown of translation invariance. The case $d=2$ is particularly
interesting in this respect. Gubser and Sondhi argued based on an
action of the Brazovskiian form \cite{Brazovkii1975} and the
Mermin--Wagner theorem
\cite{Mermin:1966fe,Hohenberg1967,Coleman:1973ci} that stripes cannot
be stable in $d=2$. However, {\selectlanguage{danish}Ambj\o rn}
\selectlanguage{english} and Catterall pointed out that this theorem
is not applicable, because here we deal with a non--local action as
the star--product shows.  In fact they did observe non--uniform
patterns in their numerical results for non--commutative
$\lambda\phi^4$ model in $d=2$ \cite{Ambjorn:2002nj}, where the two
coordinates obey the commutator relation (\ref{eq:NC-comm-star}). In
general their results agree qualitatively with the results we got in
$d=3$. The only difference was that they obtained more complicated
patterns than the two--stripe
pattern at rather small coupling $\lambda$.\\

Inspired by their results we also studied the 2d version of the
non--commutative $\lambda\phi^4$ model. In two dimensions the action
reads
\begin{equation}
  \label{eq:2d-action}
      S[\hat{\phi}]=N\trace\biggl[\frac{1}{2}\sum_\mu
      \lrx{\hat{D}_\mu\,\hat{\phi}\,\hat{D}_\mu^\dagger-\hat{\phi}}^2
      +\frac{m^2}{2}\hat{\phi}^2+\frac{\lambda}{4}\hat{\phi}^4\biggl]\,,
\end{equation}
where we used the twist--eaters (\ref{eq:lat-gamma}) as shift
operators $\hat{D}_\mu$.  We performed the similar measurements as we
did in $d=3$ beginning with the phase diagram.  Again we used the
order parameter (\ref{eq:phi-order-final}), but since we are in two
dimensions there is no sum over the time $t$.
\begin{figure}[htbp]
  \centering
  \includegraphics[width=.85\linewidth]{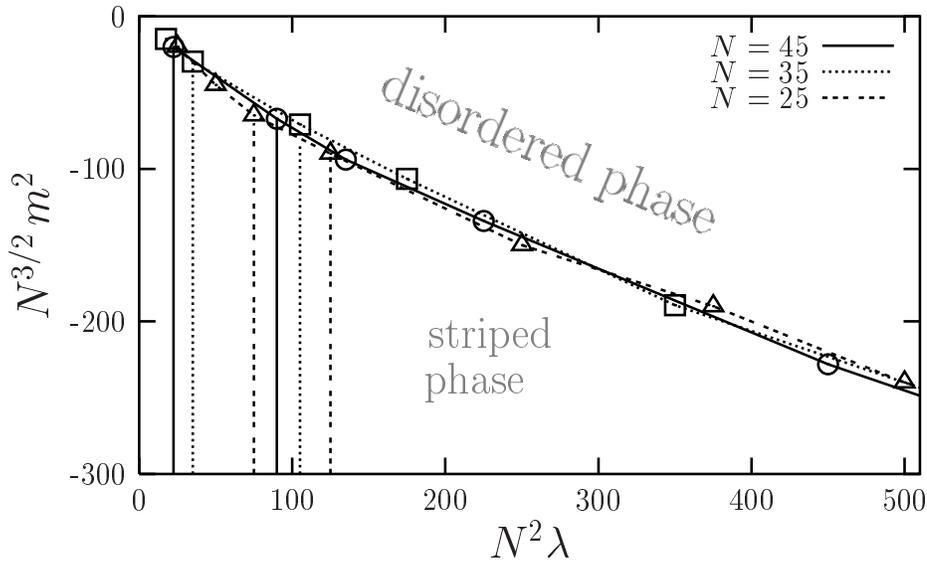}
  \caption{The phase diagram of the non--commutative 2d $\lambda\phi^4$ in the
    $m^2$ -- $\lambda$ plane.}
  \label{fig:2d-phase-diagram}
\end{figure}
The phase diagram we obtained is plotted in Figure
\ref{fig:2d-phase-diagram}. As in the 3d case the ordered regime is
split into a uniform and into a striped phase in agreement with
{\selectlanguage{danish}Ambj\o rn} and Catterall. In this case it
takes the factors $N^{3/2}$ and $N^2$ on the axes to stabilize the
phase
transitions in $N$.\\

\begin{figure}[htbp]
  \centering
  \vspace{.5cm}
  \hspace{-.5cm}\includegraphics[width=.2\linewidth]{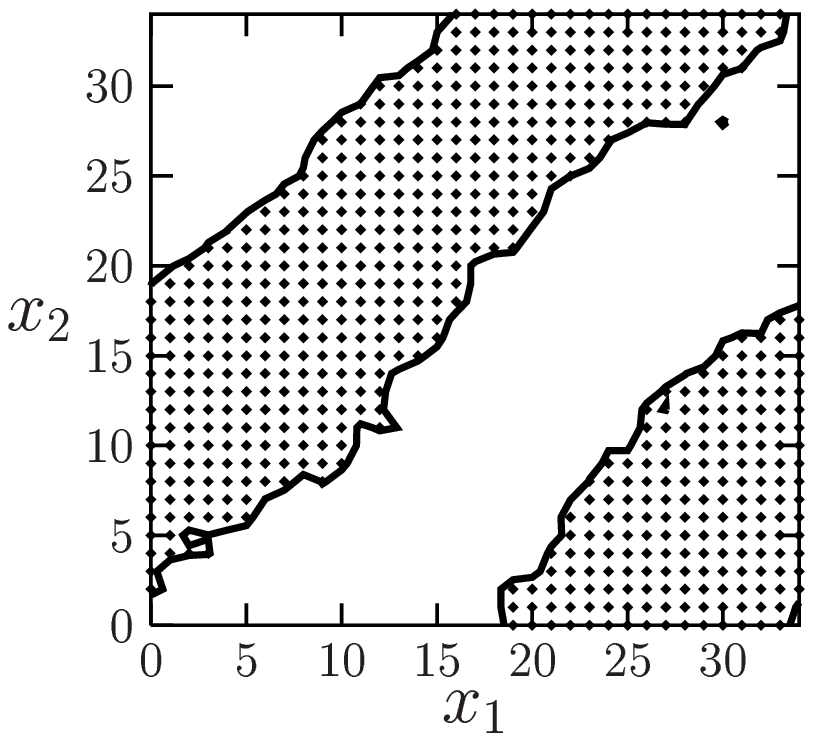} %
  \hspace{.2cm}\includegraphics[width=.2\linewidth]{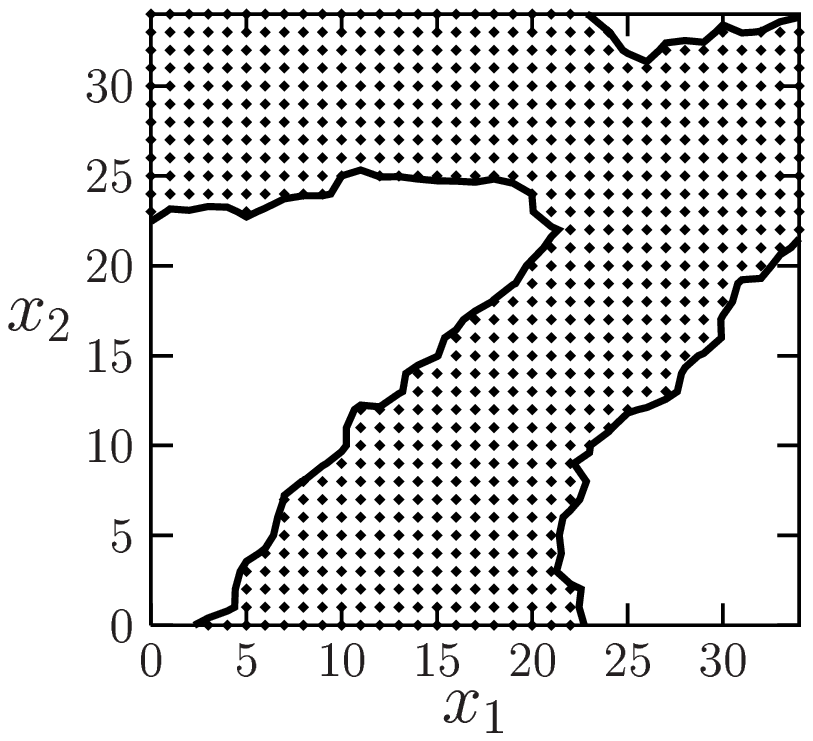} %
  \hspace{.2cm}\includegraphics[width=.2\linewidth]{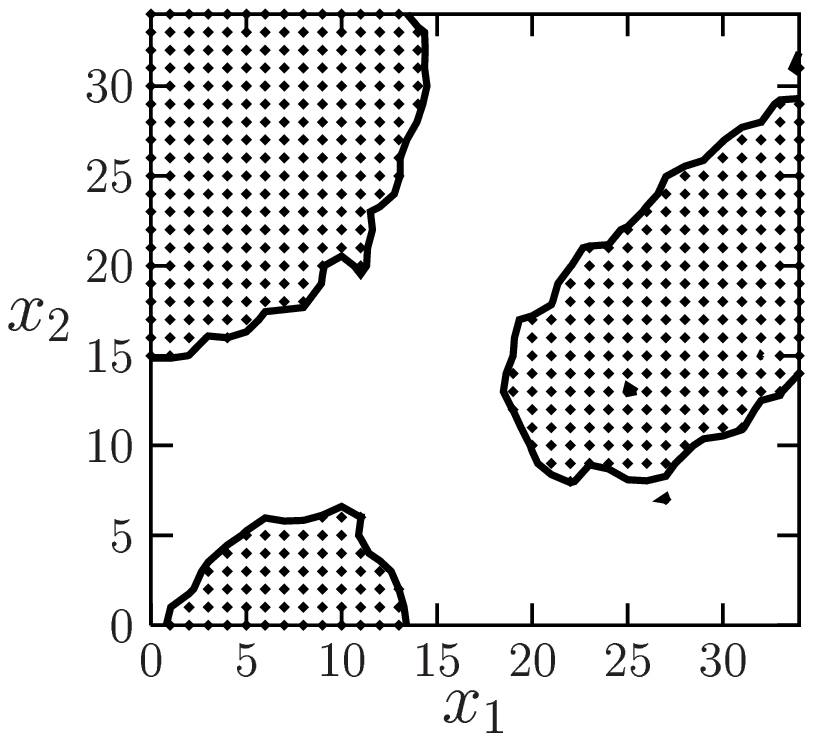} %
  \hspace{1.5cm}\includegraphics[width=.206\linewidth]{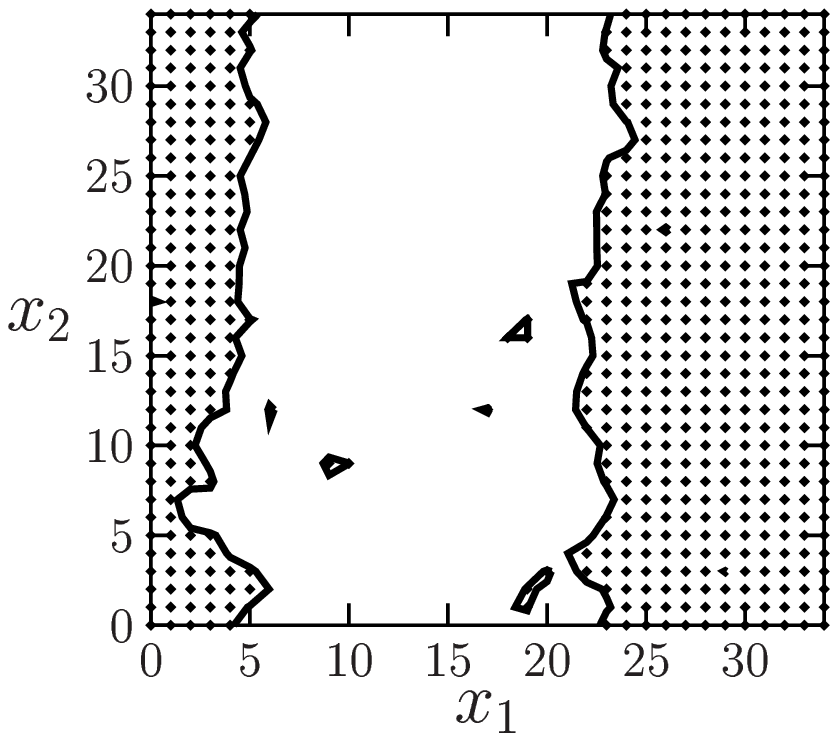} %
  \caption{Meta--stable patterns at $N=35$ and $N^2\lambda=350$ in the striped 
    phase. After a very long thermalization time all patterns turn
    into a two stripe pattern (on the right).}
  \label{fig:2d-meta-stripes} 
\end{figure}

We also analyzed the striped phase in the range of the parameters that
are plotted in Figure \ref{fig:2d-phase-diagram}. In this range we
always obtained patterns with two stripes.  Since
{\selectlanguage{danish}Ambj{\o}rn} and Catterall
\selectlanguage{english} obtained multi--stripes already at these
values of the coupling, we tried to understand the origins of these
differences. According to Ref.\ \cite{Ambjorn:2002nj} they used an
algorithm that updates the complete matrix at once, in contrast to our
algorithm where we updated pairs of matrix elements.  Therefore we
applied the same algorithm (although it has thermalization and
ergodicity problems see Section \ref{sec:numerics-2}) and indeed we
also discovered other patterns than two stripes.  Some example
snapshots at $N=35$ are plotted in Figure \ref{fig:2d-meta-stripes}.
These patterns show up after approximately $500$ simulation steps
(starting form a random configuration) and survived about $10^5$
further update steps.  After increasing the number of update steps
further to approximately $10^6$ to $10^7$ steps all patterns turn into
the two stripe pattern (Figure \ref{fig:2d-meta-stripes} on the right)
independent of the
starting configuration.
\footnote{In fact we faced the same problem in our study of the 3d model.
  In a first attempt we used the same algorithm as in Ref.\ 
  \cite{Ambjorn:2002nj} and we obtained multi stripes already at
  $N=35$.  Since we saw a dependence of the patterns on the starting
  configuration we improved the algorithm, which unmasked these patterns
  as meta--stable.}
We show here examples at $N=35$, but the situation is the same at
$N=45$.  However, the main result in Ref.\ \cite{Ambjorn:2002nj} is of
course the spontaneous breakdown of the translation invariance, and
therefore the existence of a striped phase, which was not expected in
two dimensions. In this point our results fully agree with those of
{\selectlanguage{danish}Ambj{\o}rn} \selectlanguage{english}
and Catterall.\\

We also computed the spatial correlation function defined in equation
(\ref{eq:phi-spat-corr}). As in $d=3$ we obtained in the striped phase
a strong correlation parallel to the stripes and a strong anti--correlation
vertical to the stripes.

\begin{figure}[htbp]
  \centering
  \includegraphics[width=.30\linewidth]{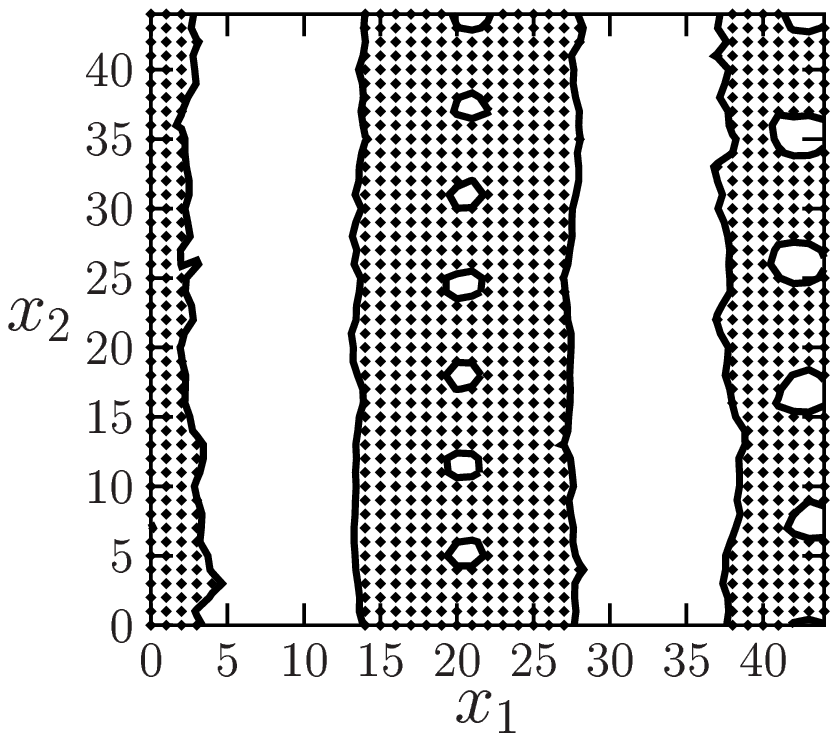} %
  \hspace{.5cm}\includegraphics[width=.46\linewidth]{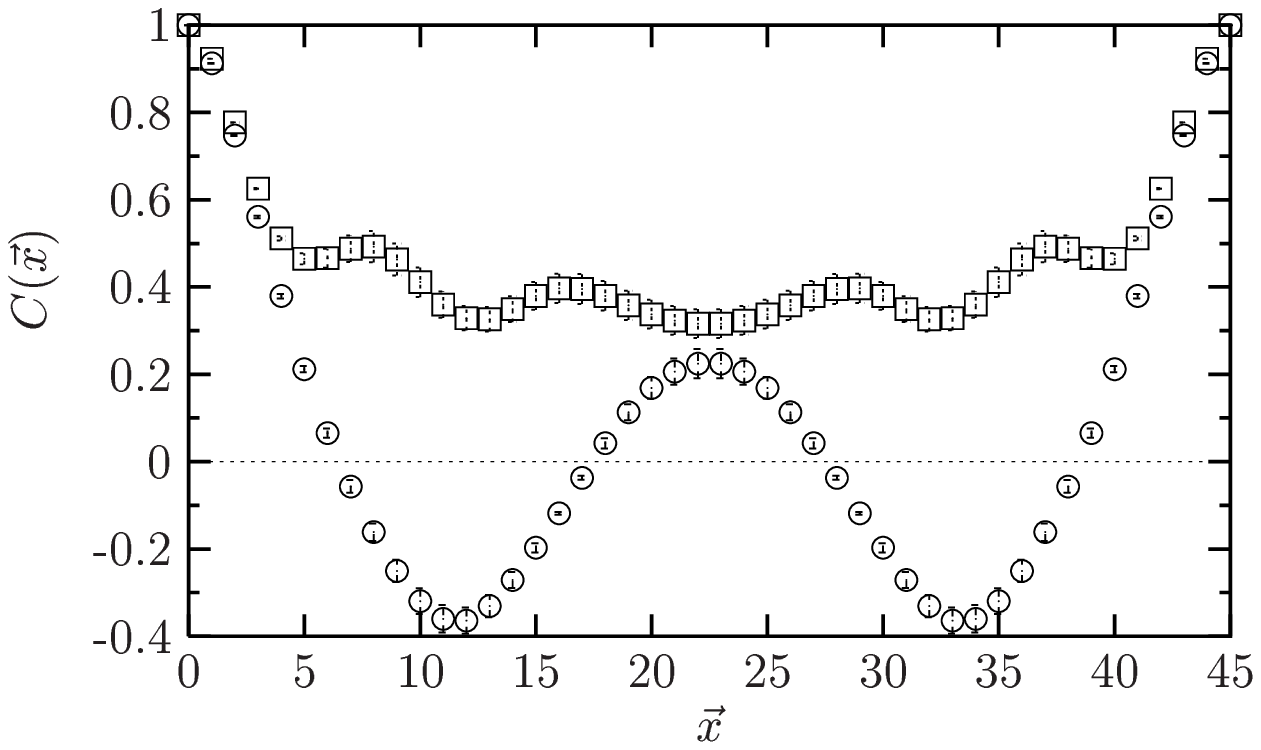} %
  \caption{An example snapshot of a configuration on the left and the correlator 
    (\ref{eq:phi-spat-corr}) against $|\vec{x}|$ on the right in the striped
    phase at $N=45$ and $N^2\lambda=20000$.}
  \label{fig:2d-multi-stripes} 
\end{figure}

For very large coupling $\lambda$ we finally obtained stable
multi--stripe patterns. One example at $N=45$ is shown in Figure
\ref{fig:2d-multi-stripes} on the left.
For all starting configurations we arrive at the same pattern, which
persists for an apparently unlimited number of update steps even for
the improved algorithm (\ref{eq:monte-phi-update}).

The pattern in Figure \ref{fig:2d-multi-stripes} has {\em four} stripes
where the stripes themselves have some substructure.  On the right we
plotted the correlation function in position space
(\ref{eq:phi-spat-corr}).  The stripes show up  as an oscillation
around zero of this correlator perpendicular to the stripes. The
substructure of the stripes leads to a slight oscillation of the
correlator parallel to the stripes.



\chapter{Summary and conclusion}
\label{conclusion}

\fancyhead{}
\fancyhead[LE,RO]{\bfseries\thepage}
\fancyhead[LO]{\bfseries Summary and conclusion}
\fancyhead[RE]{\bfseries Summary and conclusion}

We investigated non--commutative field theories in lower dimensions
non--perturba\-tively in the lattice approach.  Since the non--local
star--product arises also in the lattice formulation it is not
suitable for a direct Monte Carlo investigation. Therefore we used a
finite dimensional representation of the operator formulation
of these field theories. This leads to dimensionally reduced models.\\

In non--commutative gauge theory the reduced model (reduced to $d=0$)
is given by the twisted Eguchi Kawai model (TEK). Within this model we
investigated the continuum limit of 2d non--commutative $U(1)$ theory.
In the TEK this limit corresponds to the large $N$ double scaling
limit. To this end we studied the double scaling limit of the 1--point
function and the 2--point function of Wilson loops as well as the
2--point function of Polyakov lines.

The first conclusion from our simulation results is that we do observe
a double scaling limit as $N, \, \beta \to \infty$. This corresponds
to the continuum limit of the non--commutative $U(1)$ gauge theory,
which has therefore also finite observables. This observation
demonstrates the non--perturbative renormalizability of 2d
non--commutative $U(1)$ gauge theory.

The Wilson loop follows an area law at small physical areas, and in
this regime non--commutative gauge theory agrees with planar standard
gauge theory. However, at larger areas the Wilson loop becomes complex
and the real part (the mean values over both loop orientations) begins
to oscillate around zero.  The phase is proportional to the physical
area enclosed by the Wilson loop, irrespectively of its shape, and the
coefficient of proportionality is given by the inverse of the
non-commutativity parameter $\theta$.  This agrees with the
Aharonov--Bohm effect in the presence of a constant magnetic background
field
\begin{equation*}
B = \frac{1}{\theta} \,.
\end{equation*}
Our results support this law, which is also a key element of the
Seiberg and Witten description of non--commutative gauge theory
\cite{Seiberg:1999vs}. Moreover, the same law also occurs in condensed
matter physics: if one assumes a plane crossed by a constant magnetic
flux, then the electrons in this plane can be projected to the lowest
Landau level in a non--commutative space, where $\theta = \hbar c/eB $
\cite{Girvin:1987fp,Bellissard1994,Girvin:1999}.

The behavior of the Wilson loop at large areas
implies that we have found a qualitatively new
universality class.
At first sight, it may look surprising that the
non-commutativity --- which introduces a short--ranged non--locality
in the 
action --- changes the IR behavior of the gauge theory completely.
However, this effect does not appear unnatural 
in the presence of UV/IR mixing. It is remarkable that this
mixing effect does not occur in the perturbative expansion
of this model, hence our results for large Wilson loops reveal
a purely non--perturbative UV/IR mixing.

For the connected Wilson loop 2--point function, as well as the
2--point function of the Polyakov line, we can confirm the large $N$
scaling. These observables are in agreement with a universal wave
function renormalization, which yields a factor $\beta^{-0.6}$ for a
connected 2--point function.\\

The second model we investigated was the 3d non--commutative
$\lambda\phi^4$ theory, with a commutative time coordinate and two
non--commutative space directions. Again the lattice model is mapped to a
dimensionally reduced model, where here the reduced model is one
dimensional.

In this model we explored the phase diagram in the $m^2$ -- $\lambda$
plane. In the ordered phase (at strongly negative $m^2$) we found at
small values of $\lambda$ a uniform order of the Ising type, as in the
commutative case. At larger $\lambda$ --- which amplifies the
non--commutative effects --- we observed striped patterns. Up to
moderate values of $\lambda$ we obtained a pattern of two stripes
parallel to the axes for $N=15\dots 45$. This corresponds to a minimum
of the energy at the smallest non--zero lattice momentum.  At very
large $\lambda$ (compared to the coupling at the uniform --
non--uniform phase transition) also other patterns showed up, in
agreement with the conjecture of Gubser and Sondhi. 
However, the continuum limit has to be taken to confirm this
agreement. We observed the same behavior in $d=2$ (see Chapter
\ref{A-2d}). Also there the multiple stripes occurred first at rather
large values of $\lambda$.  The dominance of stripes implies the
spontaneous breaking of translation symmetry, which is also possible
in 2d since the action is non--local.

In the ordered regime, the spatial correlations are dictated by the
dominant pattern: uniform as in the commutative case, or striped with
strong correlations in the direction of the stripes and strong
anti--correlation vertical to them. This agrees with the prediction in
Ref.\ \cite{Gubser:2000cd}. In the disordered phase the spatial
correlators deviate at small $\lambda$ from the exponential decay.  At
large $\lambda$ the decay is again exponential as was predicted from
one loop perturbation theory in Ref.\ \cite{Minwalla:1999px}.

The correlators in momentum space do decay exponentially in time for
all momenta. This property allowed us to study the dispersion relation
in the disordered phase: at small $\lambda$ the dispersion relation
behaves qualitatively like in the commutative case, but at large
$\lambda$ there appears a jump at $|\vec{p}\,|=0$ as a non--commutative
effect. Here we observed the energy minimum at
$|\vec{p}\,|=\frac{2\pi}{N}$ for moderate $\lambda$ and at very large
$\lambda$ the minimum is at $|\vec{p}\,|=\sqrt{2}\,\frac{2\pi}{N}$.  The
results agree qualitatively with the results obtained in perturbation
theory. Also from non--perturbative studies we observed an IR
divergent behavior with the predicted $1/|\vec{p}\,|$ divergence. However, since
we obtained rather large fitting errors in the parameters of the IR
dominant term this needs further confirmation.

\newpage
\subsubsection{Outlook}

In this work we presented results of our first steps in the
non--perturbative study of non--commutative field theory.  The next
steps can be divided into next, near future and future projects.

The next project is to study the 3d $\lambda\phi^4$ model at larger
values of $N$. We expect from these studies more insight into the
stripe structure in the ordered regime as well as into the IR behavior
of the theory. Since the smallest non--zero momentum is given by
$|\vec{p}|=\frac{2\pi}{N}$, a study at larger $N$ will resolve the
interesting low momentum regime better. In addition we want to study
the continuum limit and the related question of renormalizability.

The near future project is to study non--commutative field theory in
larger dimensions, for example 4d $\lambda\phi^4$ model or 3d gauge
theory. Also a two or three dimensional $\sigma$ model would be
interesting to study. In addition there are plans to study models
including fermions for example a non--commutative Gross--Neveu model.

In the long run we want to study models, which allow phenomenological
predictions, like non--commutative pure gauge theory or
non--commutative QED in $d=4$.




\vspace{3cm}
\noindent{\bf Acknowledgements:}
First of all I would like to thank M.\ M\"uller--Preu{\ss}ker, D.\ L\"ust
and W.\ Bietenholz for their support and for the pleasurable and inspiring
atmosphere, they provided.

In particular I also thank J.\ Nishimura, who
suggested the numerical study of non--commutative field theory.
Together with W.\ Bietenholz we build this small collaboration and I want to
thank both for letting me participate in this challenging field of
research.

In addition I thank V.\ Linke. I was a member also of his group and
therefore I had the advantages of two working groups.  I am indebted
to both groups and especially to A.\ Barresi for helpful discussions
about critical phenomena and phase transitions. I also thank S.\ 
Shcheredin, A.\ Sternbeck and C.\ Urbach for inspiring discussions.


\begin{appendix}
  \addcontentsline{toc}{chapter}{Appendix}
  \renewcommand{\chaptermark}[1]%
  {\markboth{\appendixname\ \thechapter\ #1}{}}

\chapter{The numerical methods}
\label{numerics}
\lhead[\fancyplain{}{\bfseries\thepage}]%
{\fancyplain{}{\bfseries\rightmark}}
\rhead[\fancyplain{}{\bfseries\leftmark}]%
{\fancyplain{}{\bfseries\thepage}} 

In this appendix we discuss the details of the algorithms
we used to simulate the non--commutative models. We start with 
some general comments on Monte Carlo simulations.

\section{Monte Carlo simulations}
\label{sec:monte}

The idea of Monte Carlo (MC) simulations is to integrate approximately
the infinite dimensional path integrals that appear in lattice
formulation of any Euclidean field theory, with statistical methods.

Vacuum expectation values of some observables $\mathcal{O}[U]$ are
computed in the path integral approach by
\begin{equation}
  \label{eq:momte-path-integral}
  \langle\mathcal{O}\rangle=\frac{1}{Z}\int DU\mathcal{O}[U]e^{-S[U]}\,,
\end{equation}
where $S$ is the action. The action $S$ may also depend on matter fields,
\footnote{In that case also the matter fields have to be
  integrated.}
but for simplicity we restrict ourselves to gauge fields
$U_\mu(x)\in U(n)$.  $Z$ is the partition function
\begin{equation}
  \label{eq:monte-partition-function}
  Z=\int DU e^{-S[U]}\,.
\end{equation}

A set of gauge fields $\{U\}_\alpha$, one for each link on the
lattice, is a called a configuration. The idea of MC simulations is to
generate as many configurations as possible via {\em importance sampling}.
Importance sampling means that the configurations are generated
according to the probability
\begin{equation}
  \label{eq:monte-probability}
  W[U]=\frac{1}{Z}e^{-S[U]}\,.
\end{equation}

In addition to importance sampling a simulation algorithm has to
fulfill the condition of ergodicity. This means that for any two
configurations $\{U\}_\alpha$ and $\{U\}_\alpha'$ the probability for
a transition from one configuration out of the other, within finite
number of update steps, is non--zero. 

On each configuration the observable $\mathcal{O}[U]$ is calculated and
the result is the value $\mathcal{O}_\alpha[U]$. The expectation value
(\ref{eq:momte-path-integral}) is then given by
\begin{equation}
  \label{eq:monte-expectation}
  \langle\mathcal{O}\rangle=\lim_{N\to\infty}\frac{1}{N}\sum_{\alpha=1}^N\mathcal{O}_\alpha[U]\,.
\end{equation}
In practice the number of configurations $N$ is of course limited and
therefore we get an approximation of the expectation value 
\begin{equation}
  \label{eq:monte-expectation-real}
  \langle\mathcal{O}\rangle\approx\overline{\mathcal{O}}=\frac{1}{N}\sum_{\alpha=1}^N\mathcal{O}_\alpha[U]\,,
\end{equation}
with an associated statistical uncertainty. If the configurations are
statistically independent this uncertainty reads
\begin{equation}
  \label{eq:monte-error}
  \lrx{\Delta\mathcal{O}}^2={\frac{\overline{\mathcal{O}^2}-\overline{\mathcal{O}}^2}{N-1}}\,,
\end{equation}
where $\overline{\mathcal{O}^2}$ is the average of $\mathcal{O}^2$. In
practical applications the configurations generated with a Monte Carlo
simulation are always correlated and the error calculated with
equation (\ref{eq:monte-error}) is usually underestimated. This
correlation in a sequence of configurations is called autocorrelation.
For an infinitely large set of configurations the autocorrelation
time is defined as
\begin{equation}
  \label{eq:num-auto-corr}
  \tau_\text{auto}=\frac{1}{2}\sum_{\tau=-\infty}^{\infty}
  \frac{\langle\mathcal{O}_{\alpha}\mathcal{O}_{\alpha+\tau}\rangle 
    -\langle\mathcal{O}_{\alpha}\rangle\langle\mathcal{O}_{\alpha+\tau}\rangle}
  {\langle\mathcal{O}^2\rangle-\langle\mathcal{O}\rangle^2}\,.
\end{equation}
There are several methods to take the autocorrelation
time into account or to eliminate the effects of autocorrelation. Here
we discuss only the methods that were used in this work.

\subsubsection{Error estimation}

In our evaluation of the sets of observables $\{\mathcal{O}_\alpha\}$
we used two types of error estimations: the {\em binning} method and the
{\em jack--knife} method. Both methods are based on dividing the set of
observables into a number $n$ of subsets. In each subset the average
is calculated according to equation (\ref{eq:monte-expectation-real}),
where the number of configurations $N$ has to be replaced by the
number of configurations in a subset. The result is a set of averages
$\tilde{\mathcal{O}}_i,\quad i=1\dots n$.

Within the binning method one considers the averages $\tilde{\mathcal{O}}_i$
as statistically independent and therefore the error is given by
equation (\ref{eq:monte-error}), where now the set of averages is
used. The error calculated in this way varies with the number of bins
$n$. In order not to underestimate the statistical error one should
compute the errors for different values of $n$ and take the largest
error.

The jack--knife method uses the averages $\tilde{\mathcal{O}}_i$ to
construct $n$ statistically independent subsets
$\hat{\mathcal{O}}_k,\quad k=1\dots n$ in the following way,
\begin{equation}
  \label{eq:monte-jackknife}
  \hat{\mathcal{O}}_k = \frac{1}{n-1}\sum_{i\neq k}\tilde{\mathcal{O}}_i\,.
\end{equation}
Here $\hat{\mathcal{O}}_k$ is the average of the
$\tilde{\mathcal{O}}_i$ omitting the subset $\tilde{\mathcal{O}}_k$.
The error is now calculated according to equation
(\ref{eq:monte-error}), using $\hat{\mathcal{O}}_k$ as subset. This
method has the advantage that it allows to compute the error of
quantities that consist of two or more expectation values like the
connected part of 2--point functions.

\subsubsection{Steps of a Monte Carlo simulation}

Before starting a simulation one has to perform tests of the
algorithm. This is of course always important. Here it is
indispensable, since in the case discussed there are no results
available in the literature to compare qualitatively the obtained
results. The test for each algorithm will be discussed at the end of
the corresponding Section.  After these tests the simulations can be
started. Here we briefly
summarize the main steps of a typical Monte Carlo simulation.\\

\noindent A Monte Carlo simulation consist of

\begin{enumerate}
\item choosing a starting configuration. This can be either a random
  configuration or the gauge field set to certain values.
\item performing update steps I: the first configurations are in general
  not distributed according to (\ref{eq:monte-probability}), since a
  random configurations is located anywhere in configuration space.
  Therefore one performs update steps till the system is in
  equilibrium. These steps are called thermalization steps.
\item performing update steps II: having reached the equilibrium a set
  of configurations is evaluated using equation
  (\ref{eq:monte-expectation-real}). The evaluated configurations
  should be separated by the autocorrelation time.

\end{enumerate}

\section{Details of the TEK Model simulation}
\label{sec:numerics-1}

For the TEK model we chose the {\em heat bath algorithm}
\cite{Creutz:1980zw}.  In this method each link variable
$U_\mu$ is updated separately, while the others are fixed. A new
link is suggested and it is accepted with the probability
\begin{equation}
  \label{eq:num-heat-bath}
  P\propto\exp\lrx{-\widetilde{S}[U]}\,,
\end{equation}
where $\widetilde{S}[U]$ is the part of the action that depends on
$U_\mu$. In other words, it is the sum over all plaquettes which
involved $U_\mu$, times the coupling $\beta$. To enable this
disentanglement the action has to be {\em linear} in the link variables.

In the form of equation (\ref{eq:tek-tek-action}), the action cannot
be simulated with the heat bath algorithm because it is not linear
in the matrices $U_\mu$. Following Ref.\ \cite{Fabricius:1984wp} we
introduce an auxiliary field $Q$, which enters in a Gaussian form
and linearizes the action in $U_\mu$,
\begin{equation}
  \label{eq:num-action}
  \begin{split}
    S_{\text{TEK}}[U,Q]=N\beta\sum_{\mu<\nu}\trace\biggl[&Q_{\mu\nu}^\dagger Q_{\mu\nu} 
    - Q^\dagger_{\mu\nu} \lrx{t_{\mu\nu}U_\mu U_\nu+t_{\nu\mu}U_\nu U_\mu}\\
    &- \lrx{t^*_{\mu\nu}U^\dagger_\mu U^\dagger_\nu+t^*_{\nu\mu}
      U^\dagger_\nu U^\dagger_\mu} Q_{\mu\nu} \biggl]\,.
  \end{split}
\end{equation}
The auxiliary field $Q$ consists of general complex $N\times N$
matrices, and their integration reproduces the TEK action
(\ref{eq:tek-tek-action}), if $t_{\mu\nu}=\sqrt{\mathcal{Z}_{\mu\nu}}$.
The $U_\mu$ matrices are updated in $SU(2)$ subgroups
in the spirit of Cabbibo and Marinari \cite{Cabibbo:1982zn}, and the 
update of $Q_{\mu\nu}$ is done by generating Gaussian variables.

With this algorithm we generated configurations at various values
of $N$ keeping $N/\beta$ fixed. Table \ref{tab:num-stat-tek} shows
the number of configurations we generated at each value of $N$.
\begin{table}[htbp]
  \centering
  \begin{tabular}{|c|c|c|c|c|c|c|c|c|}
    \hline $N$&25&35&55&85&125&195&255&515\\\hline
    \# of configurations&9000&8400&20000&8000&3400&260&120&18\\\hline
  \end{tabular}
  \caption{Overview of the statistics in our simulation of the TEK model.}
  \label{tab:num-stat-tek}
\end{table}
All configurations are separated by $50$ update steps, which is more
than the autocorrelation time in these simulations (the maximum
autocorrelation time measured was $40$ update steps). We started the
simulations with random configurations and it took approximately $100$
update steps to reach equilibrium.

The errors of the Wilson loops (\ref{eq:Ntek-norm-wilsonloop}) and the
Polyakov lines (\ref{eq:tek-pol}) were computed with the binning
method.  To estimate the errors of the two--point functions of Wilson
loops (\ref{eq:Ntek-wiltwo}) we used the jack--knife
method, since this quantity consists of two expectation values.\\

The simulations were performed on $10$ Intel PCs with the operating
system Linux. The PCs had a clock rate of $450$ MHz. The generation
of the configurations took approximately five month.

For the measurement of the observables we had to compute high powers
of ${U}_\mu$ matrices.  Since they are not sparse matrices, in
contrast to typical situations in lattice gauge theory, there is no
simplified method to calculate these powers. One way to compute these
powers is to evaluate the eigenvalues of the matrices first. All
powers of the matrix are then just powers of the eigenvalues values. In this
case one can use for example the conjugate gradient algorithm, which
yields the eigenvalues up to a required precision.  Unfortunately with
this method we needed more matrix multiplications for a reasonable
precision than with the direct multiplication.

Therefore we decided the use a slightly modified direct multiplication
of the matrices. The aim of the modification was to use more than one
computer to calculate the powers of a given matrix without wasting too
much computer time.
\footnote{Note that we were not computing on a parallel machine.}
Assume that we want to compute ${U}_\mu^{100}$. To accelerate the
calculation we want to use five computers, where to first computes
${U}_\mu^{2}$ to ${U}_\mu^{20}$ the second ${U}_\mu^{21}$
to ${U}_\mu^{40}$ and so on. This is of course only faster than
using one computer if we manage to calculate the starting power for
each computer in a fast way.

For this we write the starting exponent in binary form. For example 
for the exponent 60
\begin{equation}
  \label{eq:num-exp}
  60 = 1\!\times\! 2^5 + 1\!\times\! 2^4 + 1\!\times\! 2^3 + 1\!\times\! 2^2 + 0\!\times\! 2^1 + 0\!\times\! 2^0\,.
\end{equation}
In the next step these different powers of ${U}_\mu$ are computed by
taking repeatedly the square of the matrix. In the above example these
are 5 matrix multiplications. The final result we get by multiplying
the powers which have a non--vanishing coefficient, here $4$
multiplications, which means that we computed ${U}_\mu^{60}$ with $9$
matrix multiplications.

After the starting power of ${U}_\mu$ is generated the following
powers are computed via direct multiplication. With this method it
took approximately two months to evaluate the configurations in Table
\ref{tab:num-stat-tek}.

\subsubsection{Tests of the algorithm}

In this model there exists a very stringent test of the implementation
of the algorithm, namely the planar limit discussed in Section
\ref{sec:TEK-cont}. In this limit there is an exact solution in
the large $N$ limit \cite{Gross:1980he}. We reproduced these results
in the planar limit with our code, confirming the validity of our
investigations. In fact the program we used was only slightly modified
compared to Ref.\ \cite{Nakajima:1998vj}. There the planar limit of the
TEK model was studied, though, with a different twist and with even
$N$, also reproducing the analytic solution of the planar limit.

\section{Details of the $\lambda\phi^4$ model simulation}
\label{sec:numerics-2}

For the scalar model we chose the standard {\em Metropolis algorithm}
\cite{Bhanot:1988sf}. The idea of this algorithm is to suggest a new
configuration, and accept this configuration with the probability
\begin{equation}
  \label{eq:monte-metropolis}
  P=\min(1,\exp(-(S_\text{new}-S_\text{old})))\,.
\end{equation}
If the action of the new configuration is smaller than the
action of the old configuration it is accepted in any case, vice versa
the new configuration is accepted with the probability $\exp(-\Delta
S)$. The explicit update was
performed by choosing an element of the Hermitian matrices
$\hat{\phi}(t)$ according to
\begin{equation}
  \label{eq:monte-phi-update}
  \hat{\phi}_{ij}(t)\to\hat{\phi}_{ij}'(t)=\hat{\phi}_{ij}(t)+\varepsilon\,\eta\andx 
  \hat{\phi}_{ij}(t) = \hat{\phi}^*_{ji}(t)\,.
\end{equation}
Here $\eta$ is a random complex number with $\langle \eta\rangle=0$
and $\varepsilon\in\mathbb{R}^+$ is an overall scale of the suggested
changes, which has to be fixed.  For every pair of complex conjugated
elements we perform the accept--reject step
(\ref{eq:monte-metropolis}).

The parameter $\varepsilon$ should be chosen in a way to get a
reasonable acceptance rate. If the acceptance rate is close to 1, the
generated configurations are strongly correlated. If it is close to
zero hardly any configurations are generated. A reasonable value would
be $0.5$, so one might try to fix $\varepsilon$ to a value providing
this acceptance rate.

However, in this study we explore a phase diagram and the optimal value
of $\varepsilon$ might depend on the different phases or it changes
close to the phase transition. Therefore we decided to implement a
{\em dynamical} value $\varepsilon$: after updating the complete lattice we
measure the acceptance rate. If the acceptance rate is above $0.6$ we
increase $\varepsilon$ by $\varepsilon/5$, if the acceptance rate
is below $0.3$ we decrease $\varepsilon$ by $\varepsilon/5$.
\footnote{Note that oscillations between two values of $\varepsilon$
  are excluded with this choice.}

We choose the real and the imaginary part of $\hat{\phi}_{ij}$ from the interval
$[-0.5,0.5]$ times $\varepsilon$.  A rea\-sonable starting value of
$\varepsilon$ is then $\varepsilon=N^{-1/4}$. In most cases
$\varepsilon$ stabilized after $3-10$ update steps, depending on the
coupling $\lambda$, at a value close to the starting value.  However,
at very large values of the coupling $\varepsilon$ stabilized at a
rather small value one order
of magnitude smaller than $N^{-1/4}$, see below.\\

\begin{figure}[htbp]
  \centering
  \includegraphics[width=.65\linewidth]{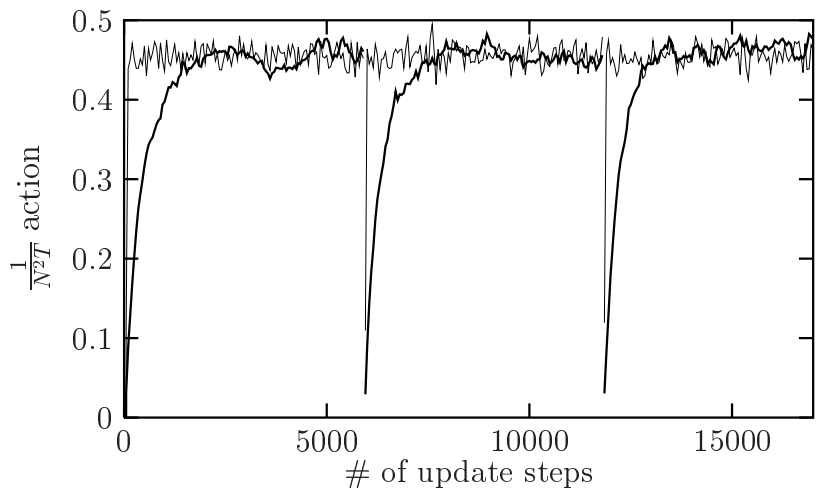}
  \caption{The action is plotted against the number of update steps close 
    to the disordered -- ordered phase transition.  The solid line
    shows the result when updating one complete matrix $\hat{\phi}(t)$
    at once, the thin line shows the result when updating every
    independent matrix element separately.}
  \label{fig:monte-therm}
\end{figure}

The update of every independent matrix element of $\hat{\phi}(t)$ is
of course more expensive than updating each matrix $\hat{\phi}(t)$ at
once. However, it improves thermalization as well as ergodicity
properties. In Figure \ref{fig:monte-therm} we show the history of the
action for three different starting configurations.
For the history shown by the thin line we used the update procedure
(\ref{eq:monte-phi-update}) and the thick lines represent the result
when we update each matrix at once. In both methods we used a
dynamical parameter $\varepsilon$. First of all we see that the system
thermalize much faster when updating the elements separately. This
already compensates the additional computer time needed for this
algorithm. We also see that with the procedure, where we update one
matrix at once, the action fluctuates much less than with the update
procedure (\ref{eq:monte-phi-update}). This means that we hardly
proceed in configuration space and the configurations are strongly
correlated.

The reason for this is that if we change a complete matrix the change
of the action is rather large and we get a reasonable acceptance rate
only if we decrease the overall factor $\varepsilon$ to a tiny value.
The configuration is hardly changed within an update step.

In practice we also measured $\varepsilon$ as an technical parameter.
It depends only slightly on $m^2$, but it strongly depends on the
coupling $\lambda$. For very large couplings we have again an
ergodicity problem. The values of $\lambda$ discussed in Chapter
\ref{phi} can be considered as the upper limit for the
algorithm described here.\\

Besides the qualitative structure of the phase diagram, no details
about the phases structure of this theory were known. As we expect a
striped phase, with general patterns, we are possibly dealing in this
phase with meta--stable states, which are usually difficult to
identify.

Therefore we {\em always} started from six independent (random)
starting configurations and initialized the random number generator
differently, i.\ e.\ we made for any set of $N,\lambda,m^2$ six
independent simulations. This is of course much more expensive than
starting only from one configuration, because we have to thermalize
all six runs. It has several benefits, however. If there are meta--stable
states, they certainly depend on the starting conditions.  With
different starting conditions we should therefore see, in that case,
different vacua. This could indicate that we should apply more
thermalization steps. On the other hand, if we thermalize the system
further, this shows that the algorithm is not ergodic, and we are
stuck in a subset of configuration space.

In fact, we saw these different meta--stable states, indicated by
different patterns, at some stage of the simulations. At moderate
values of $\lambda$, applying more thermalization steps all starting
conditions lead to the same results. At very large $\lambda$ this
tunneling from meta--stable patterns to the stable pattern occurred
only close to the ordered -- disordered phase transition (see Section
\ref{sec:phi-stripe-rev}).

A second advantage is that it clarifies the error estimation. As
discussed in the last Section, the number of bins influences the
statistical error in the binning as well as in the jack--knife method.
In our case there is a natural choice of bins. Since we generate six
statistically independent sets of configurations, we also chose six
bins for the error analysis.\\

Table \ref{tab:num-stat-phi} shows the number of configurations we
generated with the algorithm described above. The necessary statistics
to study the phase diagram is relatively small; this is shown in
the second row of table \ref{tab:num-stat-phi}.
\begin{table}[htbp]
  \centering
  \begin{tabular}{|c|c|c|c|c|c|}
    \hline $N$&&15&25&35&45\\\hline
    \multirow{2}{35mm}{\# of configurations}&phase diagram&300&180&60&30\\\cline{2-6}
    &2--point functions&10000&10000&5000&3000\\\hline
  \end{tabular}
  \caption{Overview of the statistics in our simulation of the 3d $\lambda\phi^4$ model.}
  \label{tab:num-stat-phi}
\end{table}
To compute the 2--point functions, especially those in momentum space,
we needed significantly more configurations to obtain small
statistical errors. At some values of the parameters we performed
high statistic simulations where we doubled the number of
configurations. The autocorrelation time in this simulation was in the
range $10$ -- $90$ update steps, depending on the observable and on
the point in the phase diagram under consideration. Therefore we
measured the observables from configurations separated by $100$ update
steps.

It took approximately 10 month on $30$ PCs (with a clock rate in the
range $450$ -- $1500$ MHz) to generate the configurations in Table
\ref{tab:num-stat-phi}. The evaluation of the configuration did not
increase the overall needed computer time, since we did not need high
powers of the $N\times N$ matrices $\hat{\phi}(t)$. Most of the
(usually expensive) matrix multiplications are in this evaluation
multiplications with the constant twist eaters $\Gamma_\mu$ (see
Chapter \ref{phi}). These multiplication can be performed by shifting
or modifying the phases of matrix elements of $\hat{\phi}$, which is
much faster than a direct multiplication.

\subsubsection{Tests of the algorithm}

To test the algorithm is here more sophisticated than in the TEK
model, since no exact solution is known in this case. Therefore we
computed the expectation value of the action $S$ in equation
(\ref{eq:phi-action}) to the first order in $\lambda$\,,
\begin{equation}
  \label{eq:monte-expand}
  \frac{1}{N^2T}\langle S\rangle= c_0(m)+\lambda\, c_1(m)+O(\lambda^2)\,.
\end{equation}
For small $\lambda$ our algorithm should reproduce this result. We
give now a brief account on how the coefficients $c_0$ and $c_1$ were
obtained \cite{Nishimura:2002}.

To compute $c_0$ we used the fact that the partition function is
invariant under a change of variables
$\hat{\phi}_{ij}(t)\to(1+\epsilon)\hat{\phi}_{ij}(t)=\hat{\phi}_{ij}'(t)$.
The integration measure $\mathcal{D}\hat{\phi}$ and the action
$\takenat{S[\hat{\phi}]}{\lambda=0}$ transform under this shift like
\begin{equation}
  \label{eq:num-test1}
  \mathcal{D}\hat{\phi}'=(1+\epsilon)^{N^2T}\mathcal{D}\hat{\phi}\andx
  S[\hat{\phi}']=(1+\epsilon)^2S[\hat{\phi}]\,,
\end{equation}
where the transformation of the action is only valid in first order
in $\lambda$. Since the partition function is invariant under this
transformation we obtain
\begin{equation}
  \label{eq:num-test2}
  \begin{split}
    \int\mathcal{D}\hat{\phi}\,e^{-S[\hat{\phi}]}&=\int\mathcal{D}\hat{\phi}'\,e^{-S[\hat{\phi}']}\\
  &=\int\mathcal{D}\hat{\phi}\,e^{-S[\hat{\phi}]}+
  \epsilon\lry{N^2T\!\int\mathcal{D}\hat{\phi}\,e^{-S[\hat{\phi}]}
    -2\int\mathcal{D}\hat{\phi}\,S[\hat{\phi}]\,e^{-S[\hat{\phi}]}}+O(\epsilon^2)\,.
  \end{split}
\end{equation}
To satisfy this equation the term in square brackets has to vanish.
This amounts to the condition
\begin{equation}
  \label{eq:num-test3}
  \frac{1}{N^2T}\langle S\rangle=0.5=c_0\,.
\end{equation}
Note that the result for $c_0$ is independent of $N,T$ and $m^2$.\\

For the determination of $c_1$ we extend the above consideration to
the interacting case $\lambda\neq0$. This leads to the identity
\begin{equation}
  \label{eq:num-test4}
  \begin{split}
      \frac{1}{N^2T}\bigl(\langle S_\text{free}\rangle+2\langle S_\text{interaction}\rangle\bigl)&=0.5\\
      \frac{1}{N^2T}\bigl(\langle S\rangle+\langle S_\text{interaction}\rangle\bigl)&=0.5\,.
  \end{split}
\end{equation}
Therefore $c_1$ is given by
\begin{equation}
  \label{eq:num-test5}
  c_1=-\frac{1}{4NT}\left\langle\sum_t\trace\,\hat{\phi}(t)^4\right\rangle\,.
\end{equation}
This expectation value can be calculated by transforming the partition
function into momentum space, with the discrete Fourier transformation
defined in equation (\ref{eq:lat-J3}) and using Wick's theorem to
evaluate the resulting expectation value.  The equation for the
coefficient $c_1$ is now given by a sum over momenta. The term that is
summed over is very lengthy and complex, therefore we will not show
the explicit result. However, for given parameters $N,T$ and $m^2$ it
can be computed for example with the program {\tt mathematica}.\\

\begin{figure}[htbp]
  \centering
  \includegraphics[width=.85\linewidth]{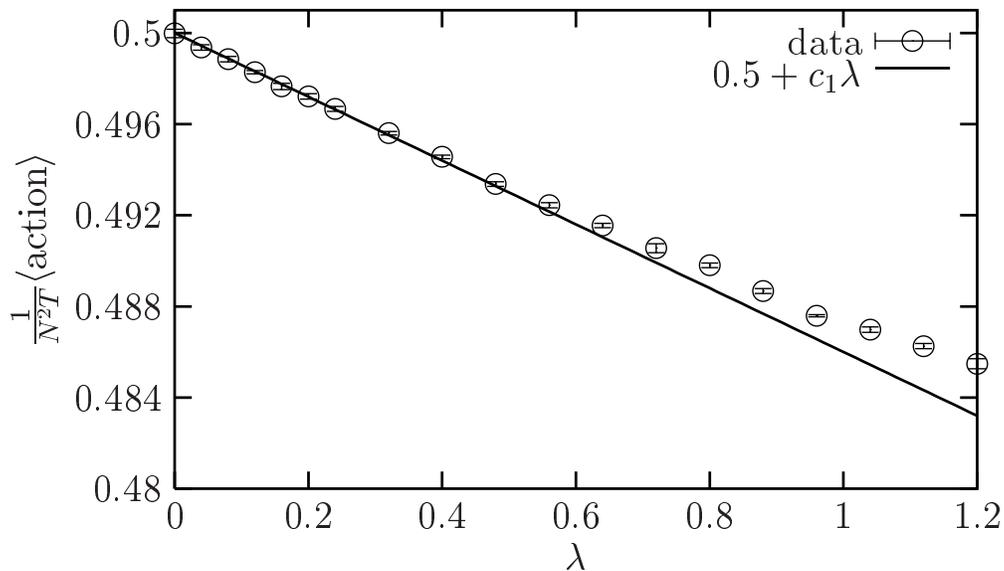}
  \label{fig:monte-expand}
  \caption{The expectation value of the action against $\lambda$, at $N=25$ and $m^2=1$. The circles
    show the results of the simulation, the straight line is the
    expectation value action according to a first order expansion in
    $\lambda$.}
\end{figure}

To test if our algorithm provides this result, we computed the action
at $N=25$ and $m^2=1$. The coefficient $c_0$ is always given by
$c_0=0.5$. For the chosen value of $m^2=1$ we find $c_1=-0.0146$.
Figure \ref{fig:monte-expand} clearly confirms that our program
reproduces this result.


\end{appendix}


\cleardoublepage

\fancyhead{}
\fancyhead[LE,RO]{\bfseries\thepage}
\fancyhead[LO]{\bfseries Bibliography}
\fancyhead[RE]



\end{document}